\documentclass[11pt,a4paper]{article}

\usepackage{amssymb}
\usepackage[latin1]{inputenc}
\usepackage[pdftex]{hyperref}
\usepackage{amsmath}

\newtheorem{thm}{Theorem}[section]
\newtheorem{lem}{Lemma}[section]
\newtheorem{prop}{Proposition}[section]
\newtheorem{cor}{Corollary}[section]

\newtheorem{defn}{Definition}[section]

\newcommand{\T}{{\rm Tr}}
\newcommand{\cA}{{\cal A}}
\newcommand{\cB}{{\cal B}}
\newcommand{\cH}{{\cal H}}
\newcommand{\cM}{{\cal M}}
\newcommand{\cL}{{\cal L}}

\newcommand{\cSU}{{\cal SU}}
\newcommand{\cU}{{\cal U}}

\newcommand{\1}{{\mathbf 1}}
\newcommand{\rc}{{\rm anti}}
\newcommand{\kA}{{\textsc A}}
\newcommand{\kB}{{\textsc B}}
\newcommand{\kC}{{\textsc C}}

\title{Anti- (Conjugate) Linearity}

\author{Armin Uhlmann}

\date{University of Leipzig, Institute for
Theoretical Physics\\
Germany, D-04009 Leipzig, PB 100920}

\begin{document}

\maketitle

key words: operators, canonical form, antilinear
(skew) hermiticity, acq--lines
\medskip

PACS-codes: 02.30 Tb, 03.10 Ud, 03.65 Fd, 03.65 Ud
\medskip

\begin{abstract} This is an introduction to antilinear
operators.
In following E.~P.~Wigner the terminus ``antilinear'' is used
as it is standard in Physics. Mathematicians prefer to say
``conjugate linear''.

By restricting to finite-dimensional complex-linear spaces, the
exposition becomes elementary in the functional analytic sense.
Nevertheless it shows the amazing differences to the linear case.

Basics of antilinearity is explained in sections 2, 3, 4, 7 and
in subsection 1.2: Spectrum, canonical Hermitian form, antilinear
rank one and two operators, the Hermitian adjoint, classification
of antilinear normal operators, (skew) conjugations, involutions,
and acq--lines, the antilinear counterparts of 1--parameter
operator groups.
Applications include the representation of the Lagrangian
Grassmannian by conjugations, its covering by acq--lines. as well
as results on equivalence relations. After remembering elementary
Tomita--Takesaki theory, antilinear maps, associated to a vector
of a two-partite quantum system, are defined. By allowing to write
modular objects as twisted products of pairs of them, they open
some new ways to express EPR and teleportation tasks.
The appendix presents a look onto the rich structure of antilinear
operator spaces.\footnote{SCIENCE CHINA Physics, Mechanics $\&$
Astronomy, {\bf 59}, 3, (2016) 630301; doi: 10.1007/s11433-015-5777-1}
 \footnote{Sincere thanks to ShaoMing Fei (Associate editor)
 for his friendly invitation and to the editorial staff
 for their patient help.}
\end{abstract}

\tableofcontents

\section{Introduction} \label{C1}
The topic of this manuscript is the phenomenon of antilinearity.
It is partly an introduction and partly a collection of examples
to show its use.

An operator $\vartheta$ acting on a complex-linear space is
called ``antilinear'' or ``conjugate linear'' if for any two
vectors $\phi_1$, $\phi_2$, and complex numbers $c_1$, $c_2$,
\begin{displaymath}
\vartheta \, (c_1 \phi_1 + c_2 \phi_2) =
c_1^{*} \vartheta \, \phi_1 + c_2^{*} \vartheta \, \phi_2
\end{displaymath}
is valid.
In Mathematics the term ``conjugate linearity'' is preferred
while in Physics the notation ``antilinearity'', as advocated
by E.~P.~Wigner, is in use. I follow the latter convention. See
also the remarks on notation below.
\medskip

I restrict myself to complex-linear spaces of finite dimension
mostly, starting in section \ref{C2} with some basic definitions
and facts. Remarkably, though quite trivial, the eigenvalues of
antilinear operators form circles centered at zero in the complex
plane. Therefore the trace and other symmetric functions of
the eigenvalues are undefined for antilinear operators. As a
"compensation" $\T \, \vartheta_1 \vartheta_2$ is an
{\em Hermitian form} with signature equal to the
dimension of the linear space on which the operators act.

While the set of linear operators is naturally an
algebra, say $\cB(\cL)$, the antilinear operators form a
$\cB(\cL)$--bimodule $\cB(\cL)_{\rc}$. Hence their direct
sum is a 2-graded algebra. An active domain of research are
operator functions, for instance power series of elements of
this algebra, see M.~Huhtanen and A.~Per\H am\H aki in
\cite{HP12, HP12a} for example. Please notice: This rich
theory is \underline{not} under consideration here.
\medskip

Beginning with section \ref{C3} the basic spaces are equipped
with scalar products making them Hilbert spaces $\cH$. As usual,
$\cB(\cH)$ denotes the algebra of linear operators. The linear
space of antilinear operators will be called $\cB(\cH)_{\rc}$.
The scalar product allows the definition of the Hermitian
adjoint $\vartheta^{\dag}$ of any antilinear operator $\vartheta$.
It is a linear operation.

Given a scalar product it becomes routine to define Hermitian
(self-adjoint), skew Hermitian, unitary, and normal antilinear
operators, including conjugations and skew conjugations.
\medskip

There is a strong connection between matrices and antilinear
operators. While matrix analysis is operator theory with a
distinguished basis, antilinear operators allow to formulate
several parts of it in a transparent and basis independent way.
The said connection is mediated by a basis
$\phi_1, \dots, \phi_d$ of $\cH$.
Depending on the chosen basis, one associates to a matrix
$a_{jk}$ an antilinear operator according to
\begin{displaymath}
\sum_j c_j \phi_j \rightarrow \sum_{jk} c_j^{*} a_{kj} \phi_k
\end{displaymath}
From a symmetric (skew symmetric) matrix one gets an antilinear
Hermitian (skew Hermitian) operator. The said operator is (skew)
Hermitian in every basis of $\cH$.

It seems tempting to translate theorems concerning symmetric
or skew symmetric matrices into the language of antilinearity.
For example, the question whether a matrix is unitary equivalent
to its transpose can be "translated" into: Is the Hermitian adjoint
$X^{\dag}$ antiunitarily equivalent to $X$ ? Some cases are
reported in section \ref{C6}.
\medskip

In exploring properties of classes of antilinear operators,
the finiteness assumption renders a lot of sophisticated
functional analysis to triviality. On the other side it makes
it much simpler to grasp the ideas coming with antilinearity!
That is particular transparent in E.~P.~Wigners classification
of antiunitary operators, \cite{Wi60}, and in its extension to
antilinear normal operators, \cite{HV66}, by F.~Herbut and
M.~Vuji{\v c}i{\' c} discussed in section \ref{C4}.
\medskip

Section \ref{C5} points to the role of antilinearity in
symplectic geometry. The maximal real Hilbert subspaces of
$\cH$ are the Lagrangian subspaces and, therefore, the points
of the Lagrangian Grassmannian $\Lambda$. The bridge to
antilinearity: Every Lagrangian subspace is the fix-point set of
a conjugation and vice versa: The manifold of all conjugations
is simplectomorphic to $\Lambda$. It will be shown how the Maslov
index of a closed curve in $\Lambda$ can be expressed by the
help of an operator-valued differential 1-form defined on the
manifold of all conjugations.
\medskip

In section \ref{C6}, as already said, some equivalence relations
are considered. There is a rich literature concerning complex
symmetric matrices. Important sources are Textbooks like
\cite{Horn90} by R.~A.~Horn and C.~R.~Johnson, and newer research
papers by L.~Balayan, S.~R.~Garcia, D.~E.~Poore, E.~Prodan,
M.~Putinar, J.~E.~Tener, and others. Only a few, hopefully typical
ones, of their results are reported. In addition I call attention
to a class of superoperators to point to some related problems.
\medskip

The definition of involutions does not depend on a scalar
product, see section \ref{C7}. But if there is one, their polar
decompositions is of interest.

The Hermitian adjoint is an involution within the space of linear
operators. Varying the scalar product varies the Hermitian
adjoint. The relations between them reflect nicely the geometric
mean of S.~L.~Woronowicz between different scalar products.
\medskip

There are well known theories in which antilinear operators play
an important role and in which the finiteness assumption is not
appropriate. To one of them, the handling of time reversal
operators, a few words will be said below. Another one is the
famous Tomita--Takesaki theory, see section \ref{C8}.
 The finite dimensional case
allows for applications to the {\em Einstein--Podolski--Rosen
effect} \cite{EPR35} as explained in section \ref{C9},

Some peculiar features of ``quantum teleportation'' are described
in section \ref{C10}. A first promise of quantum teleportation is
already in \cite{BeWi92}. The very origin of all the matter is in
the paper \cite{BBCJPW93} by C.~Bennett, G.~Brassard, C.~Crepeau,
R.~Jozsa, A.~Peres, and W.~Wootters.

The purpose of the appendix is to point at antilinear operator
spaces.
\medskip

As already said, the paper remains within finite dimensional
Hilbert and other linear spaces. But even within this strong
restriction, only a selected part of the theory could be exposed.

Concerning (skew) conjugations and their applications, see also
the review \cite{GPP14} of S.~R.~Garcia, E.~Prodan and M.~Putinar.
For much of their work there is a natural counterpart
within the language of antilinearity.
\medskip

There seems to be no monograph dedicated essentially to conjugate
linear, i.~e. antilinear operators and maps in finite dimensional
complex-linear spaces. As already stressed, there are important parts
of Linear Algebra with a ``hidden'' antilinearity. The treatment
of symmetric matrices, for example, can be ``translated'' into
that of antilinear Hermitian operators. The transpose of a matrix
can be viewed as the transform of its Hermitian adjoint by a
conjugation, and so on. These and some other examples are
explained in the main text. To reach completeness could by no
means be the aim.

\subsection{Time reversal operations and beyond}  \label{C1.1}
The first explicit use of antilinearity in Physics is in Wigner's
1932 paper {"{\"U}ber die Operation der Zeitumkehr in der
Quantenmechanik}" \cite{Wig32}. It describes the prominent role of
antilinearity in time reversal operations. In Mathematics the
idea of conjugate linearity goes probably back to E.~Cartan,
\cite{Ca31}. The time--reversal symmetry has seen a lot of
applications all over Physics \cite{Sa87} and it is well described
in many textbooks. A survey can be found in \cite{AR05}. In what
follows, only a particular feature will be in the focus.

Assume $\psi(\vec{x},t)$ satisfies a Schr\"odinger equation with
real Hamiltonian $H$,
\begin{displaymath}
\mathrm{i} \hbar \frac{\partial \psi}{\partial t} = H \, \psi, \quad
(H \psi)^{*} = H \psi^{*} \, .
\end{displaymath}
For any instant $s$ of time the time reversal operator $T_s$ is
defined by
\begin{displaymath}
(T_s \psi)(\vec{x}, t - s) =
\psi(\vec{x}, s - t)^{*} , \quad s \in \mathbb{R}  \, .
\end{displaymath}
If $\psi$ is a solution of the Schr\"odinger equation, so it is
$T_s \psi$. For physical reasons one cannot avoid antilinearity:
$H$ must be bounded from below. Hence, generally, if $H$ is an
Hamilton operator, $-H$ is not. Computing
\begin{displaymath}
(T_s T_r\,\psi)(\vec{x}, t) = \psi(\vec{x}, t - 2(s-r)) ,
\quad T_s T_r = U(2r - 2s) \; ,
\end{displaymath}
one gets the unitary evolution operator. It shifts any solution
in time by the amount $2(s-r)$. Hence time translation operators
can be written as products of two antilinear operators. This is a
salient general feature: {\em Some physically important
operations can be written as products of two antilinear ones.}

There is a further structure in the game. Using the equation above
one shows
\begin{displaymath}
T_r T_{(r+s)/2} = T_{(r+s)/2} T_s \,
, \quad r,s \in \mathbb{R} \; . \qquad  (*)
\end{displaymath}
To stress the abstraction from time--reversal symmetry, the
last equation is rewritten as
\begin{equation} \label{Ya0}
\vartheta_r \vartheta_{(r+s)/2} = \vartheta_{(r+s)/2} \vartheta_s
\, , \quad r,s \in \mathbb{R} \; .
\end{equation}
{\em Curves $t \to \vartheta_t$ satisfying (\ref{Ya0}) should be
seen as antilinear surrogates of 1-parameter groups.} The
ad hoc notation {\em ``acq-line''} will be used for them: ``ac''
stands for \underline{a}ntilinear \underline{c}onjugate, and
``q'' for \underline{q}uandle. (Linear realizations of (\ref{Ya0})
may be called {\em cq-lines},)

A set $\cM$ of invertible antilinear operators will be called an
``antilinear conjugate quandle'' or an ``ac-quandle'' if,
given $\vartheta, \vartheta' \in \cM$, it follows
$\vartheta^{-1} \vartheta' \vartheta \in \cM$. The sets of all
(skew) conjugations and of all (skew) involutions are examples
of ac-quandles,

These notations follow that of {\em rack, quandle,} and {\em kei,}
see S.~R.~Blackburn \cite{Bl12}. The fruitfulness of quandles and
analogue structures has been discovered in knot theory by D.~Joyce
\cite{Jo82}. See also E.~Nelson \cite{Ne11}.
\medskip

Returning now to an acq--line (\ref{Ya0}). Defining
$\vartheta'_s = \vartheta_{as+b}$, $a,b$ real, $a \neq 0$, one
gets a new acq--line which is a new parameterization of the
original one. Notice the change of orientation if  $a < 0$.
However, the peculiar feature is the invariance of (\ref{Ya0})
by a change $s \to s+b$. Indeed, in contrast to groups, there is
neither an identity nor an otherwise distinguished element in an
acq--line.
\medskip

Using an idea of M.~Hellmund \cite{MH15}, an important class of
acq--lines is gained:
\begin{lem} \label{Hell} [Hellmund]
Let $\vartheta_0$ antilinear, $H$ linear and Hermitian.
If $H$ commutes with $\vartheta_0$, then
\begin{equation} \label{Ya1}
t \to \vartheta_t := U(-t) \vartheta U(t), \quad U(t) = \exp itH
\end{equation}
is an acq--line. Moreover, $B = \vartheta^2_s$ does not depend on
$s$ and commutes with all $U(t)$.
\end{lem}
Proof:  $B = \vartheta_0^2$ is linear, commutes with $H$, and
hence with $U(t)$. By $\vartheta_t^2 = U(-t) \vartheta_0^2 U(t)$
one concludes $B = \vartheta_t^2$ for all $t$. Again by
$\vartheta_0 H = H \vartheta_0$ one obtains
\begin{displaymath}
U(-t) \vartheta_0 U(t)  = U(2t) \vartheta_0 = \vartheta_0 U(-2t).
\end{displaymath}
For two arguments one obtains
\begin{equation} \label{Yb1}
\vartheta_s \vartheta_t = U(2s) B U(-2t) = U(2s - 2t) B \; .
\end{equation}
Substituting either $s\to s$, $t\to (r+s)/2$ or $s \to (r+s)/2$,
$t \to r$, one gets $U(s-r)B$ in both cases. Thus
$t \to \vartheta_t$ fulfills (\ref{Ya0}).
\medskip

A further example: The CPT operators \cite{WWW52}, combinations
of particle--antiparticle conjugation, parity and time reversal.
constitute physically important ac-quandles. CPT acts on bosons
as conjugations and on fermions as skew conjugations.
R.~Jost \cite{RJ65} could prove that CPT operators are genuine
symmetries of any relativistic quantum field theory
satisfying Wightman's axioms.

A CPT operator is defined up to the choice of a point $\mathbf{x}$
in Minkowski space, which is the unique fix point of the map
$\mathbf{x}' - \mathbf{x} \mapsto \mathbf{x} - \mathbf{x}'$, where
$\mathbf{x}'$ runs through all world points. Let
$\Theta_{\mathbf{x}}$ be the CPT-operator with this kinematical
action. Then
\begin{equation} \label{Ya2}
\Theta_{\mathbf{x}} \Theta_{\mathbf{y}} = U(2\mathbf{y}-2\mathbf{x})
\end{equation}
where $U$ denotes the unitary representation of the translation
group as part of representations of the Poincar\'e
group or its covering group. Notice the relation
\begin{equation} \label{Ya3}
\Theta_{\mathbf{x}} \, \Theta_{(\mathbf{x+y})/2} =
\Theta_{(\mathbf{x+y})/2} \, \Theta_{\mathbf{y}} \, .
\end{equation}
It implies that, given two points $\mathbf{x}$ and $\mathbf{y}$,
there is an acq--line $s \mapsto \Theta_s$ with $\Theta_0 =
\Theta_{\mathbf{x}}$ and $\Theta_1 = \Theta_{\mathbf{y}}$.

In the same spirit the ``physical'' representations of
the Poincar{\`e} group, respectively its covering group, can
be gained by products of antilinear CT operations, provided the
theory is CT-symmetric. The geometric part of a CT operations is
a reflection on a space--like hyperplane in Minkowski space.

However, this is just an example within the world of Hermitian
symmetric spaces \cite{Ca36}, their automorphism groups and
Shilov boundaries. \medskip

In subsection \ref{C7.gm} there is a variant of (\ref{Ya0}):
The arithmetic mean is replaced be the geometric one.

\subsection{Choosing notations}
There are some differences in notations in Mathematics and
Physics. This is somewhat unfortunate. I mostly try to follow
notations common in the physical literature and in Quantum
Information Theory. A mathematically trained person should
not get into much troubles by a change in notations anyway.

\noindent {\bf 1.} The complex conjugate of a complex number $c$
is denoted by $c^*$ (and not by $\bar c$ ).

\noindent {\bf 2a.} If not explicitly said all linear spaces are
assumed complex-linear and of finite dimension. Sometimes real
linear spaces become important. These cases will be
explicitly noticed by saying ``real Hilbert space'', ``real
linear space''.

\noindent {\bf 2b.} Given a Hilbert space $\cH$, its scalar
product is written $\langle \phi' , \phi'' \rangle$ with ``Dirac
brackets'' \cite{Dirac30}. It is assumed linear in the {\em second}
argument $\phi''$, and antilinear (conjugate linear) in its first
one $\phi'$. This goes back to Schr\"odinger (1926).

\noindent {\bf 2b.} The symbol $(.,.)$ with arguments in
$\cB(\cL)_{\rc}$ or in $\cB(\cH)_{\rc}$ represents the
canonical form, see subsection \ref{C2.4},

\noindent {\bf 2c.} To avoid dangerous notations like
$\langle \phi | \theta$ with $\theta$ antilinear, I do {\em not
use Dirac's bra convention $\langle \phi|$.}  Schr\"odingers way
of writing elements of Hilbert spaces say, $\phi$ or $\psi$,
will be used mostly. Of course, Dirac's ``ket'' notation, say
$| \phi \rangle$ or $|12\rangle$ does not make any harm as it is
not ``dangerous'' in the sense explained in subsection \ref{C3.1}.

\noindent {\bf 2d.} The expressions
$|\phi_1\rangle \langle \phi_2|$ and
$|\phi_1\rangle \langle \phi_2|_{\rc}$ will be used as entities.
See subsection \ref{C3.2}.
\medskip

\noindent {\bf 3a.} $\cB(\cH)$, $\cB(\cH, \cH')$ stand for the set
of all linear maps from $\cH$ into $\cH$ respectively
$\cH'$. These maps are usually called {\em operators}. Similarly,
$\cB(\cH)_{\rc}$ and $\cB(\cH, \cH')_{\rc}$ denote the sets of
antilinear maps. They are also called {\em antilinear operators.}

\noindent {\bf 3b.} The Hermitian adjoint of an operator $X$,
whether linear or antilinear, will be called $X^{\dag}$
and not $X^*$.

\noindent {\bf 3c.} $X$ is called Hermitian (or self-adjoint)
if $X^{\dag} = X$ and skew Hermitian if $X^{\dag} = - X$.
The latter notation is essential for antilinear operators
as explained in \ref{C3.1}. (Because of $\dim \cH < \infty$
finer functional analytic issues become irrelevant.)

\noindent {\bf 3d.} A linear operator $A$ is called {\em positive
semi-definite} if $\langle \psi, A \psi \rangle$ is real and
non-negative for all $\psi \in \cH$. If no danger of confusion,
such an operator is simply called ``positive''.
\medskip

\noindent {\bf 4.} The terminus technicus ``superoperator'':
A linear map from $\cB(\cH)$ into itself (or from $\cB(\cH)$
into $\cB(\cH')$ will be called {\em superoperator}. Similarly
there are antilinear superoperators. While a linear map from
$\cH$ into itself is usual called ``operator'', the term
``superoperator'' shall remember that they are linear maps
between linear operators. In this picture $\cH$ is the
``floor'', $\cB(\cH)$ the ``first etagere'', while
superoperators live at the ``second etagere''.

Speaking about a superoperator implies another
understanding of positivity: A superoperator $\Phi$ is a
{\em positive} one if $X \geq {\mathbf 0}$ always implies
$\Phi(X) \geq {\mathbf 0}$.

\section{Anti- (or conjugate) linearity} \label{C2}
Here $\cL$ is a complex-linear space without a distinguished
scalar product. Some elementary facts about antilinear
operators are gathered.

If not said otherwise, we always assume $\dim \cL = d < \infty$.
$\dim \cL$ is the dimension of $\cL$ as a complex-linear space.

\subsection{Definition} \label{C2.1}
\begin{defn} \label{defn1}
An operator $\vartheta$ acting on a complex linear space $\cL$
is called {\em antilinear} or, equivalently, {\em conjugate
linear} if it obeys for complex numbers $c_j$ and vectors
$\phi_j \in \cL$ the relation
\begin{equation} \label{rule1}
\vartheta (\, c_1 \phi_1 + c_2 \phi_2 \,) =
c_1^* \vartheta \phi_1 + c_2^* \vartheta \phi_2
, \quad c_j \in \mathbb{C}  \; .
\end{equation}
\end{defn}
Important: {\em An antilinear operator
acts from the left to the right:} Whether $X$ respectively $Y$
is linear or antilinear, $XY \phi := X (Y \phi)$.

The product of $n$ linear and of $m$ antilinear operators
in arbitrary positions is linear for even $m$ and
antilinear for odd $m$.
\medskip

Let $\vartheta$ be antilinear. By setting $H = \1/2$ in lemma
\ref{Hell} one gets the acq-lines
\begin{displaymath}
s \to \vartheta_s := e^{is} \vartheta, \quad
s \to \vartheta_s^{-1} = e^{is} \vartheta^{-1}
\end{displaymath}
provided $\vartheta^{-1}$ does exist in the latter case. Notice
that the set of all invertible antilinear operators is an
ac-quandle.

\subsection{Eigenvalues} \label{C2.2}
A particular case of (\ref{rule1}) is the commutation relation
$c \vartheta = \vartheta c^*$, $c$ a complex number.
Hence, if $\phi$ is an eigenvector of $\vartheta$ with
eigenvalue $a$, one concludes
\begin{displaymath}
\vartheta \phi = a \phi \, \Rightarrow \,
\vartheta z \phi = z^* \vartheta \phi = a z^* \phi
\end{displaymath}
which can be rewritten as
\begin{equation} \label{spec1}
\vartheta \phi = a \phi \, \Rightarrow \,
\vartheta (z \phi) = a \frac{z^*}{z} (z \phi) \; .
\end{equation}
\begin{prop} \label{prop1}
The (non-zero) eigenvalues of an antilinear operator
$\vartheta$ form a set of circles with 0 as their common center.
\end{prop}

The square $\vartheta^2$ of an antilinear operator is linear.
If, as above, $\phi$ is an eigenvector of $\vartheta$ with
eigenvalue $a$, then
\begin{displaymath}
\vartheta^2 \phi = \vartheta (a \phi) = a a^* \phi \; .
\end{displaymath}
\begin{prop} \label{prop2}
If $\phi$ is an eigenvector of $\vartheta$, then the
corresponding eigenvalue of $\vartheta^2$ is real and
not negative, see also \cite{HP12a}
\end{prop}
\begin{cor} \label{cor1}
Let $\vartheta$ be diagonalizable. Then $\vartheta^2$ is
diagonalizable and its eigenvalues are real and not negative.
\end{cor}

As the example below shows, antilinear operator does not
necessarily have eigenvectors.

Because non-zero eigenvalues gather in circles, the unitary
invariants of linear operators are mostly undefined for
antilinear ones. The trace, for example, does not exist
for conjugate linear operators.

\subsubsection{dim $\cL = 2$} \label{C2.2.1}
Let $\dim \cL = 2$ and choose
two linearly independent vectors, $\phi_1$ and $\phi_2$.
Define
\begin{equation} \label{example1.1}
\theta_{\rm F} \, (c_1 \phi_1 + c_2 \phi_2) =
i( c_1^* \phi_2 - c_2^* \phi_1) \; .
\end{equation}
The $i$ is by convention. The index ``F'' is to honor E.~Fermi
for his pioneering work about spin $\frac{1}{2}$ particles.
(\ref{example1.1}) is also called ``spin flip operator''.

From the definition (\ref{example1.1}) one gets
\begin{equation} \label{example1.2}
\theta_{\rm F}^2 = - \1 \; .
\end{equation}
Clearly, the spectrum of $\theta_{\rm F}$ must be empty.

Let $z \neq 0$ be a complex number and $\phi_1, \phi_2$ linear
independent. An interesting set of antilinear operators
is defined by
\begin{equation} \label{example1.3}
\theta_z \, (c_1 \phi_1 + c_2 \phi_2) =
z c_1^* \phi_2 + z^* c_2^* \phi_1 \; .
\end{equation}
For $z = i$ one recovers $\theta_{\rm F}$. Short exercises yield
\begin{equation} \label{example1.4}
\theta_z^2 \, (c_1 \phi_1 + c_2 \phi_2) =
(z^*)^2 c_1 \phi_1 + z^2 c_2 \phi_2
\end{equation}
and
\begin{equation} \label{example1.5}
\theta_{z} \theta_{z^*} = zz^* \1, \quad
\theta_{z}^{-1} = |z|^{-2} \theta_{z^*} \; .
\end{equation}
Assume that $\phi$ is an eigenvector of $\theta_z$ with
eigenvalue $\lambda$. Then $\phi$ is an eigenvector of
$\theta_z^2$ with eigenvalue $|\lambda|^2$. It follows
$z^2 = (z^*)^2 = |\lambda|^2 > 0$, and $z$ must be real.
This allows to conclude
\begin{displaymath}
\theta_z^2 = |z|^2 \1, \quad \theta_z = z \theta_1,
\quad z \in \mathbf{C}
\end{displaymath}
whenever $\theta_z$ possesses an eigenvector. Furthermore, if
$z$ in (\ref{example1.3}) is real, there are eigenvectors,
$\phi_1 \pm \phi_2$ with eigenvalues $\pm z$.
\begin{lem} \label{qbl1}
With a linear basis $\phi_1, \phi_2$ let
$\theta \phi_1 = z \phi_2$ and $\theta \phi_2 = z^* \phi_1$.\\
a) If $z$ is real, then $\theta^2 = z^*2 \1$. $\phi_1 \pm \phi_2$
are eigenvectors with eigenvalues $\pm |z|$.\\
b) If $z$ is not real, then $\theta$ does not possess
an eigenvector.
\end{lem}

\noindent {\bf Remarks} \\
{\bf 1.} Things are perfect if $\cL$ is finite
dimensional. If $\dim \cL = \infty$,
antilinear operators become as sophisticated as in the
linear case, perhaps even more.\\
{\bf 2.} Assume there is a norm $\parallel . \parallel$ defined in
$\cL$. Then the definition of an operator norm extends to the
antilinear case.
\begin{equation} \label{norm1}
\parallel \vartheta \parallel_{op} :=  \sup
\frac{\parallel \vartheta \phi \parallel}{\parallel \phi \parallel}
\end{equation}
is the {\em norm} of $\vartheta$ with respect to the norm given
on $\cL$ or simply the {\em operator norm.} In (\ref{norm1})
$\phi$ runs through $\cL$ with the exception of its zero element.
The absolute values of the eigenvalues of $\vartheta$ are
bounded from above by $\parallel \vartheta \parallel_{op}$.\\
{\bf 3.} The operator norm (\ref{norm1}) mimics the linear
case and it obeys
\begin{equation} \label{norm2}
\parallel R_1 R_2 \parallel_{op} \, \leq \,
\parallel R_1 \parallel_{op} \cdot \parallel R_2 \parallel_{op}
\end{equation}
where $R_1$ and $R_2$ can be chosen linear or antilinear
independently one from another

\subsection{Rank-one operators} \label{C2.3}
Completely similar to the linear case, rank-one operators
can be used as building blocks to represent antilinear
operators.

The {\em rank} of an operator is the dimension of its range
(or output space),
\begin{displaymath}
{\rm rank} \, X = \dim X \cL
\end{displaymath}
whether $X$ is linear or antilinear.

Assume $\vartheta$ is an antilinear rank-one operator with
range generated by $\phi'$. Then there is a linear function
$\phi \mapsto l''(\phi) \in \mathbb{C}$ from $\cL$ into the
complex numbers such that
\begin{equation} \label{Rank1}
\vartheta \, \phi = l''(\phi)^* \phi' \; .
\end{equation}
{\bf Remarks.}\\
{\bf 1.} Let $\langle .,. \rangle$ be a scalar
product\footnote{Remind that $\langle .,. \rangle$ is assumed
antilinear in the first argument} on $\cL$.
Then there is $\phi'' \in \cL$ such that
$l''(\phi) = \langle \phi'' , \phi \rangle$ on $\cL$.
Then (\ref{Rank1}) is equivalent to
\begin{displaymath}
\vartheta \, \phi = \langle \phi , \phi'' \rangle \, \phi'
\end{displaymath}
This operator will be denoted by
$\vert \phi' . \phi'' \vert_{\rc}$ in subsection
\ref{C3.2}.

Similar to the linear case one proves:\\
{\em Let $\vartheta$ be antilinear and of rank $k$. Let
$\phi'_1, \dots, \phi'_k$ be vectors generating
$\vartheta \cL$. Then there is exactly one set of $k$ linear
functionals $l''_1, \dots, l''_k$ such that}
\begin{equation} \label{Rank2}
\vartheta \, \phi = \sum_{j=1}^k l''_j(\phi)^* \phi'_j
, \quad \phi \in \cL \; .
\end{equation}
{\bf 2.} If, as above, a scalar product is given. Then there
are vectors $\phi''_1, \dots, \phi''_k$ such that
\begin{displaymath}
\vartheta \, \phi = \sum_{j=1}^k \langle \phi , \phi''_j \rangle
\, \phi'_j \; .
\end{displaymath}
{\bf 3.}  Comparing (\ref{Rank2}) with the definition
 of $\theta_{\rm F}$, the vectors $\phi_1$
and $\phi_2$ generate $\cL$ by definition. The two linear forms
\begin{displaymath}
l_1(c_1 \phi_1 + c_2 \phi_2) = -i c_2 , \quad
l_2(c_1 \phi_1 + c_2 \phi_2) = i c_1
\end{displaymath}
are such that
$\theta_{\rm F} \phi = l'_1(\phi)^* \phi_1 + l'_2(\phi)^* \phi_2$
 \\
{\bf 4.} Often the use of a bi-orthonormal construction is
appropriate: Choose $d = \dim \cL$ linear independent elements
$\phi_k \in \cL$. There are $d$ linear functions $l^j$ such that
\begin{equation} \label{Rank3}
l^j(\phi_k) = \delta_{j,k}, \quad j,k = 1, \dots, d \; .
\end{equation}
The $d^2$ antilinear rank-one operators
\begin{equation} \label{Rank4}
\vartheta_k^j \, \phi := l^j(\phi)^* \phi_k, \quad
\phi \in \cL
\end{equation}
form a linear basis of the antilinear operators on $\cL$.
Hence, any antilinear operator on $\cL$ can
be written as
\begin{equation} \label{Rank5}
\vartheta \, \phi = \sum_{jk} a_j^k l^j(\phi)^* \phi_k , \quad
\vartheta = \sum_{jk} a_j^k \vartheta_k^j \; .
\end{equation}
The coefficients $a_j^k$ are gained by
\begin{equation} \label{Rank6}
\T \, \vartheta \vartheta^n_m = \sum a^k_j \T \,
\vartheta^j_k \vartheta^n_m = a^n_m   \; .
\end{equation}
Indeed, the trace of $\vartheta^j_k \vartheta^n_m$ is equal
to one for $j=m$, $k=n$ and vanishes otherwise.

Notice also that
\begin{displaymath}
\T \, A = \sum l^j(A \phi_j)
\end{displaymath}
is valid for linear
operators. Indeed, (\ref{Rank4}), (\ref{Rank5}) mimic standard
bi-orthonormal bases as seen from the following remark:\\
{\bf 4.} Given a scalar product on $\cL$.
As shown by {\bf 1}  above, there are $\phi^j$ such that
\begin{displaymath}
l^j(\phi) = \langle \phi^j , \phi \rangle, \quad \phi \in \cL \;.
\end{displaymath}
The $2d$ vectors $\phi_k$, $\phi^j$, are bi-orthonormal:
\begin{displaymath}
 \langle \phi^j , \phi_k \rangle = \delta_{j,k},
\quad j,k = 1, \dots, d \; .
\end{displaymath}

\subsection{The canonical Hermitian form} \label{C2.4}
There is a further important fact. {\em antilinear operators
come naturally with an Hermitian form:}
The product of two antilinear operators is linear. Its trace
\begin{equation} \label{Hform1}
(\vartheta_1 , \vartheta_2) := \T \, \vartheta_2 \vartheta_1
\end{equation}
will be called the {\em canonical Hermitian form,} or just the
{\em canonical form} on the space of antilinear operators.

The canonical form (\ref{Hform1}) is conjugate linear in the
first and linear in the second argument. Remembering
(\ref{Rank4}), it follows
\begin{equation} \label{Hform2}
(\vartheta_k^j , \vartheta_n^m)^* =
(\vartheta_n^m , \vartheta_k^j) \; .
\end{equation}
Because every antilinear operator is a linear combination
(\ref{Rank5}) of the $\vartheta_k^j$, the canonical form
(\ref{Hform1}) is Hermitian:
\begin{equation} \label{Hform3}
(\vartheta_1 , \vartheta_2)^* = (\vartheta_2 , \vartheta_1) \; .
\end{equation}
\begin{thm} \label{thrCan}
(\ref{Hform1}) is Hermitian, non-degenerate, and of signature
$\dim \cH$.\\
If $S$ is linear and invertible, respectively
$\vartheta$ invertible and antilinear, then
\begin{equation} \label{HformGL}
(\vartheta_1 , \vartheta_2) =
(S^{-1} \vartheta_1 S, S^{-1} \vartheta_2 S) =
(\vartheta^{-1} \vartheta_2 \vartheta,
\vartheta^{-1} \vartheta_1 \vartheta) \; .
\end{equation}
\end{thm}
\underline{Proof:} That (\ref{Hform1}) is Hermitian
has already be shown. Its symmetry properties: As
$\vartheta_2 \vartheta_1$ is a linear operator,
\begin{displaymath}
\T \, \vartheta_2 \vartheta_1 =
\T \, S \vartheta_2 \vartheta_1 S^{-1} =
\T \, (S \vartheta_2 S^{-1})(S \vartheta_1 S^{-1})
\end{displaymath}
is true. The case of an antilinear $\vartheta$ is similar.

To prove non-degeneracy and the asserted signature of
(\ref{Hform1}) a linear basis of the space of antilinear
operators is constructed by using (\ref{Rank3}) and
(\ref{Rank4}): Let $j,k = 1, \dots, d$ under the
condition $j < k$. The elements of the desired basis are
\begin{equation} \label{Hform4}
\vartheta_k^k , \quad \frac{\vartheta_j^k +
\vartheta_k^j}{\sqrt{2}} , \quad
\frac{\vartheta_j^k - \vartheta_k^j}{\sqrt{2}} \; .
\end{equation}
Representing a general antilinear operator by that basis,
\begin{equation} \label{Hform5a}
\vartheta = \sum_k a_{k} \vartheta_k^k
+ \sum_{j<k} a_{jk} \frac{\vartheta_j^k + \vartheta_k^j}{\sqrt{2}}
+ \sum_{j<k} b_{jk} \frac{\vartheta_j^k - \vartheta_k^j}{\sqrt{2}}
\end{equation}
one obtains by the help of (\ref{Rank6})
\begin{equation} \label{Hform5b}
(\vartheta, \vartheta) = \sum_k |a_k|^2
+ \sum_{j<k} |a_{jk}|^2
- \sum_{j<k} |b_{jk}|^2 \; .
\end{equation}
There are $d(d+1)/2$ positive and
$d(d-1)/2$ negative terms in these representations proving
\begin{equation} \label{Hform6}
\hbox{rank} \, (.,.) = d^2, \quad
\hbox{signature} \, (.,.) = \frac{d(d+1)}{2} - \frac{d(d-1)}{2}
= d \; .
\end{equation}
Now the theorem has been proved.

Converting $(A \vartheta_2, \vartheta_1) =
(\vartheta_1, A \vartheta_2)^*$
into a trace equation gives
\begin{equation} \label{Rank7}
\T \, \vartheta_1 A \vartheta_2 =
( \T \, A \vartheta_2 \vartheta_1 )^{\dag} \; .
\end{equation}
{\bf Remarks}\\
{\bf 1.} According to the theorem, there are decompositions of
the space of antilinear operators into a direct sum of two
subspaces, one with dimension $d(d+1)/2$ and one with
dimension $d(d-1)/2$. Choosing such a decomposition,
the one with the larger dimension becomes a
Hilbert space. The other one is a Hilbert space with respect
to $-(. , .)$.\\
{\bf 2.} The canonical form can be regarded as ``antilinear
polarization'' of the trace,
\begin{displaymath}
\T \, A  \quad \to \quad \T \, \vartheta_1 \vartheta_2 \; .
\end{displaymath}
It is tempting to apply this idea to any symmetric function
of the characteristic values. Examples are
$\det \, \vartheta_1 \vartheta_2$ and
\begin{displaymath}
\T \, A^2 \, \to \, \T \, A_1 A_2  \quad \to \quad \T \,
\vartheta_1 \vartheta_2 \vartheta_3 \vartheta_4 \; .
\end{displaymath}
There seems to be no systematic studies of these problems.

\subsection{Pauli operators and their antilinear partners} \label{C2.5.1}
The way from matrices to operators is mediated by
choosing a general basis of $\cL$. A particular simple
and nevertheless important example comes with $\dim \cL =2$
to which we now stick. We choose two
linear independent vectors $\phi_1$ and $\phi_2$ as the
preferred basis. With respect to the chosen basis,
{\em Pauli operators} are defined by
\begin{eqnarray} \label{Pauli1}
\sigma_1 (c_1 \phi_1 + c_2 \phi_2) & := & c_1 \phi_2 + c_2 \phi_1
 , \\
 \sigma_2 (c_1 \phi_1 + c_2 \phi_2) & := & i c_1\phi_2 - i c_2 \phi_1
 , \\
\sigma_3 (c_1 \phi_1 + c_2 \phi_2) & := & c_1 \phi_1 - c_2 \phi_2
\; .
\end{eqnarray}
They are related by
\begin{displaymath}
\sigma_j \sigma_k + \sigma_k \sigma_j = 2 \delta_{j,k} \1 ,
\quad \sigma_1 \sigma_2 \sigma_3 = i \1 \; ,
\end{displaymath}
.
These Pauli operators have antilinear counterparts:
\begin{eqnarray} \label{Pauli2}
\tau_0 (c_1 \phi_1 + c_2 \phi_2) & := &
c_1^*\phi_2 - c_2^* \phi_1 , \\
\tau_1 (c_1 \phi_1 + c_2 \phi_2) & := &
- c_1^*\phi_1 + c_2^* \phi_2  , \\
\tau_2 (c_1 \phi_1 + c_2 \phi_2) & := &
i c_1^*\phi_1 + i c_2^* \phi_2 , \\
\tau_3 (c_1 \phi_1 + c_2 \phi_2) & := &
c_1^*\phi_2 + c_2^* \phi_1 \; .
\end{eqnarray}
It is $\theta_{\rm F} = i \tau_0$ by (\ref{example1.1}).
The spectrum of $\tau_0$ is empty. The spectrum of $\tau_j$,
$j=1,2,3$, is a doubly covered circle of radius $1$. The
eigenvectors of $\tau_1$ and of $\tau_2 $ are $z \phi_1$ and
$z \phi_2$ with $z \in \mathbb{C}$, $z \neq 0$.
The eigenvectors of $\tau_3$ are the multiples of the two
vectors $\phi_1 \pm \phi_2$.

The four ``Pauli-like'' antilinear operators satisfy some
nice commutation relations. By the help of the matrix
\begin{equation} \label{Pauli3}
\{ g_{jk} \} = \begin{pmatrix} -1 & 0 & 0 & 0 \\ 0 & 1 & 0 & 0 \\
0 & 0 & 1 & 0 \\ 0 & 0 & 0 & 1 \end{pmatrix}
\end{equation}
they  can be written
\begin{equation} \label{Pauli4}
\tau_j \tau_k + \tau_k \tau_j  = 2 g_{jk} \1 ,
\quad  j,k \in \{ 0, 1, 2, 3 \} \; .
\end{equation}
Remark the invariance of these and the following
equations against an ``abelian gauge''
$\tau_m \to (\exp is) \tau_m$, $m = 0,1,2,3$.
The following can be checked explicitly:
\begin{eqnarray}
\tau_1 \tau_2 \tau_3 = - {\mathrm i} \tau_0, &\quad&
\tau_1 \tau_2 = {\mathrm i} \sigma_3 , \label{Pauli5} \\
\tau_2 \tau_3 = {\mathrm i} \sigma_1, & \quad &
\tau_3 \tau_1 = {\mathrm i} \sigma_2 \; ,
\end{eqnarray}
and, for $j = 1,2,3$,
\begin{equation} \label{Pauli6a}
 \tau_0 \sigma_j  =  \tau_j , \quad
\tau_j \tau_0 = \sigma_j   \; .
\end{equation}
From (\ref{Pauli5}), (\ref{Pauli6a}) and $\theta_{\rm F} =
i \tau_0$, see (\ref{example1.1}), one also obtains
\begin{equation} \label{Pauli6b}
\tau_j \tau_k = \sigma_j \sigma_k, \quad
\tau_j \theta_{\rm F} = \theta_{\rm F} \tau_j
\end{equation}
for all $j, k = 1, 2, 3$ .

The trace of $\tau_j \tau_k$ vanishes if $j \neq k$,
and one gets from (\ref{Pauli4})
\begin{equation} \label{Pauli4a}
(\tau_j , \tau_k ) = 2 g_{jk}, \quad j,k \in \{0,1,2,3\} \; .
\end{equation}

\noindent \underline{Example.} Consider a real vector
$\{ p_0, p_1, p_2, p_3\}$. Define
\begin{displaymath}
\tilde p := \sum_0^3 p_j \tau_j \; .
\end{displaymath}
Then
\begin{displaymath}
 (\tilde p)^2 = ( p_1^2 + p_2^2 + p_3^2 - p_0^2 ) \1 \; .
\end{displaymath}
Therefore,  if $\{ p_j \}$ is a space-like vector  with respect
to the Minkowski structure (\ref{Pauli3}), $\tilde p$ can be
diagonalized with (doubly counted) eigenvalues
$(\exp is) \sqrt{p_1^2 +p_2^2 + p_3^2 - p_0^2}$ with real
$s$. If  $\{ p_j \}$ is not space-like,  $\tilde p$ does
not have eigenvectors. $\tilde p$ is nilpotent for light-like
vectors $\{ p_j \}$.

Depending on the sign of $\sqrt{...}$,
One can define antilinear Operators $S_{\pm}$ by
\begin{equation} \label{infex1}
\tilde p =  \sqrt{p_1^2 +p_2^2 + p_3^2 - p_0^2 } \, S_+,
\quad S_+^2 = \1 \; .
\end{equation}
for space-like $\{ p_j \}$. For time-like $\{ p_j \}$ one gets
\begin{equation} \label{infex2}
\tilde p =  \sqrt{p_0^2 - p_1^2 -p_2^2 - p_3^2 } \, S_-,
\quad S_-^2 = - \1 \; .
\end{equation}
Mention the sign ambiguity in the definitions (\ref{infex1}) and
(\ref{infex2}). $S_0$ is undefined. $S_+$ is an involution,
$S_-$ a skew involutions, See section \ref{C7}.

\begin{lem} \label{c2.2.2b}
Let $\dim \cH = 2$ and $\vartheta$ antilinear.
If and only if $\vartheta^2 = \lambda \1_2$ there are
real numbers $x_0, \dots, x_3$ and a unimodular number $\epsilon$
such that
\begin{equation} \label{example2.11}
\vartheta = \epsilon \sum_{j=0}^3 x_j \tau_j \; .
\end{equation}
\end{lem}
Proof: It has been shown already that (\ref{example2.11}) implies
$\vartheta^2 = \lambda \1_2$. This conclusion does not depend on
the choice of $\epsilon$. For the other direction one assumes
general complex numbers $c_j$ instead of the $x_j$ in
(\ref{example2.11}). Then
\begin{equation} \label{lambda1}
\lambda =  (|c_1|^2 + |c_2|^2 + |c_3|^2 - |c_0|^2) \; ,
\end{equation}
$\lambda$ is real, and
\begin{displaymath}
\vartheta^2 = \lambda \1 +
\sum_{j \neq k} c_j c_k^* \tau_j \tau_k \; .
\end{displaymath}
This is equivalent to
\begin{equation} \label{example2.10}
\vartheta^2 = \lambda \1 + \sum_{j \neq k} (c_j c_k^* - c_k c_j^*)
\tau_j \tau_k \, .
\end{equation}
Assuming at first $c_0 \neq 0$, and that $\epsilon$ in
(\ref{example2.11}) has been chosen such that $c_0$ is real.
The right hand side of (\ref{example2.10}) can be written as
a linear combination of $\1$ and the Pauli operators. The
coefficients $z_j$ of the Pauli operators must vanish. Consider
for example $z_3$. To get this coefficient,
$\tau_1 \tau_2 = {\mathrm i} \sigma_3$ and $\tau_0 \tau_3 = - \sigma_3$
can be used, see (\ref{Pauli5}) and (\ref{Pauli6a}):
\begin{displaymath}
z_3 = 2 {\mathrm i} (c_1 c_2^* - c_2 c_1^*) -2(c_0 c_3^* - c_3 c_0^*)
\end{displaymath}
There are purely imaginary numbers within the parentheses.
Therefore $z_3 = 0$ implies $c_0 c_3^* = c_3 c_0$, i.~e.
$c_3$ is real. The same way one proves $c_1$ and $c_3$ real.

If $c_0 = 0$, one may assume $c_1 \neq 0$ and real. Then
$z_3 = 0$ implies $c_1 c_2^* = c_1 c_2$. In the same manner
one shows $c_3$ real.

\subsection{Matrix representation}   \label{C2.5}
There is a bijection between matrices and linear, respectively
antilinear operators, mediated by a fixed linear basis of $\cL$.
This can be done either by an isomorphism or else by an
``anti-isomorphism'' which reverses the direction of actions.
Formally the two possibilities differ by a transposition with
respect to the given basis.

Here the first possibility, the isomorphism, is used throughout!
In doing so, {\em operators have to act always from left to
right.} Hence  $R_1 R_2 \phi := R_1 (R_2 \phi)$.
\medskip

Let $\phi_j$, $j = 1, 2, \dots, d$ denote an arbitrary
linear basis of $\cL$. The are linear functionals $l^j$
satisfying $l^j(\phi_k) = \delta_{j,k}$ as in (\ref{Rank3}).
Every vector $\phi \in \cL$ can be uniquely decomposed
according to
\begin{equation} \label{matrix0a}
\phi = c_1 \phi_1 + \dots + c_d \phi_d \; , \quad
c_j = l^j(\phi) \; .
\end{equation}
To enhance clarity, the coefficients in linear combinations like
(\ref{matrix0a}) are written left of the vectors. In doing so,
the action of a linear or antilinear operator becomes
\begin{equation} \label{matrix0b}
A \, \phi = \sum c_j A \phi_j , \quad
\vartheta \, \phi = \sum c_j^* \vartheta \phi_j \; .
\end{equation}

Let $\{ A \}_{jk}$ and $\{ \vartheta \}_{jk}$ denote the matrices
which are to be associated to the linear operator $A$ and to the
antilinear operator $\vartheta$.  They are \underline{defined} by
\begin{equation} \label{matrixdef}
A \, \phi_j = \sum_k \{ A \}_{kj}  \phi_k, \quad
\vartheta \, \phi_j = \sum_k \{ \vartheta \}_{kj} \phi_k \; .
\end{equation}
It follows
\begin{equation} \label{matrix1c}
A \phi = \sum c_j \{A\}_{kj} \phi_k , \quad
\vartheta \phi = \sum c_j^* \{\vartheta \}_{kj} \phi_k
\end{equation}

To see the mechanism in converting products, consider
\begin{eqnarray}
\vartheta_1 \vartheta_2 \, \phi &=& \vartheta_1
\sum_{jk} c_j^* \{\vartheta_2 \}_{kj} \phi_k
\nonumber \\
 &=& \sum_{jk} c_j \{\vartheta_2 \}_{kj}^* \vartheta_1 \phi_k
\nonumber \\
 &=& \sum_{jkl} c_j
\{\vartheta_2 \}_{kj}^* \{\vartheta_1 \}_{lk}  \phi_l
\nonumber \\
 &=& \sum_{jkl} c_j
\{\vartheta_1 \}_{lk} \{\vartheta_2 \}_{kj}^*  \phi_l
\nonumber \\
 &=& \sum_{jl} c_j \{ \vartheta_1 \vartheta_2 \}_{lj} \phi_l
 \nonumber \; .
\end{eqnarray}
This way one checks: The product of two operators, whether
linear or antilinear, can be reproduced by matrix
multiplication. Indeed, let $A_i$ and $\vartheta_i$ be linear
and antilinear operators respectively. It is\footnote{For
comparison we start with the well known linear case.}
\begin{eqnarray} \label{matrix3a}
\{A_1 A_2\}_{ik} = \sum_j \{A_1\}_{ij} \{A_2\}_{jk},
\quad
\{\vartheta_1 \vartheta_2\}_{ik} =
\sum_j \{\vartheta_1\}_{ij} \{\vartheta_2\}_{jk}^*,
\\
\label{matrix4a}
\{\vartheta_1 A_2\}_{ik} =
\sum_j \{\vartheta_1\}_{ij} \{A_2\}_{jk}^* ,
\quad
\{A_1 \vartheta_2\}_{ik} =
\sum_j \{A_1\}_{ij} \{\vartheta_2\}_{jk}
\; .
\end{eqnarray}

Let us now look at the equations (\ref{matrix0a}),
(\ref{matrix0b}), and (\ref{matrix1c}) from the matrix point
of view. A matrix, say $M$, with matrix entries $m_{jk}$ can
be converted as well into a linear as into an antilinear
operator, provided we have distinguished a basis $\{\phi_j\}$
in advance. The need for a notational rule is obvious.

We propose to distinguish
the two cases by the following {\em notation:}
\begin{equation} \label{matrix1d}
\{M \, \vec{c}\}_k  = \sum_j m_{kj} c_j , \quad
\{M_{\rc} \, \vec{c}\}_k = \sum_j m_{kj} c_j^* \; .
\end{equation}
In these two equations $\vec{c}$ represents the column
vector built from the coefficients $c_j$
in $\phi = \sum c_j \phi_j$.

Given two complex matrices, $M'$ and $M''$, with matrix elements
$M'_{jk}$ and $M''_{jk}$, Accordingly to the rules above we can
perform four different products. The first two below
define linear, the second two antilinear operations. Their
matrix entries are
\begin{eqnarray} \label{matrix3}
(M' M'')_{jk} = \sum_i M'_{ji} M''_{ik},
\quad
(M'_{\rc} M''_{\rc})_{jk} = \sum_i M'_{ji} (M''_{ik})^*
\; , \\
\label{matrix4}
( M'_{\rc} M'' )_{jk} = \sum_i M'_{ji} (M''_{ik})^*,
\quad
(M' M''_{\rc} )_{jk} = \sum_i M'_{ji} M''_{ik}
\end{eqnarray}
Notice that the right hand sides of the equations
(\ref{matrix3}), respectively of (\ref{matrix4}), define
matrices acting linearly, respectively antilinearly,
in the sense of (\ref{matrix1d}).
\medskip

For illustration, a simple example is added:
\begin{displaymath}
\begin{pmatrix} 0 & z^* \\ z & 0 \end{pmatrix}_{\rc}
\begin{pmatrix} c_1 \\  c_2 \end{pmatrix} =
\begin{pmatrix} z^* c_2^* \\  z c_1^* \end{pmatrix}
, \quad
\begin{pmatrix} 0 & z \\ z^* & 0 \end{pmatrix}_{\rc}
\begin{pmatrix} 0 & z^* \\ z & 0 \end{pmatrix}_{\rc}
= z^*z \begin{pmatrix} 1 & 0 \\ 0 & 1 \end{pmatrix} \; .
\end{displaymath}

\section{Antilinearity in Hilbert spaces} \label{C3}
If a finite linear space $\cL$ carries a distinguished positive
definite scalar product $\langle .,. \rangle$ it becomes
a Hilbert space, now denoted by $\cH$.

 A {\em basis} of $\cH$ is a set of vectors, say $\{\phi_1,
 \dots, \phi_d \}$, $\dim \cH = d < \infty$,
satisfying $\langle \phi_j , \phi_k \rangle = \delta_{jk}$.
Sometimes, to definitely distinguish from a general linear basis,
such a basis is called an {\em Hilbert basis.}

The set of all linear operators mapping $\cH$ into itself is
denoted by $\cB(\cH)$. To refer to the linear space of all
antilinear operators from $\cH$ into $\cH$ we write
$\cB(\cH)_{\rc}$. Their dimensions as linear spaces is $d^2$.

For both, $\cB(\cH)$ and $\cB(\cH)_{\rc}$, the dominant news
is the {\em Hermitian adjoint} introduced below. In the
antilinear case it shows essential differences to the linear
operator case.

\subsection{The Hermitian adjoint} \label{C3.1}
The antilinearity requires an extra definition of the
Hermitian adjoint:
\begin{defn}[Wigner] \label{defn2}
The Hermitian adjoint, $\vartheta^{\dag}$, of
$\vartheta \in \cB(\cH)_{\rc}$ is defined by
\begin{equation} \label{rule2}
\langle \phi_1 , \vartheta^{\dag} \,  \phi_2 \rangle  = \langle
\phi_2 , \vartheta \, \phi_1 \rangle, \quad
\phi_1, \phi_2 \in \cH \; .
\end{equation}
\end{defn}
The following conclusions from (\ref{rule2}) are immediate:
\begin{equation} \label{rule3}
(\vartheta^{\dag})^{\dag} = \vartheta, \quad
(\vartheta_1 \vartheta_2)^{\dag} = \vartheta_2^{\dag}
\vartheta_1^{\dag}
\end{equation}
and similarly, with $A$ linear, we get
$(\vartheta A)^{\dag} = A^{\dag} \vartheta^{\dag}$ and
$(A \vartheta)^{\dag} = \vartheta^{\dag} A^{\dag}$.

An important fact is seen by setting $A = c \1$, namely
$(c\vartheta)^{\dag} = \vartheta^{\dag}c^* = c\vartheta^{\dag}$.
\begin{prop} \label{propH}
 $\vartheta \to \vartheta^{\dag}$
is a {\em linear operation,}
\begin{equation} \label{rule4}
\bigl( \sum_j c_j \vartheta_j \bigr)^{\dag} =
\sum c_j \vartheta^{\dag}_j  \; .
\end{equation}
\end{prop}
As with linear operators we notice (respectively define)
\begin{equation} \label{def2}
\vartheta^{\dag}  \vartheta \geq {\mathbf 0}, \quad
|\vartheta| := (\vartheta \vartheta^{\dag})^{1/2}
\geq {\mathbf 0} \; .
\end{equation}
Indeed, $\langle \phi, \vartheta^{\dag} (\vartheta \phi)\rangle
= \langle \vartheta \phi, \vartheta \phi \rangle \geq 0$ by
(\ref{defn2}). Next, a look at
\begin{displaymath}
\T \, \vartheta_2 \vartheta_1 =
[\T \, (\vartheta_2 \vartheta_1)^{\dag}]^*  =
[\T \, (\vartheta_1^{\dag} \vartheta_2|^{\dag})]^* \; .
\end{displaymath}
proves the validity of the relation
\begin{equation} \label{rule6}
(\vartheta_1 , \vartheta_2) =
(\vartheta_1^{\dag}, \vartheta_2^{\dag})
\end{equation}

Main classes of antilinear operators are defined as in the
linear case. However, their properties can be quite different.

An antilinear operator $\vartheta$ is said to be
{\em Hermitian} or {\em self-adjoint}
 if $\vartheta^{\dag} = \vartheta$.\\
$\vartheta$ is said to be {\em skew Hermitian} or
{\em skew self-adjoint} if $\vartheta^{\dag} = - \vartheta$,
\cite{Wi60}; see also \cite{Ru12}.

We denote the set of antilinear Hermitian and the set of
skew Hermitian operators by $\cB(\cH)_{\rc}^{+}$ and
by $\cB(\cH)_{\rc}^{-}$ respectively.

An antilinear $\vartheta$ can be written uniquely as a sum
$\vartheta = \vartheta^{+} + \vartheta^{-}$ of an Hermitian
and a skew Hermitian operator with
\begin{equation} \label{def3}
\vartheta \to \vartheta^{+} :=
\frac{\vartheta + \vartheta^{\dag}}{2}  , \quad
\vartheta \to \vartheta^{-} :=
\frac{\vartheta - \vartheta^{\dag}}{2} \; .
\end{equation}
Relying on (\ref{Hform2}) and (\ref{def3}) one concludes
\begin{equation} \label{rule5}
(\vartheta^{+} , \vartheta^{+}) \geq 0 , \quad
(\vartheta^{-} , \vartheta^{-}) \leq 0 , \quad
(\vartheta^{+} , \vartheta^{-}) = 0 .
\end{equation}
The essential difference to the linear case is caused by
(\ref{rule4}), saying that
{\em taking the Hermitian adjoint is a linear operation.}

In particular, equipped with the canonical form,
$\cB(\cH)_{\rc}^{+}$ becomes a Hilbert space.
Completely analogue, $-(.,.)$ is a positive definite scalar
product on $\cB(\cH)_{\rc}^{-}$.

\begin{prop} \label{prop5}
An antilinear operator $\vartheta$ is Hermitian respectively
skew Hermitian if and only if its matrix $\{\vartheta\}_{jk}$
is symmetric respectively skew symmetric with respect to
any Hilbert basis.
The Hilbert space $\cB(\cH)_{\rc}^{+}$ is of
dimension $d(d+1)/2$, the dimension of
$\cB(\cH)_{\rc}^{-}$ is equal to $d(d-1)/2$.
\end{prop}

The first assertion follows directly from (\ref{rule2}).
Clearly a symmetric (a skew symmetric) matrix depends on exactly
$d(d+1)/2$, respectively $d(d-1)/2$ complex numbers. Because
$\cB(\cH)_{\rc}^{+}$ and $\cB(\cH)_{\rc}^{-}$ are Hilbert spaces
with $(.,.)$ respectively $-(.,.)$, any chosen bases of
them provides antilinear operators satisfying
(\ref{Hform4}), (\ref{Hform5a}, and (\ref{Hform5b}.

\noindent {\bf Remarks:}\\
1.) Do not apply an antilinear
operator to a bra in the usual Dirac manner! By (\ref{rule2})
one gets absurd results: The map
$\langle \phi| \to \vartheta |\phi\rangle$ maps the dual of
$\cH$ linearly onto $\cH$.\\
2.) Notice that
\begin{equation} \label{FvN1}
\vartheta_1, \, \vartheta_2 \quad \Longrightarrow \quad
(\vartheta_2 , \vartheta_1^{\dag})
\end{equation}
is a positive definite scalar product a la {\em Frobenius} and
{\em von Neumann.}\\
3.) Matrix representation: \\
With two matrices, $M'$ and $M''$,
and a given basis a look at (\ref{matrix3}) shows
\begin{equation} \label{matrix4c}
M' (M'')^{\dag} = M'_{\rc} (M''_{\rc})^{\dag} \; .
\end{equation}
In the second expression the $\dag$-operation results in a change
$(M''_{\rc})_{jk} \to (M''_{\rc})_{kj}$. The action of
$M'_{\rc}$ provides the complex conjugation
$(M''_{\rc})_{kj} \to (M''_{\rc})_{kj}^*$ before the matrix $M'$
is multiplied on as in the linear case. Denoting by $M^{\top}$
the transpose of the matrix $M$ on gets similarly
\begin{equation} \label{matrix4d}
(M_{\rc})^{\dag} = (M^{\top})_{\rc} \; .
\end{equation}

4.) As we have seen,
knowledge about Hermitian antilinear Operators
can be translated into properties of symmetric Matrices
and vice versa, just by choosing an appropriate basis.
This procedure depends on that basis. Indeed, with the
exception of the multiples of $\1$, a linear operator cannot
be symmetric (or skew symmetric) in all Hilbert bases.

About symmetric and skew symmetric matrices see
\cite{Horn90} or any other reasonable
book on matrix algebra.
\begin{lem} \label{hdiagonal}
If the antilinear operator $\vartheta$ is Hermitian (self-adjoint)
then there exists a basis of eigenvectors, If $\vartheta$ is
skew Hermitian there does not exist any eigenvector.
\end{lem}
Proof: Let $\vartheta$ be Hermitian. Then $B = \vartheta^2$ is
positive semi-definite.  Let $\lambda \geq 0$ and $\lambda^2$
an eigenvalue of $B$. The space $\cH_{\lambda}$ of all
eigenvectors of $B$ with eigenvalue $\lambda^2$ is
$\vartheta$-invariant, (Because $\vartheta$ and $B$ commute.)
Let $\cH_{\lambda}^{+}$, respectively $\cH_{\lambda}^{-}$, the
real subspaces of $\cH_{\lambda}$, $\lambda > 0$, consisting of
$\vartheta$-eigenvectors with eigenvalues $\lambda$ respectively
$- \lambda$. It is $\cH_{\lambda}^{-} = i \, \cH_{\lambda}^{+}$.
(Because $\vartheta$ is antilinear.) Hence they have the same
real dimensions. Now for any $\phi \in \cH_{\lambda}$ there is a
decomposition $\phi = \phi^{+} + \phi^{-}$ given by
\begin{displaymath}
\phi^{\pm} = \frac{\phi \pm \lambda^{-1} \vartheta \phi}{2}
\in  \cH_{\lambda}^{\pm} \; .
\end{displaymath}
It follows
\begin{displaymath}
\cH_{\lambda} = \cH_{\lambda}^{+} + \cH_{\lambda}^{-},
\quad
\cH_{\lambda}^{-} \cap \cH_{\lambda}^{+} = \{ 0 \} \; ,
\end{displaymath}
and the real dimension of $\cH_{\lambda}^{+}$ is equal to
$\dim \cH_{\lambda}$. Therefore one can choose within every
$\cH_{\lambda}$ a basis of eigenvectors of $\vartheta$. Hence
the first assertion is true. If $\vartheta$ is skew Hermitian
then $\vartheta^2 \leq {\mathbf 0}$. Therefore, any
eigenvector $\phi$ is annihilated by $\vartheta$,
$\vartheta \, \phi = 0$.

Another way to prove the existence of a basis of eigenvectors
for any Hermitian antilinear operators is in showing: Every
$\vartheta$-irreducible subspace is 1-dimensional,
see subsection \ref{C4}.

\begin{prop} \label{diagonal}
For an antilinear operator $\vartheta$ the following properties
are equivalent:\\
{\bf a)} \, $\vartheta$ can be diagonalized.\\
{\bf b)} \, There is a linear $Z$ such that $Z \vartheta Z^{-1}$
is Hermitian.\\
{\bf c)} \, There is a positive linear operator $A$ such that
$A \vartheta A^{-1}$ is Hermitian.\\
{\bf d)} \, There is a positive linear operator $B$ such that
$\vartheta^{\dag} = B \vartheta B^{-1}$
\end{prop}
Proof: Let $\vartheta$ be diagonalizable. If $Z$ is invertible
then $Z \vartheta Z^{-1}$ can be made diagonal. (a) means
the existence of $d = \dim \cH$ linear independent
vectors $\tilde \phi_j$ such that
$\vartheta \, \tilde \phi_j = \lambda_j \tilde \phi_j$.
One can choose $Z$ such that
$\phi_1 = Z \tilde \phi_1, \dots, \phi_d = Z \tilde \phi_d$ is a
basis of $\cH$. Then $A \vartheta A^{-1} \phi_j =
\lambda_j \phi_j$. Thus $Z \vartheta Z^{-1}$ can be diagonalized
by an Hilbert basis and, therefore, it is Hermitian, proving
(a) $\mapsto$ (b). Next, $Z$ can be written $Z = U A$ with a unitary
$U$ and a positive operator $A$. Because hermiticity of an operator
is conserved by unitary transformations, (b) $\mapsto$ (c). Now
(c) $\mapsto$ (a) follows from lemma \ref{hdiagonal}. More
explicitly (c) reads
\begin{displaymath}
A \vartheta A^{-1}  = (A \vartheta A^{-1})^{\dag} =
A^{-1} \vartheta^{\dag} A \; .
\end{displaymath}
With $B = A^2$ this means $\vartheta^{\dag} = B \vartheta B^{-1}$
with $B$ positive. Hence (c) $\mapsto$ (d). On the other hand, if
$A$ is the positive root of $B$, then one inversely sees that
$A \vartheta A^{-1}$ is Hermitian.

\begin{cor}
An antilinear operator $\vartheta$ is diagonalizable if and
only if there is a  scaler product with respect to which
$\vartheta$ becomes Hermitian.
\end{cor}
Proof: In the setting above, the scalar product reads
$\langle \phi, \phi'\rangle_A = \langle \phi, A \phi'\rangle$.

\subsubsection{The field of values} \label{C3.1.0}
The {\em field of values} is the set of all expectation values
$\langle \phi, X \phi\rangle$, $\langle \phi, \phi \rangle = 1$,
of an operator $X$. According to the Toeplitz-Hausdorf theorem,
it is a compact and convex set if $X$ is a linear operator,
\cite{HJ91}. The field of values of an antilinear operator is a
disk with center at $0$.
\begin{prop} \label{propfov}
Let $2 \leq \dim \cH < \infty$ and $\vartheta$ antilinear. Then
\begin{equation} \label{fov1}
\{z \, : \, z = \langle \phi, \vartheta \phi \rangle, \,
\langle \phi , \phi \rangle = 1 \} = \{ z \, : \, |z| \leq r\}
\end{equation}
and $r$ is the operator norm of $\vartheta^{+}$, which
is the largest eigenvalue of $|\vartheta^{+}|$.
\begin{equation} \label{fov2}
r = \sup | \langle \phi, \vartheta^{+} \phi \rangle|, \quad
\vartheta^{+} = \frac{1}{2} (\vartheta + \vartheta^{\dag})
\end{equation}
and the $\sup$ runs over all unit vectors.
\end{prop}
Proof:
For the time being denote by F$(\vartheta)$ the left of
(\ref{fov1}). F$(\vartheta)$ is a connected and compact
set of complex numbers. By proposition \ref{prop1} it consists
of circles. Hence F$(\vartheta)$ is a set of the form
$0 \leq r_0 \leq |z| \leq r$. By (\ref{rule2}) the expectation
values of $\vartheta$ and $\vartheta^{\dag}$ coincide.
Hence F$(\vartheta)$ equals F$(\vartheta^{+})$. Now,
$\vartheta^+$ being Hermitian, there is a basis of eigenvectors
with non-negative eigenvalues. It follows that the largest
eigenvalue of $|\vartheta^{+}|$ is the radius $r$ in
(\ref{fov1}), i.~e. $r$ is the operator norm of
$|\vartheta^{+}|$ or, equivalently, of $\vartheta^{+}$.

If $\vartheta^{+}$ is not invertible, then $r_0 = 0$
trivially. In the remaining case, there are at least two
orthogonal unit eigenvectors, say $\phi_1$, $\phi_2$, with real
eigenvalues $s_1 < 0 < s_2$. With $0 \geq s \geq 1$ let
$\phi = \sqrt{s} \phi_1 + \sqrt{1-s} \phi_2$. Then
$\langle \phi , \vartheta \phi\rangle = s_1s + s_2 (1-s)$.
The expectation value becomes negative if $s \approx 1$ and
positive if $s \approx 0$. Hence the expectation value
becomes zero for a certain $s$. Hence $r_0 = 0$.

\subsection{Antilinear rank one operators} \label{C3.2}
Now we compare with subsection \ref{C2.2} and introduce
some convenient description of rank-one operators a la
Dirac.
Any linear function $l$ on $\cH$ can be expressed by
$l(\phi) = \langle \phi'', \phi \rangle$ with unique
$\phi'' \in \cH$. Having this in mind, the
rank-one linear and antilinear operators can be described by
\begin{equation} \label{rank1a}
(|\phi' \rangle\langle \phi''|) \, \phi :=
\langle \phi'', \phi \rangle \, \phi' , \quad
(|\phi' \rangle\langle \phi''|_{\rc}) \, \phi :=
\langle \phi, \phi'' \rangle \, \phi' \; ,
\end{equation}
which projects any vector $\phi$ onto a multiple of $\phi'$.
(We assume $\phi'$ and $\phi''$ different from the null-vector.
We do not give a rank to the null-vector of $\cH$.)

\underline{Attention:} $|\phi' \rangle\langle \phi''|$
and $|\phi' \rangle\langle \phi''|_{\rc}$ {\em are defined}
by (\ref{rank1a}). I do {\em not} use $\langle \phi''|$
decoupled from its other part as Dirac did, and I {\em do not}
give any meaning to $\langle \phi''|_{\rc}$ as an standing alone
expression. (Though one could do so as a conjugate linear
functional).

To get the Hermitian conjugate look at
\begin{displaymath}
\langle \phi_1, \,
(|\phi' \rangle\langle \phi''|)_{\rc} \phi_2 \rangle =
\langle \phi_2, \phi''\rangle \, \langle \phi_1, \phi'\rangle
\end{displaymath}
which is symmetric by interchanging $\{ \phi_1, \phi' \}$ and
$\{ \phi_2, \phi'' \}$. Hence
\begin{equation} \label{rank1b}
(|\phi' \rangle\langle \phi''|_{\rc})^{\dag} =
|\phi'' \rangle\langle \phi'|_{\rc} ,
\quad
|\phi' \rangle\langle \phi''|^{\dag} =
|\phi'' \rangle\langle \phi'| \; .
\end{equation}
Here and below the linear case is mentioned for comparison

For the time being, $A_L$, $A_R$ denote linear and $\vartheta_L$
and $\vartheta_R$ antilinear operators to write down some
useful identities:
\begin{eqnarray} \label{rank1ca}
A_L |\phi' \rangle\langle \phi''| =
|A_L \phi' \rangle\langle \phi''|
&,& \quad
|\phi' \rangle\langle \phi''| A_R =
|\phi' \rangle\langle A_R^{\dag} \phi''| \; ,
\nonumber \\
\vartheta_L |\phi' \rangle\langle \phi''| =
|\vartheta_L \phi' \rangle\langle \phi''|_{\rc}
&,& \quad
|\phi' \rangle\langle \phi''| \vartheta_R =
|\phi' \rangle\langle \vartheta_R^{\dag} \phi''|_{\rc} \; .
\end{eqnarray}
The first two are well known. Concerning the others,
one proceeds as follows.
\begin{displaymath}
\vartheta_L |\phi' \rangle\langle \phi''| \phi = \vartheta_L
\langle \phi'', \phi\rangle \phi' =
\langle \phi, \phi''\rangle \vartheta_L \phi' \; ,
\end{displaymath}
and by the definition (\ref{rank1a}) one obtains the third
relation of(\ref{rank1ca}). The proofs of the other one
and of the following equations is similar.
\begin{eqnarray} \label{rank1cb}
A_L |\phi' \rangle\langle \phi''|_{\rc} =
|A_L \phi' \rangle\langle \phi''|_{\rc}
&,& \quad
|\phi' \rangle\langle \phi''|_{\rc} A_R =
|\phi' \rangle\langle A_R^{\dag} \phi''|_{\rc} \; ,
\nonumber \\
\vartheta_L |\phi' \rangle\langle \phi''|_{\rc} =
|\vartheta_L \phi' \rangle\langle \phi''|
&,& \quad
|\phi' \rangle\langle \phi''|_{\rc} \vartheta_R =
|\phi' \rangle\langle \vartheta_R^{\dag} \phi''| \; .
\end{eqnarray}
Applying $\vartheta_L$ to the last relation and looking at the
second one in (\ref{rank1ca}) yields
\begin{equation} \label{rank1cc}
\vartheta_L |\phi' \rangle\langle \phi''| \vartheta_R^{\dag} =
|\vartheta_L \phi' \rangle\langle \vartheta_R \phi''|
\end{equation}
as one of further possibilities to combine (\ref{rank1ca}) and
(\ref{rank1cb}). The chaining
\begin{equation} \label{mranka}
|\phi_1 \rangle\langle \phi_2|_{\rc} \;
|\phi_3 \rangle\langle \phi_4|_{\rc}  =
\langle \phi_3, \phi_2 \rangle \, |\phi_1 \rangle\langle \phi_4|
\end{equation}
is straightforwardly. By definition (\ref{Hform1}) it follows
\begin{equation} \label{mrankb}
\T \,   ( |\phi_1 \rangle\langle \phi_2|_{\rc} \,
|\phi_3 \rangle\langle \phi_4|_{\rc} )  = \langle \phi_3,
\phi_2 \rangle \, \langle \phi_4, \phi_1 \rangle \; .
\end{equation}

\noindent {\bf Remark:}\\
Let $\phi_1, \dots, \phi_d$
be a basis of $\cH$. (\ref{Rank4}) becomes
\begin{equation} \label{Rank4a}
\vartheta_k^j \, \phi := \langle \phi, \phi_j \rangle \,\phi_k
= |\phi_j \rangle\langle \phi|_{\rc} \, \phi_k \; .
\end{equation}
One gets a basis of $\cB(\cH)_{\rc}$ in terms of Dirac
symbols.

\subsection{Linear and antilinear Pauli operators}
\label{C3.3}
In subsection \ref{C2.5.1} the operators $\sigma_j$, $\tau_k$,
have been defined relative to two linear independent but
otherwise arbitrary vectors. From now on, $\phi_1$, $\phi_2$
denotes a Hilbert basis of $\cH_2$. Then
{\em $\tau_1$, $\tau_2$, $\tau_3$ are Hermitian. They form
a basis of $\cB(\cH_2)_{\rc}^{+}$. $\tau_0$ is skew Hermitian.
Every element of $\cB(\cH_2)_{\rc}^{-}$ is a multiple of
$\tau_0$.}

An antilinear Hermitian operators, the square of which is $\1$
is called a {\em conjugation}. It is a {\em skew conjugation}
If its square is $-\1$. See the next section for more.

The following identities are useful.
\begin{eqnarray}
\tau_0 = |\phi_2 \rangle\langle \phi_1 |_{\rc} -
|\phi_1 \rangle\langle \phi_2 |_{\rc} ,
&\quad&
\tau_1 = |\phi_1 \rangle\langle \phi_1 |_{\rc} -
|\phi_2 \rangle\langle \phi_2 |_{\rc} ,
\label{Pauli7} \\
\tau_2 = i |\phi_1 \rangle\langle \phi_1 |_{\rc} +
i |\phi_2 \rangle\langle \phi_2 |_{\rc} ,
&\quad&
\tau_3 = |\phi_1 \rangle\langle \phi_2 \rangle_{\rc} +
|\phi_2 \rangle\langle \phi_1 \rangle_{\rc}  ,\\
|\phi_1 \rangle\langle \phi_1 |_{\rc} =
\frac{1}{2} (\tau_1 - i \tau_2) ,
&\quad&
|\phi_1 \rangle\langle \phi_2 |_{\rc} =
\frac{1}{2} (\tau_3 - \tau_0) ,
\label{Pauli7a} \\
|\phi_2 \rangle\langle \phi_1 |_{\rc} =
\frac{1}{2} (\tau_3 + \tau_0) ,
&\quad&
|\phi_2 \rangle\langle \phi_2 |_{\rc} =
- \frac{1}{2} (\tau_1 + i \tau_2) ,
\end{eqnarray}
\medskip

Miscellaneous facts about the $\tau_j$-operators will be
gathered:
Remember of $\theta_{\rm F} = i \tau_0$, see
(\ref{example1.1}), and consider the equation
\begin{equation} \label{example1.5a}
\begin{pmatrix} 0 & -i \\ i & 0 \end{pmatrix}_{\rc}
\begin{pmatrix} a_{00}^* & a_{10}^* \\ a_{01}^* & a_{11}^*
\end{pmatrix}
\begin{pmatrix} 0 & -i \\ i & 0 \end{pmatrix}_{\rc}
= -
\begin{pmatrix} a_{11} & -a_{01} \\ -a_{10} & a_{00}
\end{pmatrix}
\end{equation}
which can be verified by direct calculation. The operator
between the two spin flips is the matrix representation of
the Hermitian adjoint of a general linear operator $A$. The
matrix equation above can be converted into a basis
independent operator equation:
\begin{equation} \label{example2.1}
\tau_0 A^{\dag} \tau_0^{-1} =
(\T \, A) \, \1 - A = (\det A) A^{-1} \; .
\end{equation}
The last equality sign supposes $A$ invertible. Another form of
(\ref{example2.1}) reads
\begin{equation} \label{example2.2}
A \tau_0 A^{\dag} = (\det A) \tau_0 \; .
\end{equation}
It is obtained by multiplying (\ref{example2.1}) by $A$ from
the left and by $\tau_0$ from the right. Note that these
equations remain valid if $\tau_0$ is replaced $\theta_{\rm F}$.

An application of the above is a description of the special
unitary group SU$(2)$. Multiplying (\ref{example2.1}) from
the left by $\tau_0^{-1}$ yields
\begin{equation} \label{example2.3}
\tau_0 A = A \tau_0 \, \Leftrightarrow \,
A A^{\dag} = (\det A) \1 \; .
\end{equation}
Hence, $U \in {\rm SU}(2)$ if and only $U$ commutes with $\tau_0$
and has determinant one.

The determinants of the Pauli operators (Pauli matrices)
are $-1$. To get SU$(2)$ operators the multiplication by
$i$ is sufficient:
\begin{equation} \label{example2.4}
\{\1_2, \, i \sigma_1, \, i \sigma_2, \, i \sigma_3 \}
\in {\rm SU}(2) \; .
\end{equation}
These operators commute with $\tau_0$, and they generate the
real linear space of operators commuting with
$\tau_0$ (respectively $\theta_{\rm F}$). Hence one gets:
\begin{lem} \label{c3.2.2a}
Let $A \in \cB(\cH_2)$. The following three conditions are
equivalent:\\
({\bf i}) $A$ has a representation
\begin{equation} \label{example2.5}
A = x_0 \1_2 + i \sum_1^3 x_j \sigma_j, \quad
x_0, x_1, x_2, x_3 \in \mathbb{R} \, .
\end{equation}
({\bf ii}) It is $A = (\det A) U$ with $U \in {\rm SU}(2)$
and $\det A$ real.\\
({\bf iii}) $A$ commutes with $\tau_0$ .
\end{lem}
The set of all operators (\ref{example2.5}) is a
{\em real} $\dag$-involutive subalgebra of $\cB(\cH_2)$
isomorphic to the field of quaternions.

An antilinear operator, $\vartheta$, can be decomposed similar
to (\ref{example2.5}):
\begin{equation} \label{example2.7}
\vartheta = c_0 \tau_0 +
c_1 \tau_1 + c_2 \tau_2 + c_3 \tau_3 \; .
\end{equation}
The first term is $\vartheta^{-}$, the sum of the last three
gives $\vartheta^{+}$. Note the typical Minkowskian structure
\begin{equation} \label{example2.9}
\frac{1}{2} (\vartheta, \vartheta) =
 (|c_1|^2 + |c_2|^2 + |c_3|^2 - |c_0|^2) \; ,
\end{equation}

Lemma \ref{c3.2.2a} can be ``antilinearly'' rewritten. At first
one multiply from the left with a unimodular number $\epsilon$
to represent all unitaries. Then one replaces $\sigma_j$ by
$\tau_j \tau_0$ and multiplies from the left by $\tau_0$. One gets
\begin{lem} \label{c3.2.2c}
The following two conditions are equivalent:\\
({\bf i}) The antiunitary operator $\vartheta$ allows for a
representation
\begin{equation} \label{example2.12}
\vartheta = \epsilon (y_0 \tau_0 + i \sum_1^3 y_j), \quad
y_0, \dots y_3 \in \mathbb{R} \; .
\end{equation}
({\bf ii}) There is an antiunitary operator $\Theta$ and
a non-negative real number $\lambda$ such that
\begin{equation} \label{2.22}
 \vartheta = \lambda \, \Theta \; .
\end{equation}
By letting $\epsilon$ constant, the set of all operators
(\ref{example2.12}) becomes a $^{\dag}$-invariant real
linear space.
\end{lem}
The lemma is an impressing example that the two-dimensional
case is often special. This is seen from
\begin{lem}
If $\dim \cH \geq 3$ then a linear operator commutes with
$\cB(\cH_{\rc}^-)$ if and only if it is a real multiple of $\1$.
\end{lem}
Proof: After choosing a basis $\{ \phi_j\}$. Let
the antilinear operators  $\theta_{jk}$, $j < k$ acting on the
subspace generated by $\phi_j$ and $\phi_k$ like $\tau_0$, and
annihilating all other elements of the basis. These operators
form a general basis of $\cB(\cH_{\rc}^-)$. It is to
prove that $A \theta_{jk} = \theta_{jk} A$ for all $j < k$
if and only if the linear operator $A$ is a real multiple
of the unit operator $\1$.

The following proof assumes $\dim \cH = 3$ for transparency.
Let $A$ be a linear operator with matrix elements $a_{jk}$.
The non-zero matrix elements of $\theta_{12}$ are $+1$ and $-1$
in the positions $12$ and $21$ respectively. The equation
$A \theta_{12} = \theta_{12} A$, written in the $\{ \phi_i \}$
matrix representation, reads
\begin{displaymath}
\begin{pmatrix}
-a_{12} & a_{11} & 0 \\ -a_{22} & a_{21} & 0 \\
 -a_{32} & a_{21} & 0
\end{pmatrix} =
\begin{pmatrix}
a_{21}^* & a_{22}^* & a_{23}^* \\
-a_{11}^* & -a_{12}^* & -a_{13}^* \\  0 & 0 & 0
\end{pmatrix}   \; .
\end{displaymath}
Firstly this implies $a_{j3} = a_{3j} = 0$ for $j = 1,2$. Using
$\theta_{23}$ or $\theta_{13}$ instead of $\theta_{12}$ gives
$a_{j1} = a_{1j} = 0$ for $j = 2,3$ or
$a_{j2} = a_{2j} = 0$ for $j = 1,3$. Therefore, $A$ must be
diagonal in the chosen basis.

Secondly one gets $a_{11}^* = a_{22}$ from the pre-proposed
commutativity between $A$ and $\theta_{12}$. The same procedure
with $\theta_{23}$ and $\theta_{13}$ implies $a_{22}^* = a_{33}$
and $a_{11}^* = a_{33}$. All together one arrives at
$a_{11} = a_{22} = a_{33} \in \mathbb{R}$. Thus, $A = a \1$,
$a$ real. The extension to higher dimensions is obvious.
That $A = a \1$, $a$ real, is in the commutant of
$\cB(\cH_{\rc}^-)$ is trivial.

Concerning the antilinear Hermitian operators one can
prove the following
\begin{lem}
A linear operator commuting with all antilinear Hermitian
operators is a real multiple of $\1$ for any
$\dim \cH \geq 1$.
\end{lem}

\subsection{Antilinear maps between Hilbert spaces}
\label{C3.4}
As in the linear case the Hermitian adjoint can be defined
not only within $\cB(\cH)_{\rc}$ but also for antilinear maps
between Hilbert spaces. The following is a mini-introduction
to this topic. Something more will be said in the sections
\ref{C9} and \ref{C10}.
\medskip

Let $\cH^{\kA}$, $\cH^{\kB}$ denote two finite dimensional Hilbert
spaces and $\vartheta$ an antilinear map from $\cH^{\kA}$ into
$\cH^{\kB}$. Its {\em Hermitian adjoint,} $\vartheta^{\dag}$,
is an {\em antilinear} map from $\cH^{\kB}$ into $\cH^{\kA}$ and
it is defined by
\begin{equation} \label{rule2e}
\langle \phi^{\kB} , \vartheta \,  \phi^{\kA} \rangle  = \langle
\phi^{\kA} , \vartheta^{\dag} \, \phi^{\kB} \rangle, \quad
\phi^{\kA} \in \cH^{\kA}, \, \phi^{\kB} \in \cH'' \; .
\end{equation}
The Hermitian adjoint from $\cH^{\kB}$ into $\cH^{\kA}$ is defined
similarly. With respect to these definitions, the rules
(\ref{rule4}), and (\ref{def2}) remain valid also in this setting.
In particular $\vartheta^{\dag \dag} = \vartheta$, In this
spirit, also the second part of  (\ref{rule3}) has its counterpart:
\begin{displaymath}
(\vartheta_{21}  \vartheta_{12})^{\dag} =
\vartheta_{12}^{\dag} \vartheta_{21}^{\dag},
\quad \vartheta_{12} \in \cB(\cH^{\kB},\cH^{\kA})_{\rc}
, \quad \vartheta_{21} \in \cB(\cH^{\kA},\cH^{\kB})_{\rc}
\end{displaymath}

The linear space of all antilinear maps from $\cH^{\kA}$ into
$\cH^{\kB}$ will be denoted by $\cB(\cH^{\kA},\cH^{\kB})_{\rc}$.
In the same manner $\cB(\cH^{\kB},\cH^{\kA})_{\rc}$ is the
linear space of all antilinear maps from $\cH^{\kB}$ into
$\cH^{\kA}$. Their dimensions are
$(\dim \cH^{\kA}) (\dim \cH^{\kB})$. The Hermitian adjoint
(\ref{rule2e}) induces a {\em linear} isomorphism
\begin{equation} \label{rule2f}
\cB(\cH^{\kA},\cH^{\kB})_{\rc} \,
\overset{\dag}{\longleftrightarrow} \,
\cB(\cH^{\kB},\cH^{\kA})_{\rc} \; .
\end{equation}

The space of antilinear maps from one Hilbert space into another
one is itself Hilbertian in a natural way. Before coming to that,
it is helpful to introduce the antilinear rank one operators
acting between a pair of Hilbert spaces. Following (\ref{rank1a}),
definition and Hermitian adjoint are given by
\begin{eqnarray}
|\phi^{\kA}_1 \rangle\langle \phi^{\kB}_1|_{\rc} \, \phi^{\kB}
& := &
\langle \phi^{\kB}, \phi^{\kB}_1 \rangle \, \phi^{\kA}_1 \; ,
\label{2rank1a} \\
|\phi^{\kB}_1 \rangle\langle \phi^{\kA}_1|_{\rc} \, \phi^{\kA}
& := &
\langle \phi^{\kA}, \phi^{\kA}_1 \rangle \, \phi^{\kB}_1 \; ,
\label{2rank1b} \\
(|\phi^{\kA}_1 \rangle\langle \phi^{\kB}_1|_{\rc})^{\dag}
&  = & |\phi^{\kB}_1 \rangle\langle \phi^{\kA}_1|_{\rc} \; ,
\label{2rank1c} \\
\phi^{\kA}, \phi^{\kA}_1 \in \cH^{\kA}, & \quad &
\phi^{\kB}, \phi^{\kB}_1 \in \cH^{\kB}  \; .
\nonumber
\end{eqnarray}
The following relations can be checked:
\begin{eqnarray}
|\phi^{\kA}_1 \rangle\langle \phi^{\kB}_1|_{\rc} \,
|\phi^{\kB}_2 \rangle\langle \phi^{\kA}_2|_{\rc} ,
& = &
\langle \phi^{\kB}_2, \phi^{\kB}_1 \rangle \,
|\phi^{\kA}_1 \rangle\langle \phi^{\kA}_2| \; ,
\label{2rank2a} \\
|\phi^{\kB}_1 \rangle\langle \phi^{\kA}_1|_{\rc} \,
|\phi^{\kA}_2 \rangle\langle \phi^{\kB}_2|_{\rc}  .
& = &
\langle \phi^{\kA}_2, \phi^{\kA}_1 \rangle \,
|\phi^{\kB}_1 \rangle\langle \phi^{\kB}_2| \; .
\label{2rank2b}
\end{eqnarray}
An intermediate step in proving (\ref{2rank2a}) is in
\begin{displaymath}
|\phi^{\kA}_1 \rangle\langle \phi^{\kB}_1|_{\rc} \,
|\phi^{\kB}_2 \rangle\langle \phi^{\kA}_2|_{\rc} \phi^{\kA}
= \langle \phi^{\kA}_2, \phi^{\kA} \rangle \,
 \langle \phi^{\kB}_2, \phi^{\kB}_1 \rangle \, \phi^{\kA}_1 \; .
\end{displaymath}

Let $\vartheta$ and $\tilde \vartheta$ be chosen from
$\cB(\cH^{\kA},\cH^{\kB})_{\rc}$ arbitrarily. Assume
\begin{equation} \label{2rank3}
\vartheta = \sum a_{jk}
|\phi^{\kB}_k \rangle \langle \phi^{\kA}_j|_{\rc} ,
\quad
\tilde \vartheta = \sum b_{mn}
|\phi^{\kB}_n \rangle\langle \phi^{\kA}_m|_{\rc} ,
\end{equation}
Using (\ref{2rank2a}) one derives the identities
\begin{eqnarray}
\tilde \vartheta \vartheta^{\dag} &=& \sum a_{jk}^* b_{mn}
\langle \phi^{\kA}_j , \phi^{\kA}_m \rangle
\, |\phi^{\kB}_n \rangle \langle \phi^{\kB}_k|
\label{2rank3a} \\
\tilde \vartheta^{\dag} \vartheta &=& \sum a_{jk}^* b_{mn}
\langle \phi^{\kB}_k , \phi^{\kB}_n \rangle
\, |\phi^{\kA}_m \rangle \langle \phi^{\kA}_j|
\label{2rank3b}
\end{eqnarray}
Taking traces one gets on $\cB(\cH^{\kA},\cH^{\kB})_{\rc}$
\begin{equation} \label{2rank4}
\T \, \tilde \vartheta^{\dag} \vartheta =
\T \, \tilde \vartheta \vartheta^{\dag} = \sum a_{jk}^{*} b_{mn}
\langle \phi_j^{\kA}, \phi_m^{\kA} \rangle \,
\langle \phi_k^{\kB}, \phi_n^{\kB} \rangle \; .
\end{equation}
These relations define the {\em natural scalar product}
\begin{equation} \label{2rank4a}
\langle \tilde \vartheta, \vartheta \rangle_{ab} :=
\T \, \tilde \vartheta^{\dag} \vartheta =
\T \, \tilde \vartheta \vartheta^{\dag}
\end{equation}
so that $\cB(\cH^{\kA},\cH^{\kB})_{\rc}$ becomes an Hilbert space.

Exchanging the roles of $\kA$ and $\kB$ by setting $\vartheta' =
\vartheta^{\dag}$, $\tilde \vartheta' = \tilde \vartheta^{\dag}$
one gets the {\em natural scalar product}
\begin{equation} \label{2rank4b}
\langle \tilde \vartheta', \vartheta' \rangle_{ba} :=
\T \, \tilde \vartheta' (\vartheta')^{\dag} =
\T \, (\tilde \vartheta')^{\dag} \vartheta' , \quad
\vartheta',\tilde \vartheta' \in  \cB(\cH^{\kB},\cH^{\kA})_{\rc}
\end{equation}
on $\cB(\cH^{\kB},\cH^{\kA})_{\rc}$.

Thus (\ref{rule2f}) {\em becomes an isometry}.
\medskip

The next aim is in construction further remarkable isometries.\\
The direct product $\cH^{\kA} \otimes \cH^{\kB}$ of two finite
dimensional Hilbert spaces is the linear span of all
symbols $\phi^{\kA} \otimes \phi^{\kB}$ with
$\phi^{\kA} \in \cH^{\kA}$, $\phi^{\kB} \in \cH^{\kB}$ modulo
the defining relations
\begin{eqnarray}
c (\phi^{\kA} \otimes \phi^{\kB}) =
(c \phi^{\kA}) \otimes \phi^{\kB} & = &
\phi^{\kA} \otimes (c \phi^{\kB}) , \nonumber \\
\phi_1^{\kA} \otimes \phi^{\kB} + \phi_2^{\kA} \otimes \phi^{\kB}
& = &
(\phi_1^{\kA} + \phi^{\kA}_2) \otimes \phi^{\kB}, \nonumber \\
\phi^{\kA} \otimes \phi_1^{\kB} + \phi^{\kA} \otimes \phi_2^{\kB}
& = &
\phi^{\kA} \otimes (\phi_1^{\kB} + \phi^{\kB}_2) \nonumber \; .
\end{eqnarray}

An antilinear operator
$|\phi^{\kB} \rangle\langle \phi^{\kA}|_{\rc}$ acts according
to the rule (\ref{2rank1b}). It is indexed by a pair of vectors,
$\phi^{\kA}$ and $\phi^{\kB}$ from which it depends bilinearly.
Therefore the defining relations of
$\cH^{\kA} \otimes \cH^{\kB}$ are fulfilled. Thus the map
\begin{displaymath}
\phi^{\kA} \otimes \phi^{\kB} \, \rightarrow \,
|\phi^{\kB} \rangle\langle \phi^{\kA}|_{\rc}
\end{displaymath}
induces a linear isomorphism
\begin{equation} \label{2rank5a}
\psi \equiv  \sum c_{jk} \phi_j^{\kA} \otimes \phi_k^{\kB}
\, \rightarrow \, \vartheta \equiv
\sum c_{jk} |\phi_k^{\kB} \rangle\langle \phi_j^{\kA}|_{\rc}
\end{equation}
from $\cH^{\kA} \otimes \cH^{\kB}$ onto
$\cB(\cH^{\kA}, \cH^{\kB})_{\rc}$. The map is onto because
both linear spaces are of the same dimension.

(\ref{2rank5a}) is an isometry. To see it one starts with
$\langle \psi, \psi \rangle = \sum |c_{jk}|^2$ and computes
$\langle \vartheta, \vartheta \rangle_{ab}$ which is given by
(\ref{2rank4a}).  To do so, it is sufficient to establish
\begin{displaymath}
\langle \phi^{\kA} \otimes \phi^{\kB}, \phi^{\kA} \otimes
\phi^{\kB}\rangle =  \T \,
|\phi^{\kA} \rangle\langle \phi^{\kB}|_{\rc}^{\dag}
|\phi^{\kA} \rangle\langle \phi^{\kB}|_{\rc} \; .
\end{displaymath}

In the same way one proves that
\begin{equation} \label{2rank5b}
\sum c_{kj} \phi_k^{\kB} \otimes \phi_j^{\kA} \, \rightarrow \,
\sum c_{kj} |\phi_j^{\kA} \rangle\langle \phi_k^{\kB}|_{\rc}
\end{equation}
is a linear isometry from $\cH^{\kB} \otimes \cH^{\kA}$
onto $\cB(\cH^{\kB}, \cH^{\kA})_{\rc}$.

\begin{prop} \label{c3.3.2a}
The Hilbert spaces
\begin{equation} \label{caniso2}
\cH^{\kA} \otimes \cH^{\kB}, \quad \cB(\cH^{\kA}, \cH^{\kB})_{\rc}
, \quad \cB(\cH^{\kB}, \cH^{\kA})_{\rc}, \quad
\cH^{\kB} \otimes \cH^{\kA}
\end{equation}
are mutually canonically isometrical equivalent. The isometries
are described by
(\ref{rule2f}), (\ref{2rank5a}), (\ref{2rank5b}), and by
\begin{equation} \label{caniso3}
\phi^{\kA} \otimes \phi^{\kB} \,  \leftrightarrow \,
\phi^{\kB} \otimes \phi^{\kA} \; .
\end{equation}
\end{prop}
\medskip

\noindent {\bf Remark:}
In order to handle some quantum theoretical problems,
a more systematic treatment starts with section
\ref{C8}. Also further notations will be introduced:
The map (\ref{2rank5a}), for instance, will be written
\begin{equation} \label{caniso4}
\psi \equiv  \sum c_{jk} \phi_j^{\kA} \otimes \phi_k^{\kB}
\, \rightarrow \, s_{\psi}^{ba} := \sum c_{jk}
|\phi_k^{\kB} \rangle\langle \phi_j^{\kA}|_{\rc} \; .
\end{equation}

\section{Antilinear normal operators} \label{C4}
An antilinear operator is called {\it normal} if it commutes
with its Hermitian adjoint:
\begin{equation} \label{def4}
\vartheta^{\dag} \vartheta = \vartheta \vartheta^{\dag} \, .
\end{equation}
Starting with the decomposition
$\vartheta = \vartheta^{+} + \vartheta^{-}$,
the operators is normal if
\begin{displaymath}
(\vartheta^{+} + \vartheta^{-}) \, (\vartheta^{+} - \vartheta^{-})
=
(\vartheta^{+} - \vartheta^{-}) \, (\vartheta^{+} + \vartheta^{-})
\; .
\end{displaymath}
Multiplying out the products, we see:\\
$\vartheta$ {\em is normal if and only if}
\begin{equation} \label{normal1}
\vartheta^{+} \vartheta^{-} = \vartheta^{-} \vartheta^{+} \; .
\end{equation}
Important subclasses of the normal antilinear operators
are the already defined Hermitian and the skew Hermitian
ones.
A further essential class constitute the
{\em unitary} antilinear operators, also called\footnote{
A notation due to E.~P.~Wigner} {\em antiunitaries.}

As in the linear case an antiunitary is characterized by
\begin{displaymath}
\Theta^{\dag} = \Theta^{-1} \; .
\end{displaymath}
Antiunitaries are isometric, i.~e.
\begin{equation} \label{isom1}
\langle \Theta \phi, \Theta \phi' \rangle =
\langle \phi', \phi \rangle  \; .
\end{equation}
The set ${\cU}_{\rc}(\cH)$ of all antiunitarities is not a group.
Yet it contains with $\Theta$ also $\Theta^{-1}$. Thus its
{\em adjoint  representation}
\begin{equation} \label{isom2}
\Theta \rightarrow \Theta^{-1} X \Theta
\end{equation}
is well defined for linear and for antilinear $X$. Assume
\begin{displaymath}
\Theta_1^{\dag} \vartheta \Theta_1 = \Theta_2^{\dag} \vartheta
\Theta_2 \quad \forall \, \vartheta \in \cB(\cH)_{\rc}
\end{displaymath}
and applying (\ref{rule4}) it follows
\begin{displaymath}
 \Theta_2 = \pm \Theta_1 \; .
\end{displaymath}

Notice also that ${\cU}(\cH) \cup {\cU}_{\rc}(\cH)$ is a group.
${\cU}(\cH)$ is a normal subgroup of it.
\medskip

A {\em conjugation} is an antilinear operator which is
both, unitary and Hermitian, i.~e. $\theta$ is a conjugation
if and only if it is antilinear and satisfies
\begin{equation} \label{conj1}
 \theta^{\dag} = \theta = \theta^{-1} \; .
\end{equation}
It follows $\theta^2 = \1$. On the other hand, $\theta^2 = \1$
together with either $\theta^{\dag} = \theta $ or, alternatively,
with $\theta^{\dag} = \theta^{-1}$ implies (\ref{conj1}).

Conjugations form an important class of operators as will
be seen later on. In many papers they appear ``masked'' by
notations like $|\psi^*\rangle$ where the complex conjugation
refers to a distinguished basis.
\medskip

In the same spirit, a {\em skew conjugation} is a skew Hermitian
antiunitary operator:
\begin{equation} \label{sconj1}
 \theta^{\dag} = - \theta = \theta^{-1}, \quad
\theta^2 = - \1  \; .
\end{equation}
\medskip

\begin{prop}[Polar decomposition] \label{prop6}
Let $\vartheta$ be an antilinear operator and $\cH$ finite
dimensional. There are antiunitaries $\theta_L$, $\theta_R$
such that
\begin{equation} \label{Pdeco1}
\vartheta = \theta_L \sqrt{\vartheta^{\dag} \vartheta} =
 \sqrt{\vartheta \vartheta^{\dag}} \theta_R
\equiv |\vartheta | \theta_R \; .
\end{equation}
\end{prop}

A proof is by transforming the assertion into the
corresponding linear case:
let us choose an antiunitary $\theta_0$. Then
there is a unitary $U_0$ such that
\begin{displaymath}
\theta_0 \vartheta = U_0 \sqrt{\vartheta^{\dag} \vartheta},
\quad \theta_L := \theta_0^{-1} U_0 \; .
\end{displaymath}
The other case is similar.
\medskip

\noindent {\bf Remark:}
A {\it partial isometry} is an antilinear
operator $\theta$ for which $\theta^{\dag} \theta$ and
$\theta \theta^{\dag}$ are projection operators.
The choice of $\theta_L$ in the polar decomposition
(\ref{Pdeco1}) is unique up to its action onto the kernel
of $\vartheta^{\dag} \vartheta$. If $\vartheta^{-1}$ does
not exist, a unique polar decomposition requires $\theta_L$
(respectively $\theta_R$) to be partially isometric.
In particular, the supports of $\theta_L^{\dag} \theta_L$
and $\vartheta^{\dag} \vartheta$ must coincide.

That sharpening of the polar decomposition is obligatory
if $\dim \cH = \infty$. For such a more general treatment one
may consult \cite{GW09b}.

\begin{prop} \label{prop7}
If $\vartheta$ is normal, there is
an antiunitary $\theta$  such that
\begin{equation} \label{Pdeco3}
\vartheta = \theta |\vartheta| = |\vartheta| \theta
, \quad
\vartheta |\vartheta| = |\vartheta| \vartheta
, \quad
\vartheta^{\dag} |\vartheta| = |\vartheta| \vartheta^{\dag}
\; .
\end{equation}
\end{prop}

Indeed, from (\ref{Pdeco1}), one deduces
\begin{displaymath}
\vartheta = \theta_L |\vartheta| = |\vartheta| \theta_R ,
\quad
|\vartheta|^2 = \vartheta \vartheta^{\dag}
= \theta_L |\vartheta|^2 \vartheta_L^{-1} \; ,
\end{displaymath}
saying that $\theta_L$ commutes with $|\vartheta|^2$ and,
therefore, with $|\vartheta|$. Hence $\theta = \theta_L$
does the job.
\begin{cor} \label{cor2}
If $\vartheta$ is Hermitian, then the restriction of $\theta$
to the support of $|\vartheta|$ is a conjugation.
If $\vartheta$ is skew Hermitian, then the restriction of
$\theta$ to the support of $|\vartheta|$ is a skew conjugation.
\end{cor}
More information can be drawn from the WHV-theorem \ref{thmW4}
below.

Let $\Theta$ be an antiunitary and $X$ a linear operator. Then
\begin{equation} \label{rule7}
\T \, \Theta^{\dag} X \Theta = \T \, X^{\dag} \; .
\end{equation}
This is because
\begin{displaymath}
\langle \phi, \Theta^{\dag} X \Theta \phi \rangle =
\langle \Theta \phi, X \Theta \phi \rangle^{*} =
\langle \Theta \phi, X^{\dag} \Theta \phi \rangle \; ,
\end{displaymath}
as with $\{\phi_j\}$ also $\{\Theta \phi_j\}$ is a basis,
this proves (\ref{rule7}).
\begin{cor}
Let $\vartheta$ be antilinear and invertible. Then
\begin{equation} \label{rule7a}
\T \, \vartheta X \vartheta^{-1} = \T \, X^{\dag} \; .
\end{equation}
\end{cor}
By (\ref{Pdeco1}) there is a positive invertible operator $A$
such that $\vartheta = \Theta^{\dag} A$ and
 $\vartheta^{-1} = A^{-1} \Theta$. Inserting into (\ref{rule7})
proves (\ref{rule7a}).

Let $\vartheta = \Theta |\vartheta |$ a polar decomposition
(\ref{Pdeco1}). Replacing $X$ by $|\vartheta | X^{*}| \vartheta |$
in (\ref{rule7}) yields
\begin{equation} \label{rule7b}
\T \, \vartheta X^{*} \vartheta^{*} =
\T \, X \vartheta^{*} \vartheta \; .
\end{equation}

\subsection{Antiunitaries acting on $\cH_2$} \label{C4.1}
An applications of (\ref{example2.1}) is as following:
\begin{prop} \label{prop9}
Let $A \in \cB(\cH_2)$ and $z \neq 0$ a complex number.\\
Then $\theta_{\rm F} A \theta_{\rm F}^{-1} = z A$
if and only if $A = a U$ with $U \in {\cSU}(2)$ and
$z = a^*/a$.
In particular, $A$ commutes with $\theta_{\rm F}$ if
and only if $a \in \mathbb{R}$.
\end{prop}
With $A$ also $A^{\dag}$ commutes with $\tau_{\rm F}$. Hence
(\ref{example2.1}) provides $(\det A) A^{-1} = z^{*} A^{\dag}$.
Hence there is a complex number $a$ such that $A = a U$ with
$U$ unitary, $\det U = 1$. Then $a = z^{*} a^{*}$.
\medskip

\noindent {\bf Remark:}
There is no antilinear operator different from ${\mathbf 0}$
commuting with ${\cSU}(n)$ if $\dim \cH > 2$.

\begin{lem} \label{qbl2}
Let $\theta$ be an antiunitary operator on $\cH_2$.
Either $\theta$ is a conjugation
or $\theta$ does not possess an eigenvector.
\end{lem}
At first, the assertion is invariant with respect to
transformations
$\theta' = V \theta V^{\dag}$ where $V \in {\rm SU}(2)$.
These transformations commute with $\theta_{\rm F}$
and we get $\theta' = U' \theta_{\rm F}$.
With a properly chosen $V$ we get $U'$ diagonal with
respect to a given basis. Then
$\theta' = \epsilon_0 \theta_{\epsilon}$ with
$|\epsilon_0| = |\epsilon| = 1$
and $\theta_{\epsilon}$ is defined as in
example 1. Now the lemma \ref{qbl1}
in example 1 proves the assertion.
\begin{prop} \label{normalantiu}
For any antiunitary operator $\theta$ there is a basis
$\phi_1, \phi_2$ such that
\begin{equation} \label{example1.7}
\theta (c_1 \phi_1 + c_2 \phi_2) =
\epsilon c_1^* \phi_2 + \epsilon^* c_2^* \phi_1 \; .
\end{equation}
$\theta$ is a conjugation if and only if $\epsilon$ is real.
\end{prop}
For the last assertion see lemma \ref{qbl1}.
Using Euler's formula,
$\epsilon = \exp(i \alpha) = \cos \alpha + i sin \alpha$,
one obtains
\begin{cor} \label{Euler}
There is a representation
\begin{equation} \label{example1.8}
\theta = \cos \alpha \cdot \theta' +
\sin \alpha \cdot \theta_{\rm F}, \quad
\theta' (c_1 \phi_1 + c_2 \phi_2) = c_1^* \phi_2 + c_2^* \phi_1
\end{equation}
or, in matrix language,
\begin{equation} \label{example1.9}
\begin{pmatrix} 0 & \epsilon^* \\ \epsilon & 0 \end{pmatrix}_{\rc}
= \cos \alpha
\begin{pmatrix} 0 &  1 \\ 1 & 0 \end{pmatrix}_{\rc} +
\sin \alpha
\begin{pmatrix} 0 &  -i \\ i & 0 \end{pmatrix}_{\rc}
\end{equation}
\end{cor}

\subsection{Decompositions of normal antilinear operators}
\label{C4.2}
The structure of antilinear unitary operators has been
clarified by E.~P.~Wigner, \cite{Wi60}, who gave credit to a
method due to E.~Cartan, \cite{Ca31}.
The extension of Wigner's classification to antilinear normal
operators is due to F.~Herbut and M.~Vuji{\v c}i{\' c},
\cite{HV66}.

As all these authors we restrict ourselves to the finite
dimensional case. A complete classification for infinite
dimensional Hilbert spaces is not known to me.
\medskip

We derive the decomposition of an antilinear
normal operator into elementary parts, a ``surrogate''
of the spectral decomposition of linear normal operators.

It will be seen that the ``1-qubit case'', $\dim \cH = 2$,
provides a key ingredient in handling the problem.

At first, however, subspace decompositions of $\cH$ are
considered which are related to a general antilinear operator.
\medskip

A subspace $\cH_0$ of $\cH$ will be called
$\vartheta$-{\em normal} if it is $\vartheta$ and
$\vartheta^{\dag}$ invariant, i.~e.
$\vartheta \cH_0 \subset \cH_0$ and
$\vartheta^{\dag} \cH_0 \subset \cH_0$ is valid.
Let $(\cH_0)^{\perp}$ be the orthogonal complement of $\cH_0$. It
contains, by definition, all vectors $\varphi$ which are
orthogonal to the vectors of $\cH_0$. As in the linear case it
follows $\vartheta^{\dag} (\cH_0)^{\perp} \subset \cH_0$ from
$\vartheta \cH_0 \subset \cH_0$. Therefore, {\em the complement
of a $\vartheta$-normal subspace is $\vartheta$-normal.}

The next step is again a standard one: If a $\vartheta$-normal
subspace $\cH_0$ contains a proper $\vartheta$-normal subspace
$\cH_1$ then $\cH_2 = (\cH_1)^{\perp} \cap \cH_0$ is a further
$\vartheta$-normal subspace contained in $\cH_0$ and orthogonal
to $\cH_1$. A {\em minimal} $\vartheta$-normal subspace is a
$\vartheta$-normal subspace which does not contain any proper
$\vartheta$-normal subspace. Because $\cH$ is finite dimensional,
there are minimal $\vartheta$-normal subspaces.
\begin{prop} \label{propW1}
Let $\vartheta$ be antilinear and $\cH$ of finite dimension.
Then there is an orthogonal decomposition of $\cH$ into
minimal $\vartheta$-normal subspaces.
\end{prop}
Notice: $\cH$ may not contain any proper $\vartheta$-normal
subspace.

The intersection of two $\vartheta$-normal subspaces is
a $\vartheta$-normal subspace. If one of the two subspaces
is minimal, then their intersection is either the minimal
$\vartheta$-normal subspace itself or it consists of the zero
vector only, see also \cite{HP12}. Hence:
\begin{prop} \label{propW2}
Any $\vartheta$-normal subspace can be decomposed into mutually
orthogonal minimal $\vartheta$-normal subspaces.
\end{prop}

Now let $\vartheta$ be an antilinear {\em normal} operator.

From $\vartheta \phi = 0$ it follows $\vartheta^{\dag} \phi = 0$
and vice vera: The null space ker$[\vartheta]$ of $\vartheta$
is $\vartheta$-normal. Hence the orthogonal complement
ker$[\vartheta]^{\perp}$ of the null space is $\vartheta$-normal.
It follows, because $\cH_0^{\perp}$ is finite dimensional,
that $\vartheta$ is invertible on ker$[\vartheta]^{\perp}$
By that fact we can ignore the kernel of $\vartheta$ in what
follows and start, without loss of generality, with the
assumption of an invertible and normal $\vartheta$.

This assumption implies the uniqueness of the polar decomposition:
There is a unique antiunitary operator $\theta$ such that
$\vartheta = \theta |\vartheta| = |\vartheta| \theta$.
Therefore, $\vartheta$, $\vartheta^{\dag}$, $|\vartheta|$,
and the unitary operator $\theta^2$ is a set of mutually
commuting operators.

Hence there is a complete set of
common eigenvectors of the linear operators
$|\vartheta|$, $\vartheta^2$, and $\theta^2$. Let $\phi$ a unit
eigenvector for these operators. Now we can assume
\begin{equation} \label{Pdeco2}
|\vartheta| \phi = s \phi, \quad \theta^2 \phi= \epsilon^2 \phi,
\quad \vartheta^2 \phi = \epsilon^2 s^2 \phi
\end{equation}
with $s>0$ and a unimodular $\epsilon$, determined up to
a sign by (\ref{Pdeco2}). Both numbers can be described by
$z = \epsilon s$, again up to a sign.
This ambiguity is respected in the following notation:
\begin{equation} \label{Pdeco4}
\cH_{\pm z} = \{\phi \in \cH\, | \, |\vartheta| \phi =
s \phi, \; \theta^2 \phi = \epsilon^2 \phi,
\; z = \epsilon s \} \; .
\end{equation}
As the notation indicates, these subspaces can be characterized
also by
\begin{equation} \label{Pdeco4a}
\cH_{\pm z} = \{\phi \in \cH\, | \quad
\vartheta^2 \phi = z^2 \phi \} \; .
\end{equation}
Indeed, (\ref{Pdeco4a}) is $\vartheta^2$-invariant and the
linear operator $\vartheta^2$ is normal. Hence (\ref{Pdeco4a})
is necessarily $|\vartheta^2|$-invariant and $\pm \epsilon$
fulfills $z^2 = \epsilon^2 |z|^2$. This proves:
\begin{prop} \label{propW3}
The spaces defined by (\ref{Pdeco4a}) are $\vartheta$-normal
for normal $\vartheta$. There is a unique orthogonal decomposition
\begin{equation} \label{Pdeco5}
\cH = \bigoplus_z \cH_{\pm z} , \quad
z^2 \in {\rm spec}(\vartheta^2) \; ,
\end{equation}
Every minimal $\vartheta$-invariant subspace belongs to just one
of these subspaces.
\end{prop}
The last assertion is evident: A minimal $\vartheta$-invariant
subspace is either contained in a given $\vartheta$-invariant
subspace or their intersection contains the 0-vector only.
\medskip

The antilinear operator $\vartheta$ is reduced to
$|z| \theta$ on $\cH_{\pm z}$, i.~e. to a positive multiple
of an antiunitary operator. By this observation {\em the
classification of normal antilinear operators becomes the
classification of antiunitary operators supported
by subspaces of type $\cH_{\pm z}$.}

A subspace, invariant under the action of an antiunitary
$\theta$, is $\theta^{\dag}$-invariant: Being finite
dimensional, the subspace is mapped onto itself by $\theta$.
So it does $\theta^{\dag} = \theta^{-1}$. In particular,
any minimal $\vartheta$-invariant subspace is
$\vartheta$-normal. Combining with corollary \ref{propW2}
results in
one of the possible forms of Wigner's classification of
antiunitaries, \cite{Wi60}, and its extension to normal
antiunitary operators by Herbut and Vuji{\v c}i{\' c},
\cite{HV66}:
\begin{thm}[Wigner; Herbut, Vuji{\v c}i{\' c}] \label{thmW4}
Let $\vartheta$ be normal and $\dim \cH < \infty$.
Every $\vartheta$-invariant subspace is
$\vartheta^{\dag}$-invariant.
$\cH$ can be decomposed into an orthogonal sum of minimal
$\vartheta$-invariant subspaces.
A minimal invariant subspace is\\
either 1-dimensional, generated by an eigenvector of
$\vartheta$ and contained in a subspace $\cH_{\pm z}$ with
$z = z^*$, \\
or it is 2-dimensional, contained in a subspace
$\cH_{\pm z}$ with $z \neq z^*$ allowing for a basis
$\phi', \phi''$ such that $\vartheta \phi' = z^* \phi''$,
$\vartheta \phi'' = z \phi'$.
\end{thm}
It remains to show that a minimal subspace, say $\cH_{\rm min}$,
is either 1- or 2-dimensional. It is obvious that a
1-dimensional subspace
is minimal and generated by an eigenvector of $\vartheta$.
Now let $\dim \cH_{\rm min} > 1$ and $\phi'$ one
of its unit vectors. Define $\phi''$ by
$\vartheta \phi' = z^* \phi''$.
Now $\vartheta^2 \phi' = z \vartheta \phi''$ and,
by assumption, $\vartheta^2 \phi' = z^2 \phi'$.
Hence $\vartheta \phi'' = z \phi'$. The subspace
$\cH_{\rm min}$ should not contain an eigenvector of
$\vartheta$. Because $|z| \theta = \vartheta$ on
$\cH_{\rm min}$ we can rely on proposition \ref{normalantiu}
to exclude $z = z^*$.
\begin{cor}[Wigner; Herbut, Vuji{\v c}i{\' c}] \label{WHVa}
There is a unique orthogonal decomposition
$\cH = \cH' \oplus \cH''$ into $\vartheta$-normal subspaces with
the following properties: If restricted to $\cH'$, $\vartheta$ is
Hermitian and can be diagonalized. $\cH''$ is even dimensional
and there is no
eigenvector of $\vartheta$ in $\cH''$. In degenerate cases one of
the two subspaces is absent and the other one is the whole of
$\cH$.
\end{cor}
\begin{cor}[Wigner; Herbut, Vuji{\v c}i{\' c}] \label{WHVb}
If $\vartheta$ is normal, it allows for a block matrix
representation with blocks of dimensions not exceeding two.
The $1\times 1$ blocks contain eigenvalues of $\vartheta$,
the $2 \times 2$ block are filled with zeros in the diagonal
and with pairs $z, z^*$, $z \neq z^*$, as off-diagonal
entries.
\end{cor}

\subsubsection{Conjugations} \label{C4.2.1a}
The decomposition of conjugations a la Wigner is a simple
particular case. A conjugation $\theta$ allows for a basis
$\phi_1, \dots, \phi_d$ such that $\theta \, \phi_j = t_j \phi_j$
with positive real $t_j$ because $\theta$ is Hermitian.
As $\theta^2 = \1$, all $t_j = 1$. It will be shown in
section \ref{C5} that the real linear hull of the basis
$\{ \phi_j \}$ uniquely characterizes $\theta$, \cite{Hu12}
\medskip

The next topic are relations between two and more conjugations.
Let $\theta$ and $\theta'$ be conjugations. Then
$U = \theta' \theta$ is unitary and its $d$ eigenvalues
are unimodular numbers. The trace of $U$ is the sum of
these eigenvalues and its absolute sum is bounded by $\dim \cH$.
\begin{equation} \label{2conj2}
|\; \T \, \theta' \theta \; | \leq \dim \cH \; .
\end{equation}
If this bound is reached, the unitary $\theta' \theta$ is a
multiple of $\1$. Thus
\begin{equation} \label{2conj3}
| \; \T \, \theta' \theta \; | = \dim \cH \, \Leftrightarrow
\, \theta = \epsilon \, \theta'
\end{equation}
with $|\epsilon| = 1$.

If $\theta$ and $\theta'$ commute, $\cH_{\theta}$ is
$\theta'$-invariant. With $\phi \in \cH_{\theta}$ one gets
$\theta' (\phi \pm \theta' \phi) = \theta' \phi \pm \phi$.
Hence $\cH_{\theta}$ splits into the real Hilbert subspace
of the vectors $\phi$ satisfying
$\theta \phi = \theta' \phi = \phi$ and into the subspace of all
vectors fulfilling $\theta \phi = - \theta' \phi = \phi$,
showing
\begin{equation} \label{2conj4}
\theta \theta' = \theta' \theta \, \Rightarrow \,
\T \, \theta' \theta =
\dim (\cH_{\theta} \cap \cH_{\theta'}) -
\dim (\cH_{\theta} \cap \cH_{- \theta'})
\end{equation}
One observes that the trace of the product of two commuting
conjugations is an integer.

One can associate to a given basis $\phi_1, \phi_2, \dots$ a set
of $2^d$  mutually commuting conjugations: For any subset $E$
of basis vectors one defines $\theta_E$ by
$\theta_E \phi_j = \phi_j$ if $\phi_j \in E$, and by
$\theta_E \phi_k = -\phi_k$ if $\phi_j \notin E$. Going through
all the subsets one gets mutually commuting conjugations.
\medskip

Some further easy relations: With two conjugations $\theta_1$
and $\theta_2$, $U = \theta_1 \theta_2$ is unitary, \cite{GL65}.
If $\phi_j$,
$j = 1,\dots d$, is an eigenvector basis of $U$, let $\theta_3$
the conjugation satisfying $\theta_3 \phi_j = \phi_j$ for all $j$.
Then $\theta_3 U^{\dag} \theta_3 = U$ and
$\theta_3 \theta_2 \theta_1 \theta_3 = \theta_1 \theta_2$. Thus
\begin{lem}
Given two conjugations, $\theta_1$ and $\theta_2$, There is a
third one, $\theta_3$, such that
\begin{equation} \label{2conj5}
\theta_3 \theta_2 \theta_1 = \theta_1 \theta_2 \theta_3
\end{equation}
and $\theta := \theta_1 \theta_2 \theta_3$ is a conjugation.
\end{lem}
A further elementary fact is stated by
\begin{lem} \label{threecon1}
Every unitary operator is the product of two conjugations.\\
Every antiunitary operator is the product of three conjugations.
\end{lem}
Indeed, let $\phi_1, \dots$ denote an eigenbasis of the unitary
$U$, $U \phi_j = \epsilon_j \phi_j$. There are two conjugations
satisfying $\theta_1 \phi_j = \epsilon_j \phi_j$ and
$\theta_2 \phi_j = \phi_j$ for all $j$. Now the linear operator
$\theta_1 \theta_2$ transforms $\phi_j$ into $\epsilon \phi_j$.
Therefore it must be the linear operator $U$. The other assertion
is now trivial as any antiunitary is a product of a unitary
operator and a conjugation.
\begin{cor}
Given $2n$ conjugations $\theta_1, \dots, \theta_{2n}$. There
exists a conjugation $\theta_{2n+1}$ such that
\begin{equation} \label{2conj6}
\theta = \theta_1 \cdots \theta_{2n} \theta_{2n+1}
\end{equation}
is a conjugation.
\end{cor}
To see it write $U = \theta_1 \cdots \theta_{2n}$. Being a unitary
it is a product of two conjugation: $U = \theta \theta_{2n+1}$.
Eliminating $U$ yields (\ref{2conj6}).

\subsubsection{Skew conjugations} \label{C4.2.1b}
Let $\theta$ be a skew conjugation. Then the dimension
of the Hilbert space must be even, $d = 2n$. By Wigner's
theorem $\cH$ can be decomposed as an orthogonal direct
sum
\begin{equation} \label{sco1}
\cH = \cH_1 \oplus \dots \oplus \cH_n, \quad \dim \cH_j = 2,
\end{equation}
of irreducible $\theta$-invariant subspaces,
\begin{equation} \label{sco2}
\theta \, \cH_j = \cH_j, \quad j = 1,2,, \dots, n \; .
\end{equation}
From $\theta^2 = - \1$ follows: $\theta_j$ acts on $\cH_j$ as
a multiple $(\exp i s) \tau_0$ of $\tau_0$. Choosing in every
$\cH_j$ a unit vector $\phi_{2j}$, then
$\psi_{2j-1} := \theta \, \psi_{2j}$ is a unit vector orthogonal
to $\psi_j$, Hence the pair
$\psi_{2j}, \, \psi_{2j-1}$ is a basis of $\cH_j$ such that
\begin{equation} \label{sco3}
\theta \, \psi_{2j} = \psi_{2j-1}, \quad
\theta \, \psi_{2j-1} = - \psi_{2j} \; ,
\end{equation}
saying that $\theta$ is an orthogonal direct sum of
$\tau_0$-operators with respect to a suitably chosen basis:
\begin{prop} \label{scon}
Let $\dim \cH = 2n$ and let $\theta$ be a skew conjugation.
Then there is a basis
$\{ \phi_j \}$ of $\cH$ such that
\begin{equation} \label{sc4}
\theta \, \phi_{2k} = - \phi_{2k-1}
, \quad
\theta \, \phi_{2k-1} = \phi_{2k}
\end{equation}
for all $k = 1, \dots n$ .
\end{prop}
From the proposition follows
\begin{cor} \label{cscon}
Let $\theta$ be a skew conjugation. There are orthogonal
decompositions $\cH = \cH_a \oplus \cH_b$ such that
for all $\phi_a \in \cH_a$ and $\phi_b := \theta \, \phi_a$
one gets $\theta \, \phi_b = - \phi_a$.
\end{cor}
Indeed, with a basis fulfilling (\ref{sc4}) let $\cH_a$ be
the complex linear hull of the $n$ vectors $\phi_{2k}$
and $\cH_b$ the complex linear hull of the $n$ vectors
$\phi_{2k-1}$. Clearly, $\cH$ is an orthogonal
sum of these two Hilbert subspaces. Furthermore, for all
$\phi_a \in \cH_a$
\begin{displaymath}
\phi_a = \sum_{k=1}^n c_{2k} \phi_{2k}, \quad \phi_b := \theta
\, \phi_a = \sum_{k=1}^n c_{2k}^* \phi_{2k-1}
\end{displaymath}
with $\phi_b \in \cH_b$. Now $\theta^2 = -\1$ shows
$\theta \, \phi_b = - \phi_a$.

Analogue to lemma \ref{threecon1} one can show
\begin{lem} \label{threecon2}
Every antiunitary operator can be represented by a product of
two conjugations and a skew conjugation. The position of the
skew conjugation can be fixed in advance.
\end{lem}

\subsubsection{The number of orthogonal (skew)
conjugations} \label{C4.3}
For the purpose of just this subsection let $N_{+}(d)$ be the
maximal possible number $n$ of conjugations,
$\theta_1, \dots, \theta_n$, such that
$(\theta_j, \theta_k) = \delta_{jk} d$. Similarly $N_{-}(d)$ is
the maximal possible number of skew conjugations,
$\theta_{-1}, \dots, \theta_{-m}$, with
$(\theta_{-j}, \theta_{-k}) = - \delta_{jk} d$.
These numbers are bounded from above by (\ref{Hform4}):
\begin{equation} \label{bconj1}
N_{+}(d) \leq \frac{d(d+1)}{2}, \quad N_{-}(d) \leq
\frac{d(d-1)}{2} \; .
\end{equation}
There are bounds from below too. Let $\theta_1$ and $\theta_2$
denote antilinear operators on Hilbert spaces $\cH_1$ and $\cH_2$
of dimensions $d_1$ and $d_2$ respectively. The operator
$\theta =\theta_1 \otimes \theta_2$ is an antilinear operator on
$\cH_1 \otimes \cH_2$. If $\theta_1$ and $\theta_2$ are
conjugations, or if both are skew conjugations, then $\theta$
is a conjugation. If one of them is a conjugation and the other
a skew conjugation, $\theta$ is a skew conjugation.

It is now possible to conclude
\begin{eqnarray} \label{bconj2}
N_{+}(d_1 d_2) & \geq & N_{+}(d_1) N_{+}(d_2)
+ N_{-}(d_1) N_{-}(d_2)
 , \label{bconj2a} \\
N_{-}(d_1 d_2) & \geq & N_{+}(d_1) N_{-}(d_2)
+ N_{-}(d_1) N_{+}(d_2)
\label{bconj2b} \; .
\end{eqnarray}
Assuming now that equality holds in (\ref{bconj1}), it follows
from (\ref{bconj2a}), (\ref{bconj2b})
\begin{displaymath}
N_{+}(d_1 d_2) \geq \frac{(d_1+d_2)(d_1+d_2+1)}{2}, \quad
N_{-}(d_1 d_2) \geq \frac{(d_1+d_2)(d_1+d_2-1)}{2}
\end{displaymath}
as a straightforward calculation establishes. Again by
(\ref{bconj1}), the last two inequalities must be equalities.
\begin{prop}
The set of Hilbert spaces for which equality holds in
(\ref{bconj1}) is closed under performing direct products.

If the dimension $d$ of an Hilbert space $\cH$ is a power of 2,
$d = 2^n$, $n >1$, then there are $d(d+1)/2$ conjugations,
$\theta_1, \theta_2, \dots$, and $d(d-1)/2$ skew conjugations,
$\theta_1', \theta_2', \dots$, such that
\begin{equation} \label{bconj3}
(\theta_j, \theta_k) = d \, \delta_{jk}, \quad
(\theta'_j, \theta'_k) = - d \, \delta_{jk}, \quad
(\theta_j, \theta_k') = 0 \; .
\end{equation}
\end{prop}
The first part of the proposition has already been proved.
For its second part it suffices that the assertion is true
for $d=2$. Indeed, the conjugations $\tau_j$, $j= 0,2,3,4,$
show it.

Choosing for every $\theta_j'$, $j = 1, \dots, N_{+}(d)$
an invariant basis, $\phi_1^j, \dots, \phi_d^j$, the
orthogonality relations (\ref{bconj3}) can be rewritten by
the use of (\ref{bconj1}). It follows
\begin{cor}
There are $N_{+}(d)$ bases $\phi_1^j, \dots, \phi_d^j$ such that
\begin{equation} \label{bconj4}
\sum_{n,m} \langle \phi_n^j , \phi_m^k\rangle^2 = d \delta_{jk}
\end{equation}
is true for all $j,k$ in $\{1, 2, \dots, N_{+}(d)\}$ .
\end{cor}

\section{A look at elementary symplectic geometry} \label{C5}
Let $\theta$ be a conjugation and define
\begin{equation} \label{conj2}
\cH_{\theta} := \{ \phi \in \cH \, : \, \theta \phi = \phi \} \;.
\end{equation}
By (\ref{isom1}) we see that $\phi_1, \phi_2 \in \cH_{\theta}$
implies $\langle \phi_1, \phi_2\rangle =
\langle \phi_2, \phi_1\rangle$. Hence $\cH_{\theta}$ is a
{\em real Hilbert subspace.} The {\em real} dimension of
$\cH_{\theta}$ is $d = \dim \cH$. Therefore it is a {\em maximal
real Hilbert subspace} of $\cH$ consisting of all vectors of the
form $\phi = (\psi + \theta \psi)$, $\psi \in \cH$. Any maximal
real Hilbert subspace is of the form $\cH_{\theta}$ with a unique
conjugation $\theta$.
\begin{prop} \label{prconj}
There is a one-to-one correspondence
\begin{equation} \label{conj3}
\theta \, \Leftrightarrow \, \cH_{\theta}  \; ,
\end{equation}
between the set of conjugations and the set of maximal real
Hilbert subspaces.
\end{prop}
This simple observation opens the door from conjugations to
elementary symplectic geometry.

Note also
\begin{displaymath}
\cH = \cH_{\theta} + i \cH_{\theta}, \quad
i \cH_{\theta} = \cH_{-\theta} \; ,
\end{displaymath}

\subsection{Conjugations and Symplectic Geometry} \label{C5.1}
Symplectic Geometry is an eminent topic in its own. See
\cite{RS13}, \cite{daSilva}, \cite{Gosson}, or any monograph on
symplectic geometry. Only some elementary comments, concerning
the correspondence (\ref{conj3}) in proposition \ref{prconj},
can be given here.

$\cH$, if considered  as a real linear space, is
naturally equipped with the symplectic\footnote{A
non-degenerate skew symmetric real bilinear form.} form
\begin{equation} \label{simplect1}
\Omega(\phi, \phi') := \frac{\langle\phi, \phi'\rangle
- \langle\phi',\phi\rangle}{2i} \;.
\end{equation}
A real linear subspace of $\cH$ on which this form vanishes, is
called {\em isotropic.} If it is a maximal isotropic one, its
real dimension is $d$, and it is called a {\em Lagrangian}
one or a {\em Lagrangian plane.} By this definition {\em the
Lagrangian subspaces are just the real Hilbert subspaces
$\cH_{\theta}$ with $\theta$ a conjugation.}

The set of all Lagrange subspaces is a compact smooth manifold
called the {\em Lagrangian Grassmannian} of $\cH$.
The Lagrangian Grassmannian is usually denoted by
$\Lambda_d$, $\Lambda[\cH]$, or Lag$[\cH]$. By (\ref{conj3})
\begin{prop} \label{simprep}
$\Lambda[\cH]$ is symplectomorphic to the manifold of all
conjugations on $\cH$.
\end{prop}
Fixing a Lagrangian subspace $\cH_{\theta}$, its isometries form
an orthogonal group ${\cal O}(\cH_{\theta}) \simeq {\cal O}(d)$.
Because the unitary group ${\cU}(\cH) \simeq {\cU}(d)$ acts
transitively on the Lagrangian subspaces, as all bases are unitary
equivalent, one gets the well known isomorphism
$\Lambda_d \simeq {\cU}(d) / {\cal O}(d)$. The dimension of the
manifold $\Lambda_d$ is $(d^2 + d)/2$. A useful relation is
\begin{equation} \label{simplect1a}
U \, \cH_{\theta} = \cH_{\theta'} \, \Leftrightarrow \,
\theta' = U \theta U^{-1} \;, \quad U \in {\cU}(\cH) \; .
\end{equation}
Next is to show: {\em The acq--lines going through $\theta_0$
cover a neighborhood of $\theta_0$}.\footnote{``acq''
abbreviates ``antilinear conjugate quandle.'' See subsection
\ref{C1.1} for more.}
\begin{lem}
Given a conjugation $\theta_0$ and a tangent $\eta_0$ at
$\theta_0$. There is an Hermitian $H$ commuting with
$\theta_0$ such that the tangent of
\begin{equation} \label{acqsp0}
t \mapsto \theta_s = U_H(t) \theta_0 U_H(-t), \quad
U_H(t) = \exp{itH} \; ,
\end{equation}
at $\theta_0$ is equal to $\eta_0$.
\end{lem}
Proof: There is an Hermitian $A$ such that
\begin{displaymath}
t \mapsto \theta'_s = U_A(t) \theta_0 U_A(-t), \quad
U_A(t) = \exp{itA}
\end{displaymath}
has tangent $\eta_0$ at $\theta_0$. This implies $\eta_0 = i(A
\theta_0 + \theta_0 A)$. Set $2H :=A + \theta_0 A \theta_0$
for the path (\ref{acqsp0}). Then $\theta_0 H = H \theta_0$ and
$\eta_0 = i(H \theta_0 + \theta_0 H) = 2i H \theta_0$.
\begin{prop} \label{spacq}
Given a conjugation $\theta$ and an Hermitian operator $H$ such\\
that $H \, \cH_{\theta} \subset \cH_{\theta}$ or, equivalently,
$H \theta = \theta H$. Let
\begin{equation} \label{acqs}
t \mapsto U_H(t) = \exp {it H}, \quad \theta_t = e^{itH} \theta
e^{-itH} \; .
\end{equation}
Then $H$ commutes with $\theta_t$ for all $t \in \mathbb{R}$ and
\begin{equation} \label{acqsp1}
\frac{d \theta_t}{d t} = i(\theta_t H + H \theta_t)
= 2i H \theta_t \; .
\end{equation}
Moreover, $t \to \theta_t$ is an acq--line, i.e.
\begin{equation} \label{acqsp2}
\theta_r \theta_{(r+s)/2} = \theta_{(r+s)/2} \theta_r , \quad
\theta_r \theta_t = U_H(2r-2t)
\end{equation}
for all $r,s,t \in \mathbb{R}$.
\end{prop}
The core of the proof is lemma \ref{Hell}, saying that
(\ref{acqsp2}) follows if $\theta = \theta_0$ commutes with $H$.
The first equality sign in (\ref{acqsp1}) is true for all Hermitian
$H$, while the second one is due to the lemma above. Finally,
$\theta_t H = H \theta_t$ by definition (\ref{acqs}).
\medskip

With $t \to \theta_t$ also $t \to \theta_{bt}$ is an acq--line.
Hence, if this path returns to $\theta_0$ for some $t' \neq 0$,
one can assume $t' = \pi$ for the parameter of return.
\begin{cor}
Let $t \to \theta_t$ be as in the proposition. The equality
$\theta_0 = \theta_{\pi}$ takes place if and only if all
eigenvalues of $H$ are integers.
\end{cor}
Let $\theta_0 \phi = \phi$ and $H \phi = a \phi$. If
$\theta_{\pi} = \theta_0$ then $(\exp i \pi H) \phi = \pm \phi$.
Hence $a$ must be an integer. Hence all eigenvalues of $H$ must
be integers.
\medskip

Let $H$ and $t \to \theta_t$ be as in proposition \ref{spacq} and
$\theta_{\pi} = \theta_0$. Then there are mutually orthogonal rank
one projection operators $P_j$ and $n_j \in \mathbb{Z}$ such that
\begin{equation} \label{cycl1}
H = \sum n_j P_j, \quad P_k \theta_s = \theta_s P_k
\end{equation}
for $k = 1,\dots,d$. As one knows that (\ref{acqsp0}) is a
generator for H$_1(\Lambda, \mathbb{Z})$ if $H$ is a projection
operator of rank one, the map
\begin{equation} \label{homo}
\{ H \,: \, {\rm spec}{H} \in \mathbb{Z} \} \longrightarrow
[H] \in {\rm H}_1(\Lambda, \mathbb{Z})
\end{equation}
is onto. If $H$ is mapped by (\ref{homo}) onto $[H]$, then
\begin{equation} \label{homo1}
[H] = \sum n_j [P_j] \; .
\end{equation}

One can find a closed curve generating H$_1[\Lambda, \mathbb{Z}]$
as follows, \cite{RS13},: Let $\phi_1,\dots,\phi_d$ be a basis of
$\cH$. Define the conjugation $\theta$ by $\theta \phi_k = \phi_k$
for all $k$, the unitaries $U(t)$ by $U(t) \phi_1 = (\exp it)
\phi_1$, and by $U(t) \phi_j = \phi_j$ if $j > 1$. Then $H$ is the
projection operator $P = |\phi_1\rangle\langle \phi_1|$ onto
$\phi_1$, i.~e.
\begin{equation} \label{homo2}
t \to \theta_t := U(t) \theta U(-t), \quad U(t) = \exp itP
\end{equation}
is a generator for H$_1[\Lambda, \mathbb{Z}]$.
\medskip

\noindent \underline{An inequality}. Let $\lambda_1, \dots$ denote
the eigenvalues of $H$, The general acq--line (\ref{acqs}) in
$\cB(\cH)_{\rc}^+$ is contained in the sphere of radius $\sqrt{d}$
around the null vector. Any piece of it has a well defined length
\begin{equation} \label{cycl2}
\int_{t'}^{t''} \T \, \sqrt{\dot \theta_s^2} \, ds = 2(t''-t')
\T \, \sqrt{H^2} = 2 (t''-t') \sqrt{\sum a_j^2} \; .
\end{equation}
Indeed, $\dot \theta_s^2 = (H \theta_s + \theta_s H)^2$ which is
equal to $4H^2$ by virtue of (\ref{acqs}). Taking into account
(\ref{cycl1}) one obtains (\ref{cycl2}).

Assuming now $\theta_0 = \theta_{\pi}$, the numbers $a_j$ become
integers $n_j$. Then
\begin{equation} \label{cycl3}
\int_0^{\pi} \sqrt{\dot \theta_s^2}ds = 2\pi \sqrt{\sum n_j^2} \;.
\end{equation}
{\em The shortest closed acq--lines are of length $2\pi$. They
are generators for H$_1[\Lambda, \mathbb{Z}]$. }

\subsection{The canonical differential 1-form} \label{C5.1.1}
Seeing $\Lambda_d$ is a submanifold of the Hermitian part
$\cB(\cH)_{\rc}^{+}$ of $\cB(\cH)_{\rc}$, the differential
${\bf d}\theta$ is well defined on the Lagrangian Grassmannian:
It is the restriction onto $\Lambda$ of ${\bf d} \vartheta$,
$\vartheta \in \cB(\cH)_{\rc}^{+}$. The operator valued
differential 1-form
\begin{equation} \label{simplect2}
\nu := \theta {\bf d} \theta
\end{equation}
is skew symmetric, $\nu + \nu^{\dag} =0$, as following from
$\theta^2 = \1$ and $\theta^{\dag} = \theta$.
\begin{prop}
The differential 1-form
\begin{equation} \label{simplect4}
\tilde \nu := \frac{1}{2\pi i} \T \, \nu = \frac{1}{2\pi i}
\T \, \theta {\bf d} \theta, \quad \tilde \nu^{\dag} = \nu \;,
\end{equation}
is closed, but not exact. It is a unitary invariant. It changes
its sign by antiunitary transformations.
\end{prop}
Proof: Fix $U$. The unitary invariance follows from
$U \theta {\bf d}\, \theta U^{\dag} = (U \theta U^{\dag})
{\bf d} (U \theta U^{\dag})$ by taking the trace. If $U$ is
replaced by an antiunitary, the sign change is seen from
(\ref{rule7}). To show ${\bf d} \tilde \nu = 0$, the
representation $U \to \theta := U \theta_0 U^{\dag}$ with $U$
varying in ${\cU}(d)$ and an arbitrarily chosen conjugation
$\theta_0$ is inserted,
\begin{displaymath}
\theta {\bf d}\theta = U \theta_0 U^{\dag} {\bf d}
(U \theta_0 U^{\dag}) = U (\theta_0 U^{\dag})
({\bf d} U \theta_0) U^{\dag} + U {\bf d} U^{\dag} \; .
\end{displaymath}
Taking the trace one finds
\begin{equation} \label{simplectp1}
\T \,\theta {\bf d}\theta = 2 \T \, U {\bf d}\, U^{\dag} \; .
\end{equation}
The equality of both summands follows by setting
$X = U^{\dag} {\bf d} U$ and $\Theta = \theta_0$ in
(\ref{rule7}). The right hand side of (\ref{simplectp1})
can simplified further: In the vicinity
of  the identity map there is a unique logarithm
$iH = \ln U$ with $H = H^{\dag}$. Then
\begin{equation} \label{simplectp2}
\T \, U {\bf d} U^{\dag} = -i {\bf d} \, \T \; H ,
\quad U = \exp iH \; .
\end{equation}
Hence, by (\ref{simplectp1}), the differential 1-form
$\tilde \nu$ is closed.
\begin{prop}
The 1--form
\begin{equation} \label{homo4}
 \tilde \nu := \frac{1}{2 \pi i} \T \, \nu
\end{equation}
generates the first integer--valued cohomology group
H$^{1}(\Lambda, \mathbb{Z})$.
\end{prop}

Let $H$ and the curve $\gamma \,:\, t \to \theta_t$ be as in
proposition \ref{spacq}. Assuming $\theta_0 = \theta_{\pi}$,
$\gamma$ becomes a closed curve. By (\ref{acqsp1})
\begin{equation} \label{simplect5}
\int_{\gamma} \nu = \int_0^{\pi} \theta_s \dot \theta_s ds
= 2 \pi i \, H \; .
\end{equation}
The eigenvalues of $H$ are integers, $n_j$, for closed
acq--lines. Taking the trace yields
\begin{equation} \label{simplect6}
\int_{\gamma} \tilde \nu =
\frac{1}{2 \pi i}  \T \, \int_{\gamma} \nu = \sum n_j \; .
\end{equation}
One knows that $\gamma$ generates the first homology group if
$H$ is a rank one projection operator \cite{RS13}. Hence
$\tilde \nu$ is a generator of the first integer--valued
cohomology group. According to V.~I.~Arnold its integral over a
closed curve provides its {\em Maslov Index} \cite{Gosson}.
\begin{cor}
$\tilde \nu$ is a generator of H$^1[\Lambda, \mathbb{Z}]$. For
 a closed curve $\gamma$ in $\Lambda_d$
\begin{equation} \label{simplect7}
{\rm Maslov}[\gamma] =
\int_{\gamma} \tilde \nu \in \mathbb{Z}
\end{equation}
is the Maslov index of $\gamma$.
\end{cor}
The right hand side of (\ref{simplectp1}) can be rewritten in
terms of unitary operators. By (\ref{simplectp1})
\begin{displaymath}
\T \nu = 2 \T U {\bf d} U^{-1} = -2i {\bf d} \T H \; .
\end{displaymath}
In small enough open sets one has
$\det U = \det \exp iH = \exp \T \, iH$. Hence
\begin{displaymath}
{\bf d} \det U = {\bf d} \exp \T \, iH =
i({\bf d} \T \, H) (\det U) \;.
\end{displaymath}
One gets one of the known expressions for the Maclov index:
\begin{equation} \label{simplect7a}
{\rm Maslov}[\gamma] =
\frac{i}{\pi} \int_{\gamma} \frac{{\bf d}
\det U}{\det U} , \quad \theta = U \theta_0 U^{\dag} \: .
\end{equation}
{\bf Remark:} \\
$\nu$ is not closed for $d \geq 2$. Therefore the
closed operator valued differential 2-form
\begin{equation} \label{simplect3}
\omega := {\bf d}\, \nu = {\bf d}\, \theta \wedge {\bf d} \theta
\end{equation}
may be of interest. The case of $\Lambda_2$ will be examined:\\
According to lemma \ref{c2.2.2b}, see also (\ref{example2.11}),
a conjugation can be written as a real linear combination of the
antilinear Hermitian Pauli matrices $\tau_j$, $j=1,2,3$,
multiplied by a phase $\epsilon$, in the form
\begin{equation} \label{simplect.d}
\theta = \epsilon \sum_{j=1}^3 x_j \tau_j, \quad
\sum_{j=1}^3 x_j^2 = 1 \; ,
\end{equation}
Varying $\theta$ within these constraints defines a space
topological equivalent to the product
${\rm S}^1 \times {\rm S}^2$ of a circle and a 2-sphere.
Locally this remains true for $\Lambda_2$. However, by
(\ref{simplect.d}), every $\theta$ is represented by two couples,
$\epsilon, \vec{x}$ and $-\epsilon, -\vec{x}$. One gets the
well known fact: {\em $\Lambda_2$ is topological the product of a
1- and a 2-sphere on which the points $\epsilon, \vec{x}$ and
$-\epsilon, -\vec{x}$ are identified.} Writing
$\theta = \epsilon \theta_x$, one obtains
\begin{equation} \label{simplecte}
\nu = \epsilon {\bf d} \epsilon^* \1_2 +
\theta_x {\bf d} \theta_x \; .
\end{equation}
Using now (\ref{simplecte}), the differential form
(\ref{simplect3}) becomes $( {\bf d} \theta_x) \wedge ( {\bf d}
\theta_x)$. Therefore,
\begin{equation} \label{simplectg}
\omega =
\sum_{j \neq k} {\bf d} x_j \wedge {\bf d} x_k \tau_j \tau_k =
2 \sum_{j < k} \tau_j \tau_k \, {\bf d} x_j \wedge {\bf d} x_k\; .
\end{equation}
By (\ref{Pauli5}) one obtains
\begin{equation} \label{simplecth}
 \omega = 2i ( \sigma_3 \, {\bf d} x_1 \wedge {\bf d} x_2
 + \sigma_1 \, {\bf d} x_2 \wedge {\bf d} x_3  +
\sigma_2 \, {\bf d} x_3 \wedge {\bf d} x_1 ) \; .
\end{equation}

\section{Equivalence relations} \label{C6}
There is a lot of literature on equivalence relations
between matrices.
A good part can be found in Horn and Johnsons's
``Matrix Analysis'', \cite{Horn90}. Concerning
more recent results, much of the following is based on
papers by L.~Balayan, S.~R.~Garcia, D.~E.~Poore, M.~Putinar,
J.~E.~Tener, and W.~R.~Wogen, in particular on \cite{BG09},
\cite{GT11}, \cite{Te12}, \cite{GPT11}.

One purpose is to ``translate'' (and extend) some of their
theorems into the language of antilinearity. in which they become
basis-independent and, hopefully, of a more transparent structure,

Let $\theta$ be a conjugation.
The operator $\theta X^{\dag} \theta$ is called the
{\em $\theta$-transpose,} or simply the {\em transpose} $X^{\top}$
of $X$ if there is no danger of confusion. Then one writes
\begin{equation} \label{trpose2}
X^{\top} = \theta X^{\dag} \theta, \quad
X^{\dag} = \theta X^{\top} \theta \; .
\end{equation}
The transpose of an operator is defined relative to a basis
$\{ \phi_j \}$. There is a conjugation fulfilling
$\theta \phi_j = \phi_j$ for all its elements. The tranpose
$X^{\top}$ of a linear operator $X$ can be written
\begin{displaymath}
X \phi_j = \sum_i x_{ji} \phi_i, \quad
X^{\top} \phi_j = \sum_i x_{ij} \phi_i ,
\end{displaymath}
with respect to the given basis. It follows
\begin{displaymath}
\theta X^{\top} \theta \, \phi_j =
\theta X^{\top} \, \phi_j = X^{\dag} \phi_j,
\end{displaymath}
and $\theta X^{\top} \theta =  X^{\dag}$ as in (\ref{trpose2}).

One may look at $X \to \theta X^{\dag} \theta$ as at a
superoperator. Its fixpoints defines an important class of linear
operators, see \cite{GW09a} and \cite{GW09b}.

\subsection{Similarity, Congruence}
A bit of terminology: Let $X$, $Y$ two operators,
either both linear or both antilinear. $Y$ is called
{\em cosimilar} to $X$ if there is an invertible antilinear
operator $\vartheta$ such that $Y=\vartheta X \vartheta^{-1}$.
The definition mimics the {\em similarity relation}
$Y = AXA^{-1}$ between $X$ and $Y$ with $A$ linear and invertible.
Proposition \ref{diagonal} is an instructive example for
similarity between antilinear operators.

$X \rightarrow \vartheta X \vartheta^{-1}$ operates antilinearly
within $\cB(\cH)$ and within $\cB(\cH)_{\rc}$, while
$X \rightarrow AXA^{-1}$ operates linearly. There are more
differences: Cosimilarity is not an equivalence relation. Instead,
if $X$ is cosimilar to $Y$ and $Y$ cosimilar to $Z$ then $X$ is
similar to $Z$. Hence similarity relations could be
``factorized'' by a pair of cosimilar ones.
\medskip

One calls $Y$ {\em congruent} to $X$ if there exists a
{\em congruence relation} $Y = AXA^{\dag}$ with invertible
linear operator $A$. In the same manner as above, $Y$ is
{\em cocongruent} to $X$ if $Y = \vartheta X \vartheta^{\dag}$
takes place with an invertible antilinear $\vartheta$.

Cocongruence is not an equivalence relation: If $X$ is
cocongruent to $Y$ and $Y$ cocongruent to $Z$ then
$X$ is congruent to $Z$.
\medskip

Sometimes it is necessary to weaken these concepts to open the
door to another domain of research: Let $\vartheta$ be an
antilinear operator. Following Woronowicz, \cite{Wo76}, the
linear map
\begin{equation} \label{trpose1}
X \longrightarrow T(X) := \vartheta X^{\dag} \vartheta^{\dag}
\; , \quad X \in \cB(\cH) \; ,
\end{equation}
is called an {\em elementary copositive map} or an {\em elementary
copositive superoperator.} A map or a superoperator of $\cB(\cH)$
into itself\footnote{or into the operators of another Hilbert
space} is called {\em completely copositive} if it can be written
as a sum of elementary copositive maps.\\
These definitions mimic the {\em elementary positive maps}
$X \mapsto A^{\dag} X A$ and the {\em completely positive maps,}
which are sums of elementary positive maps,
see, a.~e., \cite{NC00}, \cite{BZ06}.

\noindent {\bf Remarks }  \\
{\bf 1.} The operator $\lambda \1$ is cosimilar to itself if
and only if $\lambda$ is real.\\
{\bf 2.} Any $X \in \cB(\cH)$ is similar to its transpose,
$X^{\top} = AXA^{-1}$. See part 3.2.3 in \cite{Horn90}.
By (\ref{trpose2}) this translates into
\begin{equation} \label{trpose3}
X^{\dag} = \vartheta X \vartheta^{-1}, \quad \vartheta \,
\hbox{ antilinear}
\end{equation}
with $\vartheta = \theta A$ and $\theta$ a conjugation as
in (\ref{trpose2}). Thus\\
{\em Every linear operator $X$ is cosimilar to $X^{\dag}$.}\\
{\bf 3.} A remarkable result of R.~A.~Horn and C.~R.~Johnson,
\cite{Horn90} theorem 4.4.9, reads: {\em Every matrix is similar
to a symmetric one.} As in the preceding example one gets:\\
{\em Every linear operator is cosimilar to an Hermitian one.}\\
Hence, given $X$, there is an invertible $A$ such that
\begin{displaymath}
\theta AXA^{-1} \theta = [AXA^{-1}]^{\dag} =
(A^{\dag})^{-1} X^{\dag} A^{\dag} \, .
\end{displaymath}
By defining $\vartheta := A^{\dag} \theta A$ one obtains a
variant of Horn and Johnson's result:\\ {\em To every $X \in
\cB(\cH)$ there is $\vartheta \in \cB(\cH)_{\rc}$ such that}
\begin{equation} \label{HJ1}
X^{\dag} = \vartheta X \vartheta^{-1}, \quad
\vartheta = \vartheta^{\dag} \,.
\end{equation}
{\bf 4.} Item (b) of proposition \ref{diagonal} states that
every diagonalisable  antilinear operator is similar to an
Hermitian one. An antilinear Hermitian operator $\vartheta'$
allows for a basis $\{ \phi_j \}$ of eigenvectors with
non-negative real eigenvalues. The conjugation $\theta$,
satisfying $\theta \phi_j = \phi_j$ for all $j$, commutes with
$\vartheta'$. If, therefore, $A \vartheta A^{-1} = \vartheta'$,
one gets, using the quoted results of Horn and Johnson,
$\vartheta' = (\theta A) \vartheta (\theta A)^{-1}$ and
$(\vartheta')^{\dag} = \vartheta'$.
\begin{lem} \label{adiagonal}
An antilinear operator is similar to an Hermitian one if and
only if it is cosimilar to an Hermitian one.
\end{lem}

\subsection{Unitary equivalence}
A matrix is said to be ``UET'' if it is unitarily equivalent to
its transpose. The questions which matrices are UET or
``UECSM'', an acronym for {\em unitarily equivalent to a complex
symmetric matrix,} goes back to P.~R.~Halmos, see \cite{Halmos}.
The abbreviations ``UET'' and  ``UECSM'' are used in the
mathematical literature.

{\em Unitary equivalence} of an operator $X$ to its transpose
means: There is a basis with respect to which the matrix
representation of $X$ is UET. By (\ref{trpose2}) this is
equivalent to the existence of an antiunitary such that
\begin{equation} \label{trpose5}
X^{\dag} = \Theta X \Theta^{-1}, \quad \Theta \,
\hbox{ antiunitary.}
\end{equation}
If there is a unitary $U$ such that $Y = UXU^{\dag}$
satisfies $Y = Y^{\top}$ in a matrix representation, then,
again by (\ref{trpose2}), there is a conjugation $\theta'$
such that $Y = \theta' Y^{\dag} \theta'$, Thus
\begin{displaymath}
UXU^{\dag} = Y = \theta' Y^{\dag} \theta' =
\theta' U X^{\dag} U^{\dag} \theta'
\end{displaymath}
and $UXU^{\dag}$ is equal to
$\theta' U X^{\dag} U^{\dag} \theta'$, i.~e.
$X^{\dag} = \theta X \theta$ with the conjugation
$\theta = U^{\dag} \theta' U$. This proves the first part of
a lemma due to Garcia and Tener \cite{GT11}.
\begin{lem} \label{GT1}
The following three items are equivalent:\\
a) \, $X$ is UECSM.\\
b) \, $X^{\dag}$ is antiunitarily equivalent to an Hermitian
operator.\\
c) \, There is a conjugation such that
$X^{\dag} = \theta X \theta$.
\end{lem}
Step c) $\to$ a) is simple: By (\ref{trpose2})
$\theta X \theta$ is the $\tau$-transpose $X^{\top}$ of $X$.
The latter is by c) equal to $X$. To be more
explicit one writes $\vartheta := \theta X^{\dag} = X \theta$.
Conjugations are Hermitian. Hence $\vartheta = \vartheta^{\dag}$.
Therefore $\langle \phi', X \theta \phi'' \rangle =
\langle \phi'', X \theta \phi' \rangle$ for any pair of vectors.
If $\phi', \phi'' \in \cH_{\theta}$, the $\theta$'s can be
skipped, and the matrix representation of $X$ is symmetric with
respect to any basis chosen from $\cH_{\theta}$.
\medskip

UET is less strong than UECSM. But sometimes they are equally
strong. An astonishing case has been settled by R.~S.~Garcia,
and J.~E.~Tener in \cite{GT11}:
\begin{thm}[Garcia, Tener] \label{GaTe}
If $\dim \cH < 8$ then a linear operator is antiunitarily
equivalent to an Hermitian one if and only if it is antiunitarily
equivalent to its Hermitian adjoint.
The assertion fails for some $X$ if $\dim \cH = 8$.
\end{thm}
The quoted authors could prove a decomposition
theorem\footnote{Please, consult their original paper,
theorem 1.1}. From it the theorem comes as a corollary.

To answer the question whether a given operator is UET or UECSM
is another issue. There are quite different approaches to obtain
criteria.

\subsection{Low dimensions}
One of the first results concerning the 2-dimensional case
is due to S.~L.~Woronowicz, see appendix of \cite{Wo76} or
\cite{St67} and the monograph \cite{St13}.
He called an operator $X$ {\em almost normal} if the rank of
$X^{\dag} X - X X^{\dag}$ is not larger than two and he proved
\begin{prop}[Woronowicz]
If $\dim \cH < \infty$ and if there are vectors
$\phi_1$, $\phi_2$ such that
\begin{equation} \label{Wo1}
X^{\dag} X - X X^{\dag} = |\phi_2><\phi_2| - |\phi_1><\phi_1|
\end{equation}
then there exists a conjugation $\theta$ fulfilling
\begin{equation} \label{Wo2}
\theta X^{\dag} \theta = X \; , \quad {\rm and} \quad
\theta \, \phi_1 = \phi_2 \; .
\end{equation}
\end{prop}
The proof is by constructing the algebra with defining relation
(\ref{Wo1}) and showing by induction along the degree of
monomials in $X$ and $X^{\dag}$ the existence of $\theta$.
\medskip

A completely other way is in characterizing unitary orbits by
values of unitary invariants.
A particular problem asks whether $X$ and its transpose
$X^{\top}$ belong to the same unitary orbit.
S.~R.~Garcia, D.~E.~Poore and J.~E.~Tener offer in \cite{GPT11}
a solution for the dimensions 3 and 4 by trace criteria.
\medskip

Let $\cU = \cU(\cH)$ and $\cU_{\rc}$ denote the unitary group
and the set of antiunitary operators respectively. The set
$\Lambda[\cH]$ is simplectomorphic to the manifold of all
One purpose is to ``translate'' (and extend) some few of their

$\cU \cup \cU_{\rc}$ is a group generated by the set $\cU_{\rc}$.
Here the interest is in the orbits of the adjoint representation
of these groups.

Let $X \in \cB(\cH)$ and denote by $\{ X \}_u$ the set of all
$UXU^{-1}$, $U \in \cU$. To get an ``orbit'' $\{ X \}_{au}$ one
enlarges $\{ X \}_u$ by the set of all operators
$\Theta X^{\dag} \Theta^{-1}$, $\Theta \in \cU_{\rc}$. Indeed,
\begin{displaymath}
X \mapsto \Theta_2 (\Theta_1 X^{\dag} \Theta_1)^{\dag} =
\Theta_2 \Theta_1 X (\Theta_2 \Theta_1)^{-1}
\end{displaymath}
is a unitary transformation. From (\ref{trpose2}) it follows
$X^{\top} \in \{ X \}_{au}$. By unitary invariance:\\
{\em The $\theta$-transpose $Y^{\top}$ is contained
in $\{ X \}_{au}$ for all $Y \in \{ X \}_{au}$ and all
conjugations $\theta$. However,} $X^{\top} \in \{ X \}_u$
{\em if and only if} $X^{\dag} \in \{ X \}_{au}$.
\medskip

If $\dim \cH =2$, an orbit $UXU^{-1}$, $U$ unitary, is completely
characterized by the numbers
\begin{equation} \label{uorbit1}
\T \, X, \quad \T \, X^2, \quad \T \, X^{\dag} X \; .
\end{equation}
They do not change by substituting $X \to X^{\top}$, i.~e. by
$X \to \theta X^{\dag} \theta$ unitary orbits transform into
itself. Hence: {\em For all $X \in \cB(\cH_2)$ one has
$X^{\top} \in \{ X \}_{au}$ and $\{ X \}_u = \{ X \}_{au}$.}
\medskip

Also the case $\dim \cH = 3$ is manageable: The unitary orbits
for $\dim \cH = 3$ are characterized by seven numbers, by the
traces of $X$, $X^2$, $X^3$, $X^{\dag}X$, $XX^{\dag}X$,
$X^{\dag} X^2 X^{\dag}$, $X^{\dag}X X^{\dag} X^2 X^{\dag}$,
\cite{Si68}, \cite{Pi62}. It turns out that only the last trace
is not invariant against $X \to X^{\top}$.
\begin{prop}[Garcia, Poore, Tener]
Assuming $\dim \cH = 3$. Then $X^{\top} \in \{ X \}_{u}$ if
and only if
\begin{equation} \label{uorbit2}
\T \,   X^{\dag} X X^{\dag} X^2 X^{\dag}
=
\T \,   X^{\dag} X^2 X^{\dag} X X^{\dag} \; ,
\end{equation}
\end{prop} \cite{GPT11}.

Unitary equivalence of two operators can be expressed by traces
in any finite dimension. However, their number increases rapidly
with increasing dimension $d$ of $\cH$. In the case
$\dim \cH = 4$  Djokovi\'c, \cite{Dj07}, could list 20 trace
relations in $X$ and $X^{\dag}$ which present a complete
description of the unitary orbits.

From them Garcia, Poore, and Tener \cite{GPT11} could extract
seven trace relations guarantying $X^{\top} \in \{ X \}_{u}$
and, hence, $X^{\dag} \in \{ X \}_{au}$.
\medskip

Already in 1970 W.~R.~Gordon \cite{Go70} could show that
$X^{\top}$ can be expressed for any $X$ in the form $UXV$ by two
unitary operators $U$ and $V$. Gordon's result translates into:

{\em Given $X$ there are two antiunitaries, $\Theta_1$
and $\Theta_2$, such that}
\begin{equation} \label{uorbit4}
X^{\dag} = \Theta_1 X \Theta_2 \; .
\end{equation}

\subsection{UET and beyond}
Remind that UET, unitary equivalence of a matrix to its
transpose, can be expressed as antiunitary equivalence of the
matrix to its Hermitian adjoint. The latter relation is basis
independent.

New criteria for being UET or UECSM have been developed by
S.~R.~Garcia, J.~E.~Tener in \cite{GT11}, by J.~E.~Tener in
\cite{Te12},  and by L.~Balayan, S.~R.~Garcia in \cite{BG09}.
Following essential ideas of the latter paper, the slightly weaker
UET assumption will be considered. The problem will be embedded
into a more general one: An antilinear variant of a
theorem of the present author \cite{Uh85}, extended and refined
by P.~M.~Alberti, \cite{Al85}. In connection with applications
to QIT, a more recent paper is by A.~Chefles, R.~Jozsa, A.~Winter,
\cite{CJW03}.

Here the starting point is a general completely copositive map
\begin{equation} \label{cocm1}
X \longrightarrow T(X) = \sum_{k=1}^r \vartheta_k X^{\dag}
 \vartheta^{\dag}_k \; , \quad X \in \cB(\cH) \, ,
\end{equation}
$\vartheta_j \in \cB(\cH)_{\rc}$, i.~e. a sum of elementary
copositive maps (\ref{trpose1}). If there is no representation
of $T$ by less than $r$ terms, the number $r$ is called the
{\em length of $T$.}

In the following
\begin{equation} \label{cmin}
\phi_1, \phi_2, \dots, \phi_d \in \cH, \quad d = \dim \cH
\end{equation}
denote a set of linear independent unit vectors.
\begin{equation} \label{cmout}
\phi'_1, \phi'_2, \dots, \phi'_d \in \cH
\end{equation}
is assumed to be a set of $d$ unit vectors.
\begin{thm} \label{copoA}
The following items are equivalent: \\
a.) There is a map (\ref{cocm1}) fulfilling
\begin{equation} \label{cocm3}
T(|\phi_j \rangle \langle \phi_j|) =
|\phi'_j \rangle \langle \phi_j'| , \quad j = 1, \dots, d \, .
\end{equation}
b.) There is a positive semi-definite matrix
\begin{equation} \label{beta}
\{\beta_{jk}\} \geq {\mathbf 0} , \quad \beta_{jj} = 1 \; .
\end{equation}
such that
\begin{equation} \label{neu1}
T( |\phi_j \rangle\langle \phi_k| ) = \beta_{jk}
|\phi'_k \rangle\langle \phi'_j|
\end{equation}
for all $j,k \in \{1, \dots, d\}$.\\
c.) There is a positive semi-definite matrix (\ref{beta})
and a positive semi-definite operator $K$ such that
\begin{equation} \label{cocm8}
\langle \phi_k, K \, \phi_j \rangle =
\beta_{jk} \langle \phi'_j , \phi'_k \rangle \; .
\end{equation}
for all $j,k \in \{1, 2, \dots, d \}$ \; .
\end{thm}
Proof: The step b.) $\Rightarrow$ a.) is trivial. (All vectors
are unit vectors by assumption, and $\beta_{jj} = 1$ necessarily.)
Consider now a.) The map (\ref{cocm1}) is constrained by
(\ref{cocm3}). Hence
\begin{displaymath}
T(|\phi_j \rangle \langle \phi_j|) =
\sum_i \vartheta |\phi_j \rangle \langle \phi_j| \vartheta^{\dag}
= |\phi'_j \rangle \langle \phi'_j| \; .
\end{displaymath}
If a sum of positive operators is of rank one, every non-zero
term must be proportional to it, i.~e. to
$|\phi'_j \rangle \langle \phi'_j|$. Using (\ref{rank1cc})
one obtains
\begin{displaymath}
| \vartheta \phi_j \rangle \langle \vartheta \phi_j| \sim
|\phi'_j \rangle \langle \phi'_j| \; .
\end{displaymath}
and $\vartheta_i \phi_j \sim \phi'_j$. Hence there are numbers
$\alpha_{jk}$ such that
\begin{equation} \label{cocm5}
\vartheta_i \, \phi_j = \alpha_{ij} \phi'_j  \; .
\end{equation}
Inserting in (\ref{cocm1}) yields
\begin{displaymath}
T(|\phi_j \rangle \langle \phi_k|) =
\sum_i |\vartheta_i \phi_k \rangle \langle \vartheta_i \phi_j|
 = \beta_{jk} |\phi'_k \rangle \langle \phi'_j| \; ,
\end{displaymath}
\begin{equation} \label{cocm9}
\beta_{jk} = \sum_i {\alpha_{ij}}^{*} \, {\alpha_{ik}} ,
\quad \beta_{jj} = 1 \; ,
\end{equation}
so that from b.) it follows a.) Coming now to the step b.)
$\Rightarrow$ c.) one observes that the right of (\ref{neu1})
is the trace of $T(|\phi_j \rangle \langle \phi_k|)$.
Generally one gets from (\ref{cocm1}), and by the help of the
(\ref{rule7b}) the identities
\begin{equation} \label{cocm2b}
\T \, T(X) = \T \, K X \; ,
\end{equation}
\begin{equation} \label{cocm2a}
K := \sum \vartheta_i^{\dag} \vartheta_i \; .
\end{equation}
This way one finds
\begin{displaymath}
\T \, T( |\phi_j \rangle\langle \phi_k| ) =
\T \, K |\phi_j \rangle\langle \phi_k| =
\langle \phi_k, K \, \phi_j \rangle \; .
\end{displaymath}
so that (c) follows from (b) Starting now from (c) one can use an
arbitrary decomposition (\ref{cocm9}) to {\em define} $r$
antilinear operators $\vartheta_j$ by (\ref{cocm5}). Then $T$,
constructed as in (\ref{cocm1}) with these antilinear operators,
satisfies (\ref{cocm3}) and (\ref{neu1}).

The map $T$ described by the theorem acts as
\begin{equation} \label{neu2}
T \, : \quad
\sum x_{jk} |\phi_j \rangle\langle \phi_k| \, \mapsto \,
\sum x_{jk} \beta_{jk} |\phi'_k \rangle\langle \phi'_j| \;.
\end{equation}
Of use is the reconstruction of the $\vartheta_i$ from
(\ref{cocm5}). By
\begin{equation} \label{gc0}
\langle \tilde \phi_j , \phi_k \rangle = \delta_{jk}
\end{equation}
the vectors $\tilde \phi_j$ are uniquely determined. Together
with the vectors $\phi_k$ they define a bi-orthogonal basis.
However, the $\tilde \phi_j$ are not necessarily normalized.
Next, for all $\phi \in \cH$,
\begin{equation} \label{neu3}
\vartheta_i \phi = \sum_j \alpha_{ij}
\langle \phi, \tilde \phi_j\rangle \, \phi'_j \; .
\end{equation}
Indeed, the left of (\ref{neu3}) defines an antilinear operator,
and for $\phi = \phi_k$ one remains with (\ref{cocm5}).
By (\ref{rank1a}) and (\ref{rank1b}) one rewrites (\ref{neu3}) as
\begin{equation} \label{neu4}
\vartheta_i = \sum_j \alpha_{ij}
|\phi'_j\rangle\langle \tilde \phi_j|_{\rc} \; .
\end{equation}
\medskip

As seen from the proof of the theorem:
\begin{cor}
The length of a representation (\ref{cocm1}) satisfying
(\ref{cocm3}) is never less than the rank of $\{ \beta_{jk}\}$.
If (\ref{cocm9}) is an orthogonal decomposition, equality
is reached.
\end{cor}
\begin{cor}
If and only if $K = \1$ the map $T$ of (\ref{cocm1}) is trace
preserving, i.~e. a {\em cochannel.}
\end{cor}
A further observation: From (\ref{beta}) one deduces
$|\beta_{jk}| \leq 1$. Therefore
\begin{equation} \label{gc7a}
| \langle \phi_j , K \phi_k \rangle | \geq
| \langle \phi'_j , \phi'_k \rangle | \; .
\end{equation}
{\em Equality} holds if and only if all $\beta_{jk}$ are
unimodular. If in addition $K = \1$ is required, one gets what
is called ``weak angle condition'' in \cite{BG09}.

\subsubsection{Length one}
It has already be shown that a completely copositive map $T$
is of length one iff $T(X) = \vartheta X^{\dag} \vartheta^{\dag}$.
If $T$ is further trace--preserving, hence a cochannel, it of
the form
\begin{equation} \label{lr1}
T(X) = \Theta X^{\dag} \Theta^{\dag}, \quad \Theta^{\dag} \Theta
= \1 \; ,
\end{equation}
(\ref{cocm5}) becomes $\Theta \phi_j = \epsilon_j \phi'_j$
for all $j$. Hence
\begin{equation} \label{lr2}
\langle \phi_k, \phi_j\rangle = \epsilon_j^{*} \epsilon_k
\langle \phi'_j, \phi'_k\rangle \; .
\end{equation}

Looking at (\ref{gc7a}) one wonders whether $|\beta_{jk}| = 1$
for all $j,k$ results in $\beta_{jk} = \epsilon^{*}_j \epsilon_k$.
At first one proves:
\begin{lem}
Assume all matrix elements of the matrix (\ref{beta})
are unimodular. Then every $3 \times 3$ main minor is of rank one.
\end{lem}
Because the determinants of the $2 \times 2$ main minors vanish,
it suffices to consider
\begin{equation} \label{lr3}
\det \, [{\rm main \, 3x3 \, minor}]
= \epsilon + \epsilon^{*} - 2 , \quad
\epsilon = \epsilon_{ij} \epsilon_{jk} \epsilon_{ki} \, .
\end{equation}
These determinants are non-negative if and only if $\epsilon = 1$.
\begin{prop} \label{ATS}
Assume (\ref{beta}). then the following conditions are
mutually equivalent.\\
a.) The matrix $\{\beta_{jk}\}$ is of rank one.\\
b.) There are unimodular numbers $\epsilon_j$ such that
    $\beta_{jk} = \epsilon_j^{*} \epsilon_k$.\\
c.) It is
\begin{equation} \label{mthree}
\beta_{ij} \beta_{jk} \beta_{ki} = 1 \; ,
 \quad \forall i, j, k \in \{1, \dots, d \}
\end{equation}
\end{prop}
{\em Proof:} \, (a) $\Leftrightarrow$ (b) is trivial. The same
is with (b) $\Rightarrow$ (c). It remains to prove (a), (b),
from (c). At first, because of $|\beta_{jk}| \leq 1$ it follows
$|\beta_{jk}| = 1$ from (c). Hence the preceding lemma implies
that every $3 \times 3$ main minor of $\{\beta_{jk}\}$ is of
rank one. Being valid for $\dim \cH = 3$ the proof proceeds
by induction. Assume (a), (b), suffices to prove  (c) if
$\dim \cH = d$. Asking for dimension $d+1$, the hypothesis
allows to start with
\begin{displaymath}
\epsilon_{ij} = \epsilon_i \epsilon_j^{*}, \quad
i, j, k \in \{1, \dots, d \}
\end{displaymath}
and $\epsilon_{1,d+1} = \epsilon_1 \epsilon_{d+1}^{*}$. Hence
\begin{displaymath}
\epsilon_{k,d+1} \epsilon_{d+1,1} \epsilon_{1k} = 1, \quad
\epsilon_{d+1,1} = \epsilon_{d+1} \epsilon_1^{*}, \quad
\epsilon_{1k} = \epsilon_1 \epsilon_k^{*}
\end{displaymath}
Inserting this for $1 < k < n+1$ into (\ref{mthree}), one
gets $\epsilon_{k (d+1)} = \epsilon_k \epsilon_{d+1}^{*}$.

Hence $T$ in theorem \ref{copoA} is of length larger than one
if and only if at least one entry of $\{ \beta_{jk} \}$ is not
unimodular.
\medskip

It is interesting to compare proposition \ref{ATS} with the
topological geometric phase of M.~Berry and B.~Simon. (For a
survey consult the monographs \cite{ChJa09}, \cite{BZ06}, and
\cite{CrUh98} See also the comment below.)
\begin{lem}
Let $K= \1$. If and only if $T$ is of length one it is
\begin{equation} \label{vthree}
\langle \phi_i,\phi_j \rangle \,\langle \phi_j,\phi_k \rangle \,
\langle \phi_k,\phi_i \rangle = \langle \phi'_j,\phi'_i \rangle
\langle \phi'_i, \phi'_k \rangle  \,
\,\langle \phi'_k,\phi'_j \rangle
\end{equation}
for all $i, j, k \in \{1, \dots, d \}$, vectors (\ref{cmin}) and
(\ref{cmout}) in theorem \ref{copoA}.
\end{lem}
For the proof one applies proposition \ref{ATS} to (\ref{cocm8})
respecting $K=\1$.
\begin{cor}
Let one of the conditions of lemma \ref{ATS} be valid and $K=\1$.
Then every operator
$X = \sum x_j |\phi_j \rangle\langle \phi_j|$ is UET, i.~e.
$\Theta X^{\dag} \Theta^{\dag} = X$ with $\Theta$ antiunitary.
\end{cor}
Indeed, one can find a cochannel $T$ of length one mapping
$|\phi_j \rangle\langle \phi_j|$ onto
$|\phi'_j \rangle\langle \phi'_j|$ for $j=1,\dots,d$. If $T$
is of the assumed form, there is a $\vartheta$ such that
$\vartheta X^{\dag} \vartheta^{\dag} = X$. By trace preserving
one gets $\vartheta^{\dag} \vartheta = \1$.\\
The lemma is a variant of Balayan and Garcia's ``Strong Angle
Test'', \cite{BG09}.

Some relations become more transparent by introducing the
projection operators
\begin{equation} \label{proj1}
P_j = |\phi_j \rangle\langle \phi_j| \quad
Q_j = |\phi'_j \rangle\langle \phi'_j| \; ,
\end{equation}
(\ref{vthree}), as an example, becomes
\begin{displaymath}
\T \, P_i P_j P_k = \T \, Q_k Q_j Q_i \; .
\end{displaymath}
One further extend (\ref{mthree}) to:
\begin{equation} \label{geoph1}
\T \, P_{i_1} \cdots P_{i_n} =
\T \, Q_{i_n} Q_{i_(n-1)} \cdots Q_{i_1} \; ,
\end{equation}
$\{i_1, \dots, i_n \} \in \{1, \dots, d \}$, for all $n$.

To see the trick consider $\T \, P_1 P_2 P_3 P_4$. One may
convert this expression into
\begin{displaymath}
\T \, P_1 P_2 P_3 P_3 P_4 P_1 = (\T \, P_1 P_2 P_3) \;
(\T \, P_3 P_4 P_1) \, .
\end{displaymath}
This way one can rewrite any of the numbers in (\ref{geoph1})
as products of form $\T \, P_i P_j P_k$ respectively
$\T \, Q_k Q_j Q_i$.

{\em Comment.} Given a pair of normalized vectors $\phi$ and
$\phi'$, the number $\langle \phi, \phi' \rangle$ is often
called {\em transition amplitude for the change} $\phi \to \phi'$.
Its absolute square is the {\em transition probability}. It is the
probability of the transition $|\phi \rangle\langle \phi| \to
|\phi' \rangle\langle \phi'|$ by a {\em von Neumann - L\"uders
measurement} which asks whether the system is either in state
$|\phi' \rangle\langle \phi'|$ or in a state orthogonal to it.
The symmetry of the transition probability prevents to
distinguish Past and Future by the von Neumann-L\"uders rule.

The phase factor of the transition amplitude becomes
physically important in cyclic processes. For instance, consider
an adiabatic process $P_1 \to P_2 \to P_3 \to P_4 \to P_1$. Its
{\em phase} $\gamma$ is defined modulo $2 \pi$ by
\begin{equation} \label{gc9}
z = |z| \exp i \gamma, \quad z = \T \, P_1 P_2 P_3 P_4
\end{equation}
if $z \neq 0$. Otherwise the phase is undefined. By an
antiunitary transformation $P_j \to \Theta P_j \Theta_j^{\dag}$
the phase changes its sign and indicates the change of the
processes orientation.

The phase $\gamma$ in (\ref{gc9}) (and in
similar constructs) is an instant of the {\em geometric} or
{\em Berry phase:} The pure states are the points of a complex
projective space of dimension $d-1$. The latter carries the
Study-Fubini metric. Let be $w$ a closed oriented curve which
is the oriented boundary of a 2-dimensional submanifold
$\mathcal F$. The geometric phase $\gamma(w)$ can be computed
by integrating the K\"ahler form of the Study-Fubini metric
over $\mathcal F$. $w$ in example (\ref{gc9}) is constructed
by connecting any two consecutive points of the sequence
$P_1 \to P_2 \to P_3 \to P_4 \to P_1$ by a short geodesic of
the Study-Fubini metric. A detailed discussion is in
(\cite{ChJa09}). Basic constructions can be found in \cite{BZ06}.

\subsubsection{Examples}
{\bf 1.} Solutions of $X = \Theta X^{\dag} \Theta^{\dag}$ for
some 2--dimensional cases are summarized.\\
{\bf 1a.} The case $\Theta = \tau_0$ can be read of from
lemma \ref{c3.2.2a}, saying: Exactly all real multiples
of unitary operators with determinant one are solutions.\\
{\bf 1b.} Let $\Theta = \tau_1$. Then $\sigma_2$ and $\sigma_3$
remain unchanged while $T(\sigma_1) = - \sigma_1$, see
subsection \ref{C2.5.1}. The set of solutions is the real
linear space generated by $\1$, $\sigma_2$, and $\sigma_3$.\\
{\bf 1c.} The border between the two cases above consists of
operators $\vartheta$ with $\vartheta^2 = 0$. With a basis
$\phi_1$, $\phi_2$ the square of the antilinear operator
$\vartheta = |\phi_1 \rangle\langle \phi_2|$ vanishes.
By (\ref{mranka})
\begin{displaymath}
\vartheta \vartheta^{\dag} = |\phi_1 \rangle\langle \phi_1|,
\quad
\vartheta^{\dag} \vartheta = |\phi_2 \rangle\langle \phi_2|,
\end{displaymath}
so that their sum is the identity map, and
\begin{displaymath}
T(X) = \vartheta X^{\dag} \vartheta^{\dag} +
\vartheta^{\dag} X^{\dag} \vartheta
\end{displaymath}
is a cochannel of length two. $T$ acts
according to
\begin{equation} \label{nilpo1}
\sum c_{jk} |\phi_j \rangle\langle \phi_k| \mapsto c_{11}
|\phi_2\rangle  \langle \phi_2| +
c_{22} |\phi_1 \rangle\langle \phi_1| \; .
\end{equation}
Therefore any fixed point is a multiples of $|\phi_2\rangle\langle
\phi_2| + |\phi_1 \rangle\langle \phi_1|$.\\
{\bf 2.} It follows from theorem \ref{copoA} that there exists
a representation $T(X) = \Theta X^{\dag} \Theta^{\dag}$, $\Theta$
antiunitary, if and only if
\begin{equation} \label{gc3}
\Theta \phi_j = \epsilon_j \phi'_j, \quad |\epsilon_j| = 1 ,
\quad j = 1, \dots, d \; .
\end{equation}
This is equivalent to
\begin{equation} \label{gc4}
\langle \phi_k, \phi_j\rangle = \epsilon_j^{*} \epsilon_k
\langle \phi'_j, \phi'_k\rangle \; .
\end{equation}
\begin{lem}
Let (\ref{gc3}) be valid. If the unit vectors $\phi_j$ and
$\phi'_j$, $j = 1, \dots, d$, are bi-orthogonal,
\begin{equation} \label{gc6}
\langle \phi_j , \phi'_k \rangle = z_j \delta_{jk} \; ,
\end{equation}
then there is a conjugation $\theta$ such that
\begin{equation} \label{gc12}
T(X) = \theta X^{\dag} \theta^{\dag} \; .
\end{equation}
\end{lem}
Proof: For a bi-orthogonal system of vectors (\ref{gc6}) one has
for all $\phi$
\begin{equation} \label{gc8}
\phi = \sum_j \frac{\langle \phi_j , \phi \rangle}{z_j} \phi'_j
= \sum_j \frac{\langle \phi'_j , \phi \rangle}{z_j} \phi_j \; .
\end{equation}
Now (\ref{gc3}) is transformed by the second relation into
\begin{displaymath}
\Theta \phi_k = \epsilon_k \phi'_k = \epsilon_k \sum_j
\frac{\langle \phi'_j , \phi'_k \rangle}{z_j^{*}} \phi_j \; .
\end{displaymath}
 Using antilinearity and again (\ref{gc3}) one obtains
\begin{displaymath}
\Theta^2 \phi_k = \sum_j \epsilon_k^{*} \epsilon_j
\frac{\langle \phi'_k , \phi'_j \rangle}{z_j} \phi'_j \; .
\end{displaymath}
Positioning the unimodular numbers within the scalar product
and using again (twice) (\ref{gc3}) it follows
\begin{displaymath}
\Theta^2 \phi_k = \sum_j \frac{\langle \Theta \phi_k ,
\Theta \phi_j \rangle}{z_j} \phi'_j \; .
\end{displaymath}
Knowing already that $\Theta$ is antiunitary, this means
\begin{displaymath}
\Theta^2 \phi_k = \sum_j \frac{\langle \phi_j ,
\phi_k \rangle}{z_j} \phi'_j = \phi_k \; ,
\end{displaymath}
the latter equality follows from the first relation
in (\ref{gc8}). Because $\Theta^2$ is linear, and the
$\phi_k$ span $\cH$, one gets $\Theta^2 = \1$.

The essence of the proof is due to Balayan and Garcia
\cite{BG09}. See also lemma \ref{GT1}.\\
{\bf 3.} Also the following lemma is due to Balayan and Garcia,
\cite{BG09}.
\begin{lem}
Let $X$ be a linear operator with non-degenerate spectrum.
If $X$ is antiunitarily equivalent to $X^{\dag}$ then
there is a conjugation $\theta$ such that
$X = \theta X^{\dag} \theta$.
\end{lem}
Proof:
(\ref{gc4}) is valid. To apply the preceding lemma, the
eigenvectors of $X$ and $X^{\dag}$ should constitute a
bi-orthogonal vector system. Indeed, this is the case:
Let $\phi_j$, $\phi'_k$ denote the eigenvectors and
$\lambda_j$, $\lambda'_k$ the eigenvalues of $X$ and of
$X^{\dag}$ respectively. Then
\begin{displaymath}
\lambda_j \langle \phi'_k, \phi_j \rangle =
\langle \phi'_k, X \phi_j \rangle ) =
\langle X^{\dag} \phi'_k, \phi_j \rangle =
(\lambda'_k)^{*} \langle \phi'_k, \phi_j \rangle
\end{displaymath}
and $\lambda_j \neq \lambda'_k$ for $j \neq k$ shows the
asserted bi-orthogonality.
\medskip

In case $X$ is UET, there is an antiunitary $\Theta$ such that
\begin{equation} \label{gc2}
\Theta \, X = X^{\dag} \, \Theta, \quad
\Theta^{\dag} \Theta = \1 \; .
\end{equation}
Applying (\ref{gc2}) to an eigenvector $\phi_j$ of $X$ yields
$\lambda^{*} \Theta \phi_j = X^{\dag} \Theta \phi_j$. Thus
$\Theta \phi_j$ is an eigenvector of $X^{\dag}$ with eigenvalue
$\lambda^{*}_j$. Because the spectrum is non-degenerate by
assumption, $\Theta \phi_j$ must be proportional to $\phi'_j$.
As $\Theta$ is isometric, the factor must be unimodular.

On the other hand, (\ref{gc4}) is sufficient for the existence
of a antiunitary $\Theta$ satisfying (\ref{gc3}) and (\ref{gc2}).

\section{Involutions} \label{C7}
An antilinear operator $S$ is an {\em involution} if $S^2 = \1$.
It is a {\em skew involution} if $S^2 = - \1$ .

These definitions do depend on the linear structure of $\cH$ only
and not on its scalar product. On the other hand, the Hilbert
structure will be used to polar decompose involutions.

Clearly, the classes of involutions and of skew involutions are
larger than those of (skew) conjugations. While the unitary
transformations act transitively on the set of conjugations, any
two involutions are similar (see below).

\subsection{Polar decomposition} \label{C7.1}
Involutions and skew involutions come with characteristic
polar decompositions. At first, with $S$ also $S^{\dag}$ is a
(skew) involution. Polar decomposing $S$ defines an antiunitary
operator $\theta$ such that
\begin{equation} \label{invol1}
S = |S| \theta = \pm \theta^{\dag} |S|^{-1} ,
\quad |S| := (S S^{\dag})^{1/2} \; .
\end{equation}
The sign reflects $S = \pm S^{-1}$. Here and below $\pm$ means that
$S$ is an involution if $+$ is valid and a skew involution if
$-$ takes place. From $(S^{\dag}S) (SS^{\dag}) = \1$ it follows
$(S^{\dag}S)^{1/2} = |S|^{-1}$ and, by going to the Hermitian
adjoint in (\ref{invol1}),
\begin{equation} \label{invol2}
|S^{\dag}| = (S^{\dag}S)^{1/2} = |S|^{-1} \; .
\end{equation}
\begin{prop}
Assume $S$ is an involution or a skew involution. Then $\theta$
in (\ref{invol1}) is a conjugation or a skew conjugations
respectively such that
\begin{equation} \label{invol6}
S = |S| \theta = \theta |S|^{-1} , \quad
\theta^2 = S^2 = \pm \1 \, ,
\end{equation}
\end{prop}
As a consequence one gets
\begin{equation} \label{invol6a}
S^{\dag} = |S|^{-1} \theta = \theta |S|  \; .
\end{equation}
For the proof let $\phi_1, \phi_2, \dots$ be an eigenvector basis
of $|S|$ and $\lambda_1, \lambda_2, \dots$ the corresponding
eigenvalues. (\ref{invol1}) can be rewritten as
\begin{displaymath}
\langle \phi_j, S \phi_k\rangle =
\lambda_j \langle \phi_j, \theta \phi_k\rangle = \pm
\lambda_k^{-1} \langle \phi_j, \theta^{\dag} \phi_k\rangle =
\pm \lambda_k^{-1} \langle \phi_k, \theta \phi_j\rangle \; ,
\end{displaymath}
and we conclude
\begin{equation} \label{invol3}
\lambda_j \lambda_k \, \langle \phi_j, \theta \phi_k\rangle =
\pm \langle \phi_k, \theta \phi_j\rangle
\end{equation}
for all $j,k = 1,\dots, \dim \cH$. As all $\lambda_j$ are
different from zero, there is the alternative: Either
$\lambda_j \lambda_k = 1$ and
$\langle \phi_j, \theta \phi_k\rangle$ is multiplied by $\pm \1$
if $j$ and $k$ are exchanged, or both expectation values in
(\ref{invol3}) vanish. Hence $\langle \phi_j, \theta \phi_k\rangle$
is for all $j$, $k$ symmetric respectively skew symmetric. Hence
$\theta$ is Hermitian respectively skew Hermitian. A
(skew) Hermitian antiunitary operator is a (skew) conjugation.
Now the assertion has been proved.\\
There is something more to be said. The spaces
\begin{equation} \label{invol4}
\cH[\lambda] :=
\{ \phi \in \cH \, : \, |S| \phi = \lambda \phi \}
\end{equation}
are mutually orthogonal and, by the proof above,
\begin{equation} \label{invol5}
 \theta \, \cH[\lambda] = \cH[\lambda^{-1}] \; .
\end{equation}
\begin{thm}
Let $S = |S| \theta = \theta |S|^{-1}$ be the polar decomposition
of either an involution or of a skew involution.
If $\lambda$ is an eigenvalue of $|S|$ then so is $\lambda^{-1}$.
There is an orthogonal decomposition of $\cH$ into
subspaces $\cH[\lambda]$  the vectors of  which are
eigenvectors of $|S|$ with eigenvalue $\lambda$.
$\theta$ maps $\cH[\lambda]$ onto $\cH[\lambda^{-1}]$.
\end{thm}
If $S$ is either a conjugation or a skew conjugation then
$|S| = \1$ and $\cH[1] = \cH$.\\
 If at least one eigenvalue of $|S|$, say $\lambda$, is
different from $1$, then $\lambda \neq \lambda^{-1}$ and
$\lambda + \lambda^{-1} > 2$, i.~e. $S$ cannot be a
conjugation or a skew conjugation.\\
In general, the eigenvectors of $|S|$ are grouped in one
subspace with eigenvalue $1$ and in pairs of subspaces with
eigenvalues $\lambda \neq 1$ and $\lambda^{-1}$, This proves
\begin{lem}
Let $S$ be either an involution or a skew involution. Then
\begin{equation} \label{invol6b}
\T \, |S|   = \T \, |S|^{-1}  \geq \dim \cH
\end{equation}
and equality holds if and only if $S$ is a conjugation
or a skew conjugation.
\end{lem}

\subsection{Similarity of involutions}
Let $S$ be an involution. Then, like (\ref{conj2}),
\begin{equation} \label{invol7}
\cH_S := \{ \phi \in \cH \, : \, S \phi = \phi \}
\end{equation}
is a real linear subspace of $\cH$ of maximal dimension.
It is generally not a Hilbertian one:
\begin{lem}
Let $S$ be an involution. The following items are equivalent.\\
{\bf 1.} $S$ is a conjugation.\\
{\bf 2.} $S$ is normal.\\
{\bf 3.} $\cH_S$ is a real Hilbert subspace.\\
{\bf 4.} $S^{\dag} \cH_S \subseteq \cH_S$ \; .
\end{lem}
From (\ref{invol1}) one infers $|S| = \1$ if $S$ is normal.
(\ref{invol1}) then  provides $\theta = \theta^{\dag}$
and $S$ must a be conjugation. The inverse statement can easily
be seen: An involution is normal if and only if it is a
conjugation. Next, with any pair $\phi_1$, $\phi_2$ of vectors
in $\cH_S$,
\begin{displaymath}
\langle \phi_1, \phi_2 \rangle = \langle \phi_1, S \phi_2 \rangle
= \langle \phi_2, S^{\dag} \phi_1 \rangle \; .
\end{displaymath}
If $S^{\dag} \cH_S \subseteq \cH_S$ then $S^{\dag} \cH_S =\cH_S$
and $S^{\dag} \phi_1 = \phi_1$. Consequently, the restriction to
$\cH_S$ of the scalar product becomes symmetric and, hence,
Hilbertian. It also follows $S - S^{\dag} = 0$ on $\cH_j$.
This proves $S = S^{\dag}$ on the whole of $\cH$.

Of course there are scalar products making $\cH_S$ Hilbertian
and $S$ a conjugation: Let
$\tilde \phi_1, \dots, \tilde \phi_d$ be a linear basis of
$\cH_S$.  There is a scalar product associated to it, say
$\langle,.,\rangle^{\sim}$, defined by
\begin{equation} \label{invol7a}
\langle \sum a_k \tilde \phi_k , \sum b_l \tilde \phi_l
\rangle^{\sim}
:= \sum a_j^* b_j \; .
\end{equation}
Equipped with {\em this} scalar product $\cH_S$ becomes a real
Hilbert space. Hence there is an invertible positive operator
$A \in \cB(\cH)$ such that
\begin{equation} \label{invol7b}
\langle \phi' , \phi'' \rangle^{\sim} =
\langle \phi' , A \phi'' \rangle \; .
\end{equation}
Computing the Hermitian adjoint $X^{\sim}$ of an operator $X$
with respect to (\ref{invol7b}),
\begin{displaymath}
\langle \phi' , X \phi'' \rangle^{\sim} =
\langle \phi' , A X \phi'' \rangle =
\langle X^{\dag} A \phi' , \phi'' \rangle =
\langle X~ \phi' , \phi'' \rangle^{\sim} \; ,
\end{displaymath}
yields $X^{\dag} A = A X^{\sim}$. Thus
\begin{equation} \label{invol7c}
  X \, \Rightarrow \, X^{\sim} = A^{-1} X^{\dag} A
\end{equation}
is the Hermitian adjoint coming with the scalar product
(\ref{invol7a}), i.~e.
\begin{equation} \label{invol7d}
\langle \phi' , \phi'' \rangle_A :=
\langle \phi' , \phi'' \rangle^{\sim} =
\langle \phi' , A \phi'' \rangle \; .
\end{equation}
In the same manner one proves (\ref{invol7c}) for antilinear
operators $\vartheta$,
\begin{equation} \label{invol7ca}
  \vartheta \, \Rightarrow \, \vartheta^{\sim} =
A^{-1} \vartheta^{\dag} A \; .
\end{equation}

Looking again at the definition (\ref{invol7a}), there is an
invertible linear operator $Z$ such that
$Z \phi_j = \tilde \phi_j$, $j = 1, \dots, d$.
The definition of $Z$ and (\ref{invol7a}) implies
\begin{equation} \label{invol7e}
  Z^{\dag} Z = A \; .
\end{equation}
Reminding now $S \tilde \phi_j = \tilde \phi_j$ and, defining the
conjugation $\theta$ by $\theta \phi_j = \phi_j$, one gets
$S Z \phi_j = \tilde \phi_j = Z \phi_j$.
Hence $Z^{-1} S Z = \theta$, and $S$ is similar to $\theta$.
As $S$ stands for any involution, and because similarity is an
equivalence relation, one gets
\begin{prop}
Any two involutions are similar. The set of all involutions
forms a GL$(\cH)$--orbit. Any two involutions are cosimilar.
\end{prop}
To prove cosimilarity, one multiplies both sides of
$S_1 = Z^{-1} S_2 Z$ by $S_1$ and defines $\vartheta = ZS_1$.
It follows $S_1 = \vartheta^{-1} S_2 \vartheta$.

\subsection{Involutions and the geometric mean} \label{C7.gm}
 The geometric mean was introduced by W.~Pusz and
S.~L.~Woronowicz \cite{PWW75}. See also \cite{CrUh98} and
\cite{Bh07} for its specification to positive operators.

Many of its properties and various applications are known.
In accordance with the aim of the present paper only its
connection with Hermitian adjoints, i.e. with certain
involutions, is presented.

Given two semi-definite positive Hermitian forms, say
$\langle.,.\rangle_1$ and $\langle.,.\rangle_2$, on an
arbitrary complex-linear spaces $\cL$. The theorem of Pusz
and Woronowicz states that within the set of semi-definite
Hermitian forms $\langle .,. \rangle_x$ satisfying
\begin{equation} \label{PuW1}
\langle \phi_a, \phi_a \rangle_1 \,
\langle \phi_b, \phi_b \rangle_2 \geq
|\langle \phi_a, \phi_b \rangle_x|^2
\end{equation}
for all $\phi_a, \phi_b \in \cL$, there is a unique largest one,
say $\langle .,. \rangle_{12}$, such that
\begin{equation} \label{PuW2}
\langle \phi, \phi \rangle_{12} \geq
\langle \phi, \phi \rangle_x, \quad \forall \phi \in \cH  \, .
\end{equation}
In the following we impose two strong restrictions:\\
(1) $\cL$ is a finite dimensional Hilbert space $\cH$ with
scalar product $\langle.,.\rangle$,\\
(2) Only positive definite Hermitian forms are considered.

Using $\langle.,.\rangle$ as reference, scalar products can be
labelled by invertible positive operators,
$A \, \to \, \langle.,.\rangle_A$, according to
\begin{equation} \label{gm1}
\langle \phi_1, \phi_2 \rangle_A  :=
\langle \phi_1,  A \phi_2 \rangle , \quad A > {\mathbf 0} \, .
\end{equation}
All possible scalar products in $\cH$ can be gained this way.

Let $\langle.,.\rangle_C$ be the geometric mean of
$\langle.,.\rangle_A$ and $\langle.,.\rangle_B$. With T.~Ando one
then writes $C = A \# B$ for the geometric mean of $A$ and $B$.
\begin{equation} \label{Ando1}
C = A \# B = A^{1/2} (A^{-1/2} B A^{-1/2})^{1/2} A^{1/2} \; ,
\end{equation}
where the right hand side is the unique solution of
\begin{equation} \label{Ando2}
B = C A^{-1} C, \quad C > {\mathbf 0} \; .
\end{equation}
For more see \cite{An78, An79, An87, LL01},

Now let $S_A$ be the Hermitian adjoint belonging to
$\langle \phi_1, \phi_2 \rangle_A$. It is an involution acting
on the linear space $\cB(\cH)$, and it is explicitly given by
\begin{equation} \label{gm3}
S_A(X) = A^{-1} X^{\dag} A \, .
\end{equation}
Indeed, one checks
\begin{displaymath}
\langle\phi_1, X \phi_2\rangle_A =
\langle X^{\dag} A \phi_1, \phi_2 \rangle =
\langle (A^{-1} X^{\dag} A) \phi_1, A \phi_2 \rangle \; .
\end{displaymath}
In conclusion one has the implication
\begin{equation} \label{gm2}
A \Longleftrightarrow \langle .,. \rangle_A \Longrightarrow S_A \; .
\end{equation}

For two scalar products, indexed by $A$ and $C$ accordingly, one gets
\begin{equation} \label{gm4}
S_A S_C (X) = A^{-1} ( C^{-1} X^{\dag} C)^{\dag} A
= A^{-1} C X C^{-1} A \; .
\end{equation}
Now given $A$ and $B$, the next aim is to solve the equation
\begin{equation} \label{gm5}
S_A S_C  =  S_C S_B \; ,
\end{equation}
for $C$. By the help of (\ref{gm4}) one gets
\begin{displaymath}
C^{-1} A C^{-1} B \, X = X \, C^{-1} A C^{-1} B  \; .
\end{displaymath}
Being valid for all $X$, one obtains
\begin{equation} \label{gm6}
C^{-1} A C^{-1} B = \lambda^2 \1 , \quad \lambda > 0 ; .
\end{equation}
Demanding $C > 0$ the solution for $C$ is unique and results in
$\lambda C = A \# B$ by (\ref{Ando2}). $\lambda$ drops out in
(\ref{gm5}), and it follows
\begin{equation} \label{gm7}
S_A S_{A \# B}  =  S_{A \# B} S_B \; ,
\end{equation}
{\bf Remark:} Up to a positive numerical factor, $C$
is uniquely
defined by the pair $S_A$, $S_B$ of involutions. By using the
reference Hilbert scalar product, and by restricting the positive
operators to those with determinant $1$, the constant $\lambda$
can be fixed to $\lambda = 1$. This is consistent because of
\begin{equation} \label{gm8}
\det A \# B = \sqrt{\det A \, \det B} \, .
\end{equation}

It can be shown that through any two positive operators $A$ and
$B$ there is an {\em acq--line}\footnote{a line of antilinear
conjugate quandles} $r \to S_r$, satisfying
\begin{equation} \label{gm11}
S_t S_{(t+r)/2} = S_{(t+r)/2} S_r, \quad r,t \in \mathbb{R} \; ,
\end{equation}
and going through $S_A$ for $s=0$ and through $S_B$ for $s=1$.

The prove starts by defining $S_q := C_q^{-1} X^{\dag} C_q$ and
\begin{equation} \label{gm9}
C_q = A^{1/2} Y^q A^{1/2}, \quad Y := A^{-1/2} B A^{-1/2} \; .
\end{equation}
One gets
\begin{displaymath}
S_t S_s = C_t^{-1} (C_s^{-1} X^{\dag} C_s)^{\dag} C_t
= C_t^{-1} C_s X C_s^{-1} C_t \, ,
\end{displaymath}
which should be equal to $S_s S_r$. Because
\begin{equation} \label{gm10}
C_s^{-1} C_t = A^{-1/2} Y^{t-s} A^{1/2} \; ,
\end{equation}
it suffices to choose $t = (r+s)/2$ to satisfy
\begin{displaymath}
C_r C_{(t+r)/2} = C_{(t+r)/2} C_t \; .
\end{displaymath}
$A$ is determined by $S_A$ up to a positive number, $C_t$
becomes unique by assuming $\det A = \det B = 1$. Remember also
$C_{(t+r)/2} = C_r \# C_t$.

\section{Modular objects} \label{C8}
In this subsection the most elementary part of the
Tomita--Takesaki theory is presented; a theory which has been
appreciated as the ``second revolution'' in the treatment of
von Neumann algebras by H.~J.~Borchers. (The first one is the
factor classification by von Neumann and Murray.) It is said
that Tomita needed about 10 years to prove the polar decomposition
of the modular involution for a general von Neumann algebra.

In contrast, just this task is quite simple for von Neumann
algebras acting on a finite dimensional Hilbert space, thereby
getting contact to the concept of entanglement.
There are some standard notations \cite{Haag93} in the
Tomita--Takesaki Theory, which are also used here.
\medskip

Let $\cH$ be a Hilbert space and $\dim \cH = d < \infty$.
An {\em operator system} $\cA$ on $\cH$ is a complex linear
subspace of $\cB(\cH)$ which contains the identity operator $\1$
of $\cH$, and which contains with any operator $A$ its Hermitian
adjoint $A^{\dag}$, \cite{Paulsen02}. A {\em von Neumann
subalgebra} of $\cB(\cH)$ is an operator system which contains with two
operators also their product, i.~e. if $A, B \in \cA$ then
$AB \in \cA$.\\
In the following $\cA$ is von Neumann subalgebra.

It is $\dim \cA \leq d^2$. Presently, the main interest is
focused on the case $\dim \cA = d$. Then, as a further input,
a vector $\psi$ is chosen from $\cH$ such that
\begin{equation} \label{TT1}
\cA \, \psi = \cH, \quad \dim \cA = \dim \cH = d \; .
\end{equation}
The left hand side is the set of all vectors which can be written
as $A \psi$ with $A \in \cA$. This set has dimension $d$ by
assumption. It follows, as we are within finite dimensional linear
spaces, that there is a bijection between the vectors of $\cH$
and the operators of $\cA$,
\begin{equation} \label{TT2}
\varphi \longleftrightarrow A \, : \qquad \varphi = A \psi
, \quad A \in \cA \; .
\end{equation}
{\bf Notational remark:} Generally, a vector $\psi$  is called
{\em cyclic} with respect to $\cA$ if $\cA \, \psi = \cH$. $\psi$
is called {\em separating} if $A \psi = 0$ and $A \in \cA$ implies
$A = {\mathbf 0}$. Hence (\ref{TT2}) is  true if $\psi$ is cyclic
and separating. In QIT one says instead: {\em $\psi$ is completely
entangled.} This  roots in the fact that there is a unique positive
linear form $\omega$, $\omega(A) = \T \; DA$, with $D \in \cA$
positive and {\em invertible}, such that
\begin{displaymath}
\langle \psi, \, A \, \psi \rangle = \T \, DA := \omega(A)
\end{displaymath}
for all $A \in \cA$. (\ref{TT2}) is an elementary example of a
{\em Gelfand isomorphism} within a {\em GNS construction.} The
letters GNS are the initials of Gelfand, Ne'imark, and Segal.
\medskip

The correspondence between vectors and operators, seen in
(\ref{TT2}), is used to define an involution $S_{\psi}$ by
\begin{equation} \label{TT3}
S_{\psi} A \psi = A^{\dag} \psi, \quad A \in \cA \; .
\end{equation}
Because by (\ref{TT2}) $S_{\psi}$ is a well defined
antilinear operator. It clearly obeys $S_{\psi}^2 = \1$. Thus\\
The involution $S_{\psi}$ is called the {\em modular
involution based on $\psi$.}

Now the polar decomposition of an involution comes into play,
and in this connection some Tomita--Takesaki terminology will be
introduced: The polar decomposition of $S_{\psi}$ provides a
conjugation $J_{\psi}$, the {\em modular conjugation,} and a
positive operator $\Delta_{\psi}$, called the {\em modular
operator.} Their definitions start with
\begin{equation} \label{TT4}
\Delta_{\psi} := S_{\psi}^{\dag} S_{\psi} , \quad
\Delta_{\psi}^{-1} = S_{\psi} S_{\psi}^{\dag} \; .
\end{equation}
The second equation appears from rewriting (\ref{invol2}).
The polar decomposition (\ref{invol6}) reads in the present
setting
\begin{equation} \label{TT5}
S_{\psi} = J_{\psi} \Delta_{\psi}^{1/2}
= \Delta_{\psi}^{-1/2} J_{\psi}, \quad
S_{\psi}^{\dag} = \Delta_{\psi}^{1/2} J_{\psi} \; .
\end{equation}
One observes $J \Delta^{1/2} J = \Delta^{-1/2}$ so that
\begin{equation} \label{TT21}
\Delta_{\psi}^t J_{\psi} = J_{\psi} \Delta_{\psi}^{-t}
, \quad t \in \mathbf{R} \; .
\end{equation}
{\bf Remark:} It is in use to write $F_{\psi}$ instead of
$S_{\psi}^{\dag}$, see (\ref{TT8}) below.
\medskip

The set of operators $B$, commuting with all $A \in \cA$, is
denoted by $\cA'$. $\cA'$ is the {\em commutant} of $\cA$.
The commutant is a von Neumann subalgebra of $\cB(\cH)$.

\begin{prop}
Let $\cA$ be a von Neumann subalgebra of $\cB(\cH)$ and assume
(\ref{TT1}) and (\ref{TT2}) are valid. Then
\begin{equation} \label{TT6}
A \in \cA \, \Rightarrow \, S_{\psi} A S_{\psi} \in \cA' \; ,
\end{equation}
and this is a one-to-one map from $\cA$ onto $\cA'$.
\end{prop}
For the proof let us write $S$ for $S_{\psi}$ to simplify the
notation. Because of (\ref{TT2}) the assertion (\ref{TT6})
can be written
\begin{displaymath}
S A_1 S A_2 A \psi = A_2 S A_1 S A \psi \; .
\end{displaymath}
for all $A_1, A_2, A \in \cA$. The left side of the relation
becomes
$S A_1 A^{\dag} A_2^{\dag} \psi = A_2 A A_1^{\dag}$ by repeated
application of (\ref{TT3}). The right is handled similar:
$A_2 S A_1 A^{\dag} \psi = A_2 A A_1^{\dag} \psi$, and (\ref{TT4})
becomes evident. It implies that $\cA'$ is at least
$d$-dimensional. However, if $\dim \cA' > d$, there is an
operator $B \in \cA'$ such that $B \psi = 0$ implying
$B A \psi = 0$ for all $A \in \cA$. By (\ref{TT1}), (\ref{TT2}),
all vectors from $\cH$ are null vectors of $B$, hence $B = {\mathbf 0}$,
and the assertion is proved.

The proof shows that $\cA$ and $\cA'$ are isomorphic
algebras: $A \to S_{\psi} A S_{\psi}$ is an isomorphism.
Both algebras can be handled completely similar with respect
to the chosen vector $\psi$. In particular, there is an
involution $F_{\psi}$ such that
\begin{equation} \label{TT7}
F_{\psi} B \psi = B^{\dag} \psi, \quad B \in \cA' \; .
\end{equation}
The same arguments as above show that $B \to F_{\psi} B F_{\psi}$
is an isomorphism from $\cA'$ onto $\cA$. The next task is the
proof of
\begin{equation} \label{TT8}
F_{\psi}^{\dag} = S_{\psi}, \quad S_{\psi}^{\dag} = F_{\psi} \; .
\end{equation}
Let $\psi_1, \psi_2 \in \cH$. There are $A \in \cA$ and
$B \in \cA'$ such that $\psi_1 = A \psi_0$ and
$\psi_2 = B \psi_0$. It follows $AB = BA$ and
\begin{displaymath}
\langle \psi_2, \psi_1 \rangle =
\langle B\psi_0, A \psi_0 \rangle =
\langle A^{\dag}\psi_0, B^{\dag} \psi_0 \rangle =
\langle SA \psi_0 ,FB \psi_0 \rangle \; ,
\end{displaymath}
and by (\ref{rule2}) the last term is
$\langle B \psi_0, F^{\dag} S A \psi_0 \rangle$ or
$\langle \psi_2, F^{\dag} S \psi_1 \rangle$. Being true for all
$\psi_1, \psi_2 \in \cH$, we must have $F^{\dag}S = \1$ and,
because $S^2 = \1$, the first equation in (\ref{TT6}) is true.
The second follows by taking the Hermitian adjoint.

Knowing (\ref{TT6}), from $SAS \in \cA'$ one gets
$S^{\dag} S \cA SS^{\dag} = \cA$. Now (\ref{TT4}) provides
$\Delta A \Delta^{-1} \in \cA$ for all $A \in \cA$. A similar
conclusion can be drawn for $\cA'$. All together, by functional
calculus, one gets, as $\Delta$ is strictly positive,
\begin{prop}
Let $A \in \cA$ and $B \in \cA'$. Then
\begin{equation} \label{TT9}
\Delta_{\psi}^{it} A \Delta_{\psi}^{-it} \in \cA, \quad
\Delta_{\psi}^{it} B \Delta_{\psi}^{-it} \in \cA'
\end{equation}
for all real numbers $t$. The 1-parameter unitary group
$t \to \Delta_{\psi}^{it}$, $s \in \mathbb{R}$, is the
so-called {\em modular automorphism group.}
\end{prop}

Combining (\ref{TT9}) with (\ref{TT6}) one gets
\begin{equation} \label{TT10}
J_{\psi} \cA J_{\psi} = \cA', \qquad
J_{\psi} \cA' J_{\psi} = \cA   \; ,
\end{equation}
saying that $JAJ \in \cA'$ if and only if $A \in \cA$ and vice
vera. (\ref{TT10}) is more comfortable than (\ref{TT6}) as $J$
is a conjugation, $S$ ``only'' an involution.

\subsection{Maximal commutative subalgebras} \label{C8.1}
Simple though illustrating is an application of the formalism
to a maximal commutative von Neumann subalgebra $\mathcal{C}$
of $\cB(\cH)$. There is a basis $\phi_1, \dots, \phi_d$ of $\cH$,
that  simultaneously diagonalize all operators of $\mathcal{C}$.
The operator $S_{\psi}$ can be based on a vector of the form
\begin{equation} \label{TT11}
\psi = \sum \epsilon_j \phi_j, \quad |\epsilon_j| = 1 \; .
\end{equation}
Let $a_1, \dots, a_d$ be the eigenvalues of $A \in \mathcal{C}$.
Then (\ref{TT3}) becomes
\begin{equation} \label{TT12}
S_{\psi} A \psi = \sum a_j^* \epsilon_j \phi_j \; .
\end{equation}
In particular $S_{\psi} \phi_k = \epsilon_k \phi_k$ for all $k$.
Applying rule (\ref{rule2}) results in
\begin{prop}
For a maximal commuting von Neumann subalgebra of $\cB(\cH)$
\begin{equation} \label{TT13}
S_{\psi} = S_{\psi}^* = J_{\psi}, \quad \Delta_{\psi} = \1
\end{equation}
is valid and (\ref{TT11}) implies (\ref{TT12}). The modular
conjugations are parameterized by the points of a $d$-Torus with
points $\{\epsilon_1, \dots, \epsilon_d\}$. Given $J_{\psi'}$ and
$J_{\psi''}$, there are solutions $S_{\psi}$ of
\begin{equation} \label{TT14}
  S_{\psi'}  S_{\psi} =  S_{\psi}  S_{\psi''} \; .
\end{equation}
Representing all unimodular number in the form $\exp it$,
(\ref{TT14}) is equivalent to
\begin{equation} \label{TT14a}
s_j = \frac{s_j' + s_j''}{2} \, \mod \pi , \quad j=1,\dots d \; .
\end{equation}
\end{prop}
The last assertion comes from rewriting (\ref{TT14}) as a set of
the $d$ conditions $\epsilon_j' = (\epsilon_j)^2 \epsilon_j''$.

\subsection{Bipartite quantum systems} \label{C8.2}
To see the meaning of the modular objects in bipartite quantum
systems,
\begin{equation} \label{TT15}
\cH^{\kA \kB} = \cH^{\kA} \otimes \cH^{\kB}
, \quad \dim \cH^{\kA} = \dim \cH^{\kB} = d \, ,
\end{equation}
is the starting assumption. The decomposition provides two
von Neumann subalgebras,
\begin{equation} \label{TT16}
\cA := \cB(\cH^{\kA}) \otimes \1^{\kB}
, \quad
\cA' = \1^{\kA} \otimes \cB(\cH^{\kB})
\end{equation}
of $\cB(\cH^{\textsc{AB}})$. $\cA'$ is the commutant of
$\cA$. $\cA$ is the commutant of $\cA'$.
A vector $\psi$, allowing for a correspondence in the sense of
(\ref{TT2}), must be separating, i.~e. $A \in \cA$ and
$A \psi = 0$ only if $A = {\mathbf 0}$. Because of (\ref{TT15})
it follows $\dim \cA \, \psi = d^2$ and $\psi$ is cyclic.
In Quantum Information Theory one would call $\psi$
{\em completely entangled}, i.~e. the Schmidt number of $\psi$
is $d = \dim \cH^{\kA}$.

Let $\psi \in \cH^{\kA \kB}$ denote a completely entangled
vector. Then there are bases
$\phi_1^{\kA}, \phi_2^{\kA}, \dots$ of $\cH^{\kA}$ and
$\phi_1^{\kB}, \phi_2^{\kB}, \dots$ of $\cH^{\kB}$
such that
\begin{equation} \label{TT17}
\psi = \sum_{j=1}^d \sqrt{p_j} \phi_j^{\kA} \otimes
\phi_j^{\kB} , \quad p_j > 0 \; ,
\end{equation}
is a {\em Schmidt decomposition} of $\psi$. The reduced
operators attached to $\psi$ are the partial traces of
$| \psi \rangle\langle \psi|$ onto the two subsystems. They are
\begin{equation} \label{TTred}
\rho^{\kA} =  \sum p_j |\phi_j^{\kA} \rangle\langle \phi^{\kA}_j|
,\quad \rho^{\kB} =
\sum p_j |\phi_j^{\kB} \rangle\langle \phi^{\kB}_j| \; ,
\end{equation}
see \cite{NC00}, \cite{BZ06}. In case $\psi$ is a unit vector
then the partial traces are of trace one and allow for an
interpretation as density operators.

To determine $S_{\psi}$, the modular involution, it suffices to
know its action on a basis of $\cH^{\kA} \otimes 1^{\kB}$. A
good choice reads
$|\phi_j^{\kA} \rangle\langle \phi_k^{\kA}| \otimes \1^{\kB}$.
Then the problem reduces to
\begin{displaymath}
S_{\psi}(| \phi_j^{\kA} \rangle\langle \phi_k^{\kA}|
\otimes \1^{\kB}) \psi = |\phi_k^{\kA} \rangle\langle
\phi_j^{\kA}| \otimes \1^{\kB} \psi \; .
\end{displaymath}
Performing the calculation results in
\begin{equation} \label{TT18}
S_{\psi} \sqrt{p_k} \phi_j^{\kA} \otimes \phi_k^{\kB}
= \sqrt{p_j} \phi_k^{\kA} \otimes \phi_j^{\kB} \; .
\end{equation}
$S_{\psi}$ is the unique antilinear operator satisfying
(\ref{TT18}). Abbreviation
$|jk\rangle := \phi_j^{\kA} \otimes \phi_k^{\kB}$, one gets
\begin{equation} \label{TT19}
S_{\psi} |jk\rangle = \sqrt{\frac{p_j}{p_k}} \, |kj\rangle, \quad
S_{\psi}^{\dag}|jk\rangle =\sqrt{\frac{p_k}{p_j}} \, |kj\rangle\;.
\end{equation}
It is now evident how to polar decompose $S_{\psi}$ as in
(\ref{TT4}), (\ref{TT5}). Having in mind that
$\Delta_{\psi} = S^{\dag}S$ is linear and $J_{\psi}$ antilinear,
it follows
\begin{equation} \label{TT20}
\Delta_{\psi} |jk\rangle = \frac{p_j}{p_k} |jk\rangle, \quad
J_{\psi} |jk\rangle = |kj\rangle \; ,
\end{equation}
From these relations one gets evidence of
\begin{equation} \label{TT20a}
\Delta_{\psi} = \rho^{\kA} \otimes (\rho^{\kB})^{-1} \; .
\end{equation}
Again from (\ref{TT19}) one can construct a basis of $\cH_{\psi}$.
\begin{equation} \label{TT22}
 \sqrt{p_k} |jk\rangle + \sqrt{p_j} |kj\rangle, \quad
 i \sqrt{p_k} |jk\rangle - i \sqrt{p_j} |kj\rangle, \quad
\forall j \leq k \; .
\end{equation}

There are links to the geometric mean, see \ref{C7.gm}.
\begin{prop} \label{Ymodop}
Let $S_k = J \Delta_k$ be defined according to (\ref{TT5}).
There are modular objects fulfilling $S = J \Delta$ such that
\begin{equation} \label{Y3}
  S_1 \: S = S \, S_2  \; .
\end{equation}
The modular operator $\Delta$ of $S$ satisfies
\begin{equation} \label{Y4}
\Delta \, \Delta_1^{-1} \Delta = \Delta_2, \quad
\Delta = \Delta_1 \# \Delta_2 \; .
\end{equation}
The geometric mean $\Delta$ of $\Delta_1$, $\Delta_2$, is
uniquely defined by (\ref{Y3}).
\end{prop}
Proof: Multiplying (\ref{Y3}) from the left by $J$, one obtains
$\Delta_2=\Delta J \Delta_1 J \Delta=\Delta \Delta_1^{-1}\Delta$,
where $J \Delta_1 J = \Delta_1^{-1}$ has been used. To compare
with (\ref{gm7}) one rewrites the first expression in (\ref{Y4})
as $(\Delta_1^{-1/2} \Delta \Delta_1^{-1/2})^2 = \Delta_1^{-1/2}
\Delta_2 \Delta_1^{-1/2}$ and solves for $\Delta$ :
\begin{equation} \label{Y5}
\Delta = \Delta_1^{1/2}
(\Delta_1^{-1/2} \Delta_2 \Delta_1^{-1/2})^{1/2} \Delta_1^{1/2}
\end{equation}

A useful observation says that {\em $\psi$ is maximally entangled
if and only if $\Delta_{\psi}$ is proportional
to $\1^{\kA \kB}$.} Then $J_{\psi} = S_{\psi}$ and all Schmidt
numbers are mutually equal. All maximal entangled vectors can be
gained by applying  unitary operators of the product form
$U^{\kA} \otimes U^{\kB}$ to a chosen maximally entangles $\psi$.
In fact it suffices to apply a unitary of the form
$U^{\kA} \otimes 1^{\kB}$.

The set of all $S_{\psi}$ satisfying $S_{\psi} = J_{\psi}$ is a
submanifold of the symplectic space of all conjugations contained
in $\cB{\cH^{\kA \kB}}_{\rc}$.\\
Question: {\em Is this submanifold symplectic?} \\
Question: {\em Is there for any pair $\psi'$, $\psi''$ of
maximally entangled vectors at least one further maximally
entangled vector $\psi$ such that}
\begin{equation} \label{Y6}
J_{\psi'} J_{\psi} = J_{\psi} J_{\psi''}
\end{equation}
{\em can be fulfilled ?}\\
Up to my knowledge the answers to these question are unknown.

By (\ref{invol6}) or, equivalently, (\ref{TT5}) one shows
\begin{equation} \label{TT23}
S_{\psi} (\Delta_{\psi}^t \cH_{\psi}) = \Delta_{\psi}^{-t-1/2}
\cH_{\psi}, \quad S^{\dag}_{\psi} (\Delta_{\psi}^t \cH_{\psi})
= \Delta_{\psi}^{t+1/2} \cH_{\psi} \; ,
\end{equation}
Notice that $\cH_{\psi} = \cH_S$ if $S = S_{\psi}$.

For real $t$ the linear spaces $\Delta^t \cH_{\psi}$ are real too.
A bases for them can be generated by applying $\Delta^t$ to the
basis (\ref{TT22}) of $\cH_J$.
so that, applying definition (\ref{invol7}) to the involutions
$S_{\psi}$ and $F_{\psi} = S_{\psi}^{\dag}$,
\begin{equation} \label{TT24}
\cH_{S_{\psi}} = \Delta_{\psi}^{-1/4} \cH_{J_{\psi}}, \quad
\cH_{F_{\psi}} = \Delta_{\psi}^{1/4} \cH_{J_{\psi}} \; .
\end{equation}

\section{Antilinearity and the
Einstein-Podolski-Rosen Effect} \label{C9}
This section aims at a particular aspect of the
Einstein-Podolski-Rosen or ``EPR'' effect \cite{EPR35}:
{\em The appearance of antilinearity.}

While the paper of Einstein et al generated some wild
philosophical discussion, Schr\"odinger \cite{Schr35a},
\cite{Schr35b}, introduced the concept of entanglement. For a
general overview one may consult \cite{Peres93}, \cite{NC00},
or any QIT textbook.

Far from being complete in any sense, the present paper introduces
to some antilinear facets of the EPR effect. The idea, to look at
these particular antilinearities, is already in \cite{CrUh98},
more elaborated, in \cite{Uh00a}, \cite{Uh00b}, \cite{CrUh09}, and,
more recently, in \cite{BNT12}. Related in spirit is \cite{BO00}.

To begin with, it needs some mathematical and notational
preliminaries. As in (\ref{TT15}) the basic structure is a
bipartite quantum system
\begin{displaymath}
\cH^{\textsc{AB}} = \cH^{\kA} \otimes \cH^{\kB} \; ,
\end{displaymath}
however without requiring equal dimensionality of the factors. In
the language of QIT: The subsystem given by $\cH^{\kA}$ is in the
hands of {\em Alice,} who is responsible for the actions on the
A--system. In the same spirit {\em Bob} is the owner of the
B--system. The two subsystems may or may be not spatially
separated.

 Fixing $\psi \in \cH^{\kA \kB}$ there are
Schmidt decompositions, (as in (\ref{TT17}) but normalized,)
\begin{displaymath}
\psi = \sum_{j=1}^k \sqrt{p_j} \phi_j^{\kA} \otimes
\phi_j^{\kB} , \quad p_j > 0 , \quad \sum p_j = 1 \; .
\end{displaymath}
$k$ is the Schmidt rank of $\psi$.
\begin{displaymath}
\rho^{\kA} := \T_{\kB}|\psi \rangle\langle \psi| =
\sum_{j=1}^k p_j |\phi_j^{\kA} \rangle\langle
\phi^{\kA}_j|
\end{displaymath}
is the partial trace over the B--system. $\rho^{\kA}$ encodes
``what can be seen'' of $\psi$ from Alice's system. By Exchanging
A with B one gets the partial trace over the A--system.

A further convention is as follows: $X^{ba}$ indicates a map from
$\cH^{\kA}$ into $\cH^{\kB}$. Similarly, the superscript ``$ab$''
in $X^{ab}$ points to a map from $\cH^{\kB}$ into $\cH^{\kA}$.
This contrasts the notion, say $X^{\kA \kB}$, for an operator
acting on $\cH^{\kA \kB}$. See also subsection \ref{C3.4}.
\medskip

The EPR phenomenon says: An action done by (say) Alice generally
influences Bob's system. This is the reason for saying ``the state
of the composed system {\em entangles} the two subsystems''.

As previously we assume the bipartite quantum system in a state
described by the unit vector  $\psi \in \cH^{\kA \kB}$. Alice asks
wether her state is $|\phi^{\kA} \rangle\langle \phi^{\kA}|$ or
not. (The general case of a mixed state will be described
in subsection \ref{C9.3}.)

In the {\em composed system} Alice's question can be seen as
a L\"uders' measurement \cite{Lued51} using the projection operator
$P = |\phi^{\kA} \rangle\langle \phi^{\kA}| \otimes \1^{\kB}$ and
its orthogonal complement $P^{\perp} = 1^{\kA \kB} - P$.
If Alice gets the answer ``yes'', then the vector
$\psi' = P\psi$ and the state $w^{-1}|\psi'\rangle\langle \psi'|$
is prepared. Here $w = \langle \psi, P \psi \rangle$ is Alice's
success probability for getting the answer YES. Now
\begin{displaymath}
P \psi =
(|\phi^{\kA} \rangle\langle \phi^{\kA}|\otimes \1^{\kB}) \psi \;,
\end{displaymath}
and in the composed system the vector $\psi$ is transformed into
a product vector of the form $\phi^{\kA} \otimes \phi^{\kB}$.
Clearly, $\phi^{\kB}$ depends on $\phi^{\kA}$ and $\psi$ only.
Fixing $\psi$ but varying $\phi^{\kA}$ defines a map
$s_{\psi}^{ba}$ from $\cH^{\kA}$ into $\cH^{\kB}$,
\begin{equation} \label{EPR1}
(|\phi^{\kA} \rangle\langle \phi^{\kA}|\otimes \1^{\kB})
\psi = \phi^{\kA} \otimes  s_{\psi}^{ba} \phi^{\kA} ,
\quad \phi^{\kA} \in \cH^{\kA} \; .
\end{equation}
Thus, if Alice is successful in preparing $\phi^{\kA}$, then
$\phi^{\kB}_{\rm out} := s_{\psi}^{ba} \phi^{\kA}$ is prepared
in Bob's system.

Of course one can exchange the roles of Alice and Bob: There is a
map $s_{\psi}^{ab}$ showing the state in Alice's system caused by
a successfully preparing $\phi^{\kB}$ by Bob.
\begin{equation} \label{EPR2}
(\1^{\kA}\otimes |\phi^{\kB} \rangle\langle \phi^{\kB}|)
\psi =  s_{\psi}^{ab} \phi^{\kB} \otimes \phi^{\kB}  ,
\quad \phi^{\kB} \in \cH^{\kB} \; .
\end{equation}
\begin{prop}
$s_{\psi}^{ba}$ and $s_{\psi}^{ab}$ are antilinear maps from
$\cH^{\kA}$ into $\cH^{\kB}$ and from $\cH^{\kB}$ into
$\cH^{\kA}$ respectively. They depend linearly on $\psi$.
If $\psi$ is represented by
\begin{equation} \label{EPR3}
\psi = \sum c_{jk} \phi_j^{\kA} \otimes \phi_k^{\kB}
\end{equation}
then
\begin{equation} \label{EPR3a}
s_{\psi}^{ba} \phi^{\kA} = \sum c_{jk}
\langle \phi^{\kA}, \phi_j^{\kA} \rangle \phi_k^{\kB}
, \quad
s_{\psi}^{ab} \phi^{\kB} = \sum c_{jk}
\langle \phi^{\kB}, \phi_k^{\kB} \rangle \phi_j^{\kA} \; .
\end{equation}
\end{prop}
Proof: It is immediate from (\ref{EPR1}) that
$\psi \to s_{\psi}^{ba}$ is linear in $\psi$ and maps $\cH^{\kA}$
antilinearly into $\cH^{\kB}$, i.~e. it is contained in
 $\cH^{\kA \kB}$. The map realizes the mapping (\ref{caniso3})
advertised in subsection \ref{C3.4}. (Though without demanding
equality of the subsystem's dimensions,)  By linearity in
$\psi$ it suffices to show the equivalence of (\ref{EPR1}) and
(\ref{EPR3a}) if $\psi = \tilde \phi^{\kA} \otimes \phi^{\kB}$
is a product vector. (\ref{EPR3a}) becomes
\begin{displaymath}
s_{\psi}^{ba} \phi^{\kA} =
\langle \phi^{\kA}, \tilde \phi^{\kA} \rangle \phi^{\kB} \; ,
\end{displaymath}
which is consistent with (\ref{EPR1}). Exchanging the roles
of Alice and Bob in the preceding discussion gives the other
part of the assertion. Also compare with (\ref{2rank5a}).

As already indicated in subsection \ref{C3.4}, the maps
(\ref{EPR3a}) are partial isometries: Rewriting (\ref{caniso4})
results in
\begin{equation} \label{EPR3c}
\langle \psi, \varphi \rangle
= \T \, s_{\varphi}^{ab} (s_{\psi}^{ab})^{\dag}
= \T \, s_{\varphi}^{ba} (s_{\psi}^{ba})^{\dag} \; .
\end{equation}
The ordering within the traces is due to the antilinearity of
the maps involved.

An important relation reads
\begin{equation} \label{EPR6}
\langle \phi^{\kA} \otimes \phi^{\kB}, \psi \rangle
= \langle \phi^{\kA}, s_{\psi}^{ab} \phi^{\kB}\rangle
= \langle \phi^{\kB}, s_{\psi}^{ba} \phi^{\kA}\rangle \; .
\end{equation}
To prove the first equality, one calculates by (\ref{EPR1})
the transition amplitudes
\begin{displaymath}
\langle \phi^{\kA} , \phi^{\kA} \rangle \,
\langle \phi^{\kA}, s_{\psi}^{ab} \phi^{\kB}\rangle =
\langle \phi^{\kA} \otimes \phi^{\kB} , (|\phi^{\kA}
\rangle\langle \phi^{\kA}|\otimes \1^{\kB}) \psi \rangle
= \langle \phi^{\kA} , \phi^{\kA} \rangle \,
\langle \phi^{\kA} \otimes \phi^{\kB} , \psi \rangle
\end{displaymath}
saying the first equality in (\ref{EPR6}) is true. The other one
is seen by exchanging the roles of Alice and Bob.

A consequence of the proposition is
\begin{equation} \label{EPR4}
(s_{\psi}^{ba})^{\dag} = s_{\psi}^{ab}, \quad
(s_{\psi}^{ab})^{\dag} = s_{\psi}^{ba} \; .
\end{equation}
Another one is the reconstruction of $\psi$ from $s_{\psi}^{ba}$:
\begin{prop}
Let
\begin{equation} \label{EPRyy}
 \sum_j |\phi_j^{\kA} \rangle\langle \phi_j^{\kA} |
= \1^{\kA}
, \quad
 \sum_k |\phi_k^{\kB} \rangle\langle \phi_k^{\kB} |
= \1^{\kB}
\end{equation}
be decompositions of the unity of $\cH^{\kA}$ and
$\cH^{\kA}$ respectively. Then
\begin{equation} \label{EPRy}
\psi =
\sum_j \phi^{\kA}_j \otimes s^{ba}_{\psi} \phi^{\kA}_j
=
\sum_k s^{ab}_{\psi} \phi^{\kB}_k \otimes \phi^{\kB}_k  \; .
\end{equation}
\end{prop}
 To see it, one starts with
\begin{displaymath}
\psi = \sum \langle \phi^{\kA}_j \otimes \phi^{\kB}_k ,
\psi \rangle \, \phi^{\kA}_j \otimes \phi^{\kB}_k
\end{displaymath}
and applies (\ref{EPR6}) to get, for instance,
\begin{displaymath}
\psi = \sum \langle \phi^{\kA}_j, s_{\psi}^{ab} \phi^{\kB}_k
\rangle \, \phi^{\kA}_j \otimes \phi^{\kB}_k \; .
\end{displaymath}
Summing up over $j$ yields
$\sum s_{\psi}^{ab} \phi^{\kB}_k \otimes \phi^{\kB}_k$.
The other case is similar.

By the help of (\ref{2rank1b}) one may rewrite (\ref{EPR1})
and (\ref{EPR2}) by
\begin{equation} \label{EPRz}
s_{\psi}^{ba} = \sum c_{jk} |\phi_k^{\kB} \rangle\langle
\phi_j^{\kA}|_{\rc}
\, \hbox{ if } \,
\psi = \sum c_{jk} \phi_j^{\kA} \otimes \phi_k^{\kB} \; .
\end{equation}

If $\psi$ is in the Schmidt form
\begin{equation} \label{EPR8a}
\psi = \sum \sqrt{p_j} \phi_j^{\kA} \otimes \phi_j^{\kB} \; ,
\end{equation}
a simplification occurs and one gets
\begin{equation} \label{EPR8}
s_{\psi}^{ba} \phi_j^{\kA} = \sqrt{p_j} \phi_j^{\kB}
, \qquad
s_{\psi}^{ab} \phi_j^{\kB} = \sqrt{p_j} \phi_j^{\kA} \: .
\end{equation}
From it one easily sees $s_{\psi}^{ab}s_{\psi}^{ba} \phi_j^{\kA}
= p_j \phi_j^{\kA}$ or. as an operator equations,
\begin{equation} \label{EPR9}
s_{\psi}^{ab}s_{\psi}^{ba} = \T_{\kB}
|\psi \rangle\langle \psi| := \rho^{\kA} , \quad
s_{\psi}^{ba}s_{\psi}^{ab} = \T_{\kA}
|\psi \rangle\langle \psi| := \rho^{\kB} ,
\end{equation}
in agreement with (\ref{TT15}).

Assuming $\phi^{\kA}$ is a unit vector,
the probability that Alice is successfully preparing
$\phi^{\kA}$ is equal to
$\langle \phi^{\kA}, \rho^{\kA} \phi^{\kA}\rangle$.
On Bob's system the transition
\begin{displaymath}
\rho^{\kB} \to \phi_{\rm out}^{\kB} = s_{\psi}^{ba}
\phi^{\kA}
\end{displaymath}
occurs with the same probability. Indeed,
\begin{equation} \label{EPR10}
\langle \phi^{\kA}, \rho^{\kA} \phi^{\kA} \rangle
=
\langle\phi^{\kA}, (s^{ba})^{\dag} s^{ba}\phi^{\kA} \rangle
=
\langle \phi_{\rm out}^{\kB}, \phi_{\rm out}^{\kB}\rangle
\; .
\end{equation}
Exchanging the roles of Alice and Bob and setting
$\phi_{\rm out}^{\kA} = s_{\psi}^{ab} \phi^{\kB}$
one gets
\begin{displaymath}
\langle \phi^{\kB}, \rho^{\kB} \phi^{\kB} \rangle =
\langle \phi_{\rm out}^{\kA}, \phi_{\rm out}^{\kA}\rangle \; .
\end{displaymath}
Finally, the change by applying a general local transformation
\begin{equation} \label{EPR11a}
\psi \to \varphi := (X^{\kA} \otimes X^{\kB}) \, \psi
\end{equation}
is described by
\begin{equation} \label{EPR11b}
s_{\varphi}^{ab}  = X^{\kA} s_{\psi}^{ab} (X^{\kB})^{\dag}
, \quad
s_{\varphi}^{ba}  = X^{\kB} s_{\psi}^{ba} (X^{\kA})^{\dag}
 \; .
\end{equation}

\subsection{Polar decomposition} \label{C9.1}
Given $\psi$, let us denote the supporting subspaces of
$\rho^{\kA}$ and $\rho^{\kB}$ in $\cH^{\kA}$ and in $\cH^{\kB}$
by $\cH^{\kA}_{\psi}$ and $\cH^{\kB}_{\psi}$ respectively. The
dimensions of the supporting subspaces coincides and they are
equal to the Schmidt rank of $\psi$,
\begin{equation} \label{EPRP1}
\hbox{Schmidt rank} [\psi] =
\hbox{rank}[\rho^{\kA}] = \hbox{rank}[\rho^{\kB}] \; .
\end{equation}
It follows from (\ref{EPR9}) that $s_{\psi}^{ba}$ maps
$\cH^{\kA}_{\psi}$ onto $\cH^{\kB}_{\psi}$.
The vectors $\phi' \in \cH^{\kA}$ which are orthogonal to
$\cH^{\kA}_{\psi}$ are annihilated: $s_{\psi}^{ba} \phi' = 0$.
Similar statements are true for $s_{\psi}^{ab}$.
\begin{thm}[Polar decomposition of the $s$-maps]
There are antilinear partial isometries
$j_{\psi}^{ba}$ and $j_{\psi}^{ab}$ fulfilling
\begin{equation} \label{EPRP5}
j_{\psi}^{ba} = (j_{\psi}^{ab})^{\dag} , \quad
j_{\psi}^{ab} = (j_{\psi}^{ba})^{\dag}
\end{equation}
and
\begin{equation} \label{EPRP2}
( j_{\psi}^{ba} )^{\dag} j_{\psi}^{ba} = P_{\psi}^{\textsc{{A}}}
, \quad
( j_{\psi}^{ab} )^{\dag} j_{\psi}^{ab} = P_{\psi}^{\textsc{{B}}}
\end{equation}
where $P_{\psi}^{\textsc{{A}}}$, respectively
$P_{\psi}^{\textsc{{B}}}$, is the projection operator onto
$\cH^{\kA}_{\psi}$ respectively onto $\cH^{\kB}_{\psi}$,
such that
\begin{equation} \label{EPRP3}
s_{\psi}^{ba} =  j_{\psi}^{ba} \sqrt{\rho_{\psi}^{\kA}}
= \sqrt{\rho_{\psi}^{\kB}} j_{\psi}^{ba} \; .
\end{equation}
\end{thm}
Proof: Let $m$ be the Schmidt rank of the unit vector $\psi$, and
assume $\psi$ in a Schmidt form (\ref{EPR8a}) with $m$ terms.
The antilinear operators defined by
\begin{equation} \label{EPRP4}
j_{\psi}^{ba} \phi^{\kA} := \sum_j^m
\langle \phi^{\kA} , \phi^{\kA}_j \rangle \, \phi^{\kA}_j
, \quad
j_{\psi}^{ab} \phi^{\kB} := \sum_j^m
\langle \phi^{\kB} , \phi^{\kB}_j \rangle \, \phi^{\kB}_j
\end{equation}
satisfy (\ref{EPRP5}). Hence
\begin{displaymath}
( j_{\psi}^{ba} )^{\dag} j_{\psi}^{ba} =
j_{\psi}^{ab}j_{\psi}^{ba} = P_{\psi}^{\textsc{{A}}}
, \quad
( j_{\psi}^{ab} )^{\dag} j_{\psi}^{ab} =
j_{\psi}^{ba}j_{\psi}^{ab} = P_{\psi}^{\textsc{{B}}}  \: .
\end{displaymath}
By (\ref{EPRP4}) it becomes obvious for $1 \leq j \leq m$ that
\begin{equation} \label{EPRP4a}
j_{\psi}^{ba} \phi_j^{\kA} = \phi_j^{\kB}
, \qquad
j_{\psi}^{ab} \phi_j^{\kB} = \phi_j^{\kA} \: ,
\end{equation}
while $j_{\psi}^{ab}$ annihilates the orthogonal complement of
the supporting space of $\rho^{\kB}$. $j_{\psi}^{ba}$ behaves
similarly. Now (\ref{EPR8}) is used to show
\begin{displaymath}
s_{\psi}^{ba} \phi_j^{\kA}
= j_{\psi}^{ba} \sqrt{\rho^{\kA}} \phi_j^{\kA}
= \sqrt{\rho^{\kB}} j_{\psi}^{ba} \phi_j^{\kA}
\end{displaymath}
for bases which serve to Schmidt compose $\psi$. Both sides
of both equations uniquely define antilinear operators, and
\begin{displaymath}
s_{\psi}^{ba} = j_{\psi}^{ba} \sqrt{\rho^{\kA}}
= \sqrt{\rho^{\kB}} j_{\psi}^{ba}
\end{displaymath}
has been established.

Taking the Hermitian adjoint one obtains
\begin{equation} \label{EPRP3a}
s_{\psi}^{ab} =  j_{\psi}^{ab} \sqrt{\rho_{\psi}^{\kB}}
= \sqrt{\rho_{\psi}^{\kA}} j_{\psi}^{ab} \; .
\end{equation}
Taking in account the support property (\ref{EPRP2}), one obtains
\begin{equation} \label{EPRP6}
j_{\psi}^{ab} \sqrt{\rho_{\psi}^{\kB}} j_{\psi}^{ba} =
\sqrt{\rho_{\psi}^{\kA}}
 , \quad
j_{\psi}^{ba} \sqrt{\rho_{\psi}^{\kA}} j_{\psi}^{ab} =
\sqrt{\rho_{\psi}^{\kB}} \; .
\end{equation}
Let $f(x)$ be real and continuous on $0 \leq x $. It follows
\begin{equation} \label{EPRP7}
j_{\psi}^{ab} f(\rho_{\psi}^{\kB}) j_{\psi}^{ba} =
\rho_{\psi}^{\kA}
 , \quad
j_{\psi}^{ba} f(\rho_{\psi}^{\kA})  j_{\psi}^{ab} =
\rho_{\psi}^{\kB} \; .
\end{equation}
by functional calculus. In doing so, the $0$-th powers of
$\rho_{\psi}^{\kA}$ and of $\rho_{\psi}^{\kB}$ should be
set to $P_{\psi}^{\kA}$ and to $P_{\psi}^{\kB}$ respectively.

\subsection{Representing modular objects} \label{C9.2}
In this subsection $\cH^{\kA}$ and $\cH^{\kB}$ are supposed to
be of equal dimensions, and $\psi \in \cH^{\kA \kB}$ to be
completely entangled, i.~e. of maximal Schmidt rank,
\begin{equation} \label{EPRM1}
\dim \cH^{\kA} = \dim \cH^{\kB} = d, \quad
(\rho^{\kA})^{-1}, \, (\rho^{\kB})^{-1} \,
\hbox{ do exist.}
\end{equation}
Then $P_{\psi}^{\kA} = \1^{\kA}$ and $P_{\psi}^{\kB} = \1^{\kB}$,
and (\ref{EPRP5}), (\ref{EPRP2})become
\begin{equation} \label{EPRM2}
( j_{\psi}^{ba} )^{\dag} = j_{\psi}^{ab} = (j_{\psi}^{ba})^{-1}
, \quad
( j_{\psi}^{ab} )^{\dag} = j_{\psi}^{ba} = (j_{\psi}^{ab})^{-1}
\; .
\end{equation}
As it turns out, the ``modular objects'' considered in subsection
\ref{C9.1} of section \ref{C9} can be represented by the maps
$s_{\psi}^{ba}$, $s_{\psi}^{ab}$, $j_{\psi}^{ba}$, and
$j_{\psi}^{ab}$. This will be shown in the next but next part.
In the next one {\em twisted direct products} of antilinear maps
will be introduced.

\subsubsection{Twisted direct products} \label{C9.2.1}
The direct product $\vartheta' \otimes \vartheta''$ of two
antilinear operators or maps is well defined. It is antilinear
and it is acting on product vectors as
\begin{displaymath}
(\vartheta' \otimes \vartheta'')
\phi' \otimes \phi'' =
(\vartheta' \phi') \otimes (\vartheta'' \phi'' ) \: .
\end{displaymath}
The main difference to the direct product of two linear operators
is in the rule
\begin{displaymath}
c (\vartheta' \otimes \vartheta'') =
(c \vartheta') \otimes \vartheta'' =
(\vartheta' c^*) \otimes \vartheta'' =
\vartheta' \otimes (c \vartheta'') =
\vartheta' \otimes (\vartheta'' c^*) =
(\vartheta' \otimes \vartheta'') c^*
\end{displaymath}
Notice: There is no mathematical consistent direct product of a
linear and an antilinear operator within the category of
complex linear spaces.

The {\em twisted direct product} will be denoted by a
``twisted cross'' $\tilde \otimes$. Let $\vartheta^{ba}$ be an
antilinear map from $\cH^{\kA}$ into $\cH^{\kB}$ and
$\vartheta^{ab}$ an antilinear map from $\cH^{\kB}$ into
$\cH^{\kA}$. Then $\vartheta^{ab} \tilde \otimes \vartheta^{ba}$
is an antilinear map defined by
\begin{equation} \label{EPRM5}
(\vartheta^{ab} \tilde \otimes \vartheta^{ba})
(\phi^{\kA} \otimes \phi^{\kB}) =
(\vartheta^{ab} \phi^{\kB}) \otimes
(\vartheta^{ba} \phi^{\kA}) \; .
\end{equation}
One of the remarkable features of the twisted direct product is
the rule
\begin{equation} \label{EPRM6}
(\vartheta^{ab} \tilde \otimes \vartheta^{ba})^2 =
(\vartheta^{ab} \vartheta^{ba}) \otimes
(\vartheta^{ba} \vartheta^{ab}) \; ,
\end{equation}
saying that the square of a twisted cross product
(or the product of any two of them) is an ``ordinary'' cross
product.

\subsubsection{Modular objects} \label{C9.2.2}
For the following we use the Schmidt form (\ref{TT17}),
(\ref{EPR8a}), of $\psi$
\begin{displaymath}
\psi = \sum \sqrt{p_j} \phi_j^{\kA} \otimes \phi_j^{\kB} \; .
\end{displaymath}
Thus (\ref{EPR8}) and (\ref{EPRP4a}) are valid:
\begin{displaymath}
s_{\psi}^{ba} \phi_j^{\kA} = \sqrt{p_j} \phi_j^{\kB}
, \quad
s_{\psi}^{ab} \phi_j^{\kB} = \sqrt{p_j} \phi_j^{\kA}
, \quad
j_{\psi}^{ba} \phi_j^{\kA} = \phi_j^{\kB}
, \quad
j_{\psi}^{ab} \phi_j^{\kB} = \phi_j^{\kA} \: ,
\end{displaymath}
As a first consequence
\begin{equation} \label{EPRM7a}
(j_{\psi}^{ab} \tilde \otimes j_{\psi}^{ba})
\phi_j^{\kA} \otimes \phi_k^{\kB} =
j_{\psi}^{ab} \phi_k^{{\kB}} \otimes j_{\psi}^{ba} \phi_j^{\kA}
= p_k \phi_k^{\kA} \otimes \phi_j^{\kB} \; .
\end{equation}
On the right one finds the action of the modular operator
$J_{\psi}$ onto $\phi_j^{\kA} \otimes \phi_k^{\kB}$.
But if the actions of two antilinear operators agree
on a basis, they are equal one to another:
\begin{equation} \label{EPRM7b}
J_{\psi} = j_{\psi}^{ab} \tilde \otimes j_{\psi}^{ba} \; .
\end{equation}
A similar reasoning, following (\ref{EPRM6}), results in
\begin{equation} \label{EPRM8a}
(s_{\psi}^{ab} \tilde \otimes s_{\psi}^{ba})
\phi_j^{\kA} \otimes \phi_k^{\kB} =
(s_{\psi}^{ab} \phi_k^{\kB}) \otimes
s_{\psi}^{ba} \phi_j^{\kA} = \sqrt{p_j p_k}
\phi_k^{\kA} \otimes \phi_j^{\kB}
\end{equation}
and, using (\ref{EPRM7b}), in
\begin{equation} \label{EPRM8b}
s_{\psi}^{ab} \tilde \otimes s_{\psi}^{ba} =
(\rho^{\kA} \otimes \rho^{\kB})^{1/2} J_{\psi} \; .
\end{equation}
This equation is symmetric with respect to the two subsystems.
One observes
\begin{displaymath}
(s_{\psi}^{ab} \tilde \otimes s_{\psi}^{ba}) J_{\psi} =
(\1^{\kA} \otimes \rho^{\textsc B}) \,
((\rho^{\kA})^{1/2}  \otimes (\rho^{\kB})^{-1/2}) \; .
\end{displaymath}
Comparing with (\ref{TT20a}), the last cross product at
the right of the equation is identified with the square
root of the modular operator. Hence,
\begin{equation} \label{EPRM8c}
s_{\psi}^{ab} \tilde \otimes s_{\psi}^{ba} =
(\1^{\kA} \otimes \rho^{\textsc B}) \,\Delta_{\psi}^{1/2} J_{\psi}
= (\1^{\kA} \otimes \rho^{\textsc B}) S_{\psi} \; .
\end{equation}
The last equality is gained from (\ref{TT5}).

\subsection{From vectors to states} \label{C9.3}
\begin{thm} \label{2iso1}
To any $\rho \in \cB(\cH^{\kA \kB})$ there are linear maps
$\Phi_{\rho}^{ba}$ and $\Phi_{\rho}^{ab}$ from $\cB(\cH^{\kA})$
into $\cB(\cH^{\kB})$ and from $\cB(\cH^{\kB})$ into
$\cB(\cH^{\kA})$ such that
\begin{equation} \label{sEPR6}
\T \, (X^{\kA} \otimes X^{\kB}) \rho =
\T \, X^{\kB} \Phi^{ba}_{\rho}(X^{\kA})  =
\T \, X^{\kA} \Phi^{ab}_{\rho}(X^{\kB})
\end{equation}
is valid for all $X^{\kA} \in \cH^{\kA}$ and for all
$X^{\kB} \in \cH^{\kB}$.
\end{thm}
Proof: By fixing $X^{\kA}$ the left of (\ref{sEPR6}) becomes a
linear form on $\cB(\cH^{\kB})$. Hence it can be expressed
uniquely as $\T \, X^{\kB} Y$, $Y \in \cB(\cH^{\kB})$. $Y$ depends
linearly on $X^{\kA}$ and maps $\cB(\cH^{\kA})$ into
$\cB(\cH^{\kB})$. By denoting $\Phi^{ba}_{\rho}(X^{\kA}) = Y$,
the asserted properties are satisfied. Exchanging the roles of
$X^{\kA}$ and of $X^{\kB}$ gives the other equation.

The map $\rho \mapsto \Phi^{ba}_{\rho}$ is linear. Because our
spaces are all finite dimensional, the map is onto. Therefore
(\ref{sEPR6}) induces isomorphisms
\begin{equation} \label{sEPR7}
\cB(\cH^{\cA} ,\cH^{\cB}) \leftrightarrow \cH^{\kA \kB}
\leftrightarrow \cB(\cH^{\cA} ,\cH^{\cB}) \; .
\end{equation}

\begin{thm} \label{2iso2}
Let $\rho$ in (\ref{sEPR6}) be positive semi-definite. Then
$\Phi_{\rho}^{ba}$ and $\Phi_{\rho}^{ab}$ in (\ref{sEPR6}) are
completely copositive linear mappings. If
\begin{equation} \label{sEPR2}
\rho = \sum |\psi_j \rangle\langle \psi_j|, \quad
s_j^{ba} := s_{\psi_j}^{ba}
\end{equation}
with vectors $\psi_j  \in \cH^{\textsc{AB}}$, then
\begin{eqnarray}
\Phi^{ba}_{\rho}(X^{\kA}) = \sum s_j^{ba} (X^{\kA})^{\dag} s_j^{ab}
\; , \label{sEPR3a} \\
\Phi^{ab}_{\rho}(X^{\kB}) = \sum s_j^{ab} (X^{\kB})^{\dag} s_j^{ba}
\; , \label{sEPR3b}
\end{eqnarray}
where $X^{\kA} \in \cB(\cH^{\kA})$, $X^{\kB} \in \cB(\cH^{\kB})$.
\end{thm}
Proof: By (\ref{sEPR6}) the maps are linearly dependent on $\rho$.
Therefore it suffices to prove (\ref{sEPR3a}) and (\ref{sEPR3b})
in case $\rho$ is of the form $|\psi \rangle\langle \psi|$,
$\psi$ a unit vector. With this specification the left of
(\ref{sEPR6}) becomes $\langle \psi, (X \otimes Y) \psi \rangle$.
Assuming $\psi$ Schmidt decomposed, (\ref{EPR8a}), one can rely
on (\ref{EPR8}) so that
\begin{displaymath}
\psi = \sum \sqrt{p_j} \phi_j^{\kA} \otimes \phi_j^{\kB}, \quad
s_{\psi}^{ba} \phi_j^{\kA} = \sqrt{p_j} \phi_j^{\kB}, \quad
s_{\psi}^{ab} \phi_j^{\kB} = \sqrt{p_j} \phi_j^{\kA} \: .
\end{displaymath}
Now $\psi$ can be represented by
\begin{displaymath}
\psi = \sum s_{\psi}^{ab} \phi_j^{\kB} \otimes \phi_j^{\kB}
= \sum \phi_j^{\kA} \otimes s_{\psi}^{ba} \phi_j^{\kA} \; .
\end{displaymath}
One of these relations will be inserted into the left of
(\ref{sEPR6}) to get
\begin{displaymath}
\langle \psi, (X \otimes Y) \psi \rangle = \sum \langle
s_{\psi}^{ab} \phi_j^{\kB} \otimes \phi_j^{\kB}, (X^{\kA} \otimes
X^{\kB}) s_{\psi}^{ab} \phi_k^{\kB} \otimes \phi_k^{\kB} \rangle
\end{displaymath}
and the better structured equation
\begin{displaymath}
\langle \psi, (X \otimes Y) \psi \rangle = \sum \langle
s_{\psi}^{ab} \phi_j^{\kB}, X^{\kA} s_{\psi}^{ab} \phi_k^{\kB}
\rangle \, \langle \phi_j^{\kB}, X^{\kB} \phi_k^{\kB} \rangle
\end{displaymath}
Next (\ref{rule2e}) will be applied to the antilinear operator
$X^{\kA} s_{\psi}^{ab}$, i.~e.
\begin{displaymath}
\langle \psi, (X \otimes Y) \psi \rangle = \sum \langle
\phi_k^{\kB}, s_{\psi}^{ba}(X^{\kA})^{\dag} s_{\psi}^{ab}
\phi_j^{\kB} \rangle \,
\langle \phi_j^{\kB}, X^{\kB} \phi_k^{\kB} \rangle \; .
\end{displaymath}
Summing up over $k$ yields
\begin{displaymath}
\langle \psi, (X \otimes Y) \psi \rangle = \sum_j \langle
(X^{\kB})^{\dag} \phi_j^{\kB} ,
s_{\psi}^{ba}(X^{\kA})^{\dag} s_{\psi}^{ab} \phi_j^{\kB} \rangle
\end{displaymath}
and this nothing than
\begin{displaymath}
\langle \psi, (X \otimes Y) \psi \rangle = \T \, Y
s_{\psi}^{ba}(X^{\kA})^{\dag} s_{\psi}^{ab} \; .
\end{displaymath}
The other part of (\ref{sEPR6}) is verified by a similar
exercise.\\
{\bf Remarks:} (a) The Hermitian conjugate ensures linearity.\\
(b) One may compare the theorem with Jamio{\l}kowski's isomorphism
\cite{Jamiol72}, \cite{BZ06}, to notice the difference enforced by
complete copositivity. The latter comes with an
``hidden antilinearity''.

\section{Antilinearity in quantum teleportation} \label{C10}
The quantum teleportation protocol was discovered by
{\it Bennett et al} \cite{BBCJPW93}. The protocol has been
extended in various directions and applied to build more complex
quantum information tasks. An overview is in {\it Nielsen and
Chuan} \cite{NC00} and most other QIT textbooks. There are many
papers concerning quantum teleportation. See \cite{Br98},
\cite{Horo98}, \cite{MH99}, \cite{We00}, \cite{AF00}, and
\cite{RHFB01} for example,

Quantum teleportation consists of some preliminaries and a
description how to do certain operations.

Generally, the basic structure is a tripartite quantum system
\begin{equation} \label{tele1}
\cH^{\textsc{ABC}} =
\cH^{\kA} \otimes \cH^{\kB} \otimes \cH^{\kC} \; .
\end{equation}
One starts with two vectors
\begin{equation} \label{tele2}
\psi \in \cH^{\kA} \otimes \cH^{\kB}
 , \quad
\varphi \in \cH^{\kB} \otimes \cH^{\kC} \; ,
\end{equation}
and given {\em input} vectors
\begin{equation} \label{tele3}
\phi^{\rm in} \in \cH^{\kA} , \quad \varphi^{\rm in} :=
\phi^{\rm in} \otimes \varphi \in \cH^{\textsc{ABC}} \; .
\end{equation}
Notice $d^{\kA} = \dim \cH^{\kA}$,
$d^{\kB \kC}$ is $\dim \cH^{\kB \kC}$, and so on.

In what follows the main emphasis is to what will be called
``teleportation map'': The vectors (\ref{tele2}) induce the maps
$s_{\psi}^{ba}$ and $s_{\varphi}^{cb}$ from $\cH^{\kA}$ into
$\cH^{\kB}$ and from $\cH^{\kB}$ into $\cH^{\kC}$ respectively.
Therefore there is a mapping
\begin{equation} \label{tele4}
t^{ca} \equiv t^{ca}_{\varphi, \psi} :=
 s_{\varphi}^{cb} \, s_{\psi}^{ba}
\end{equation}
which will be called the {\em teleportation map} associated to
the vectors (\ref{tele2}). This notation will be justified below.

$t^{ca}$ transports any $\phi^{\rm in} \in \cH^{\kA}$ to an
output vector $\phi^{\rm out} \in \cH^{\kC}$,
\begin{equation} \label{tele5}
\phi^{\rm in} \,\to \, \phi^{\rm out} :=
t^{ca} \varphi^{\rm in} \;.
\end{equation}
As a product of two antilinear maps, the teleportation map is
linear. The teleportation map (\ref{tele4}) depends linearly on
$\varphi$ and antilinearly on $\psi$.

For the Hermitian adjoint one gets
\begin{equation} \label{tele6}
(t^{ca}_{\varphi, \psi})^{\dag} = t^{ac}_{\psi, \varphi} =
s_{\psi}^{ab} \,  s_{\varphi}^{bc}   \; .
\end{equation}

$\varphi$ is given in advance.
The symmetry between (\ref{tele4}) and (\ref{tele6}) will be
broken in the teleportation protocol by a measurement by which
$\psi$ in (\ref{tele2}) and (\ref{tele4}) will be either
prepared or not. {\em This way} the irreversibility of quantum
teleportation comes into the game: The teleportation map only
applies if $\psi$ has been prepared. See the next subsection.

The teleportation map has been introduced by the present author,
see in \cite{Uh00b} and \cite{CrUh09}. A more recent work on
these questions is in Bertlmann et al \cite{BNT12}.

\subsection{Quantum teleportation} \label{C10.1}
Let the A-system be in a {\em not necessarily known} pure state,
represented by $\phi^{\rm in}$. In the BC-system a pure state,
is given in advance by the vector $\varphi =
\varphi^{\kB \kC}$. This vector {\em must be known.} These
assumption are allowed by the mutual independence of the A- and
the BC-system.

The next step is to assume a basis
$\{ \psi_1^{\textsc{AB}}, \dots, \psi_n^{\textsc{AB}} \}$
defining a von Neumann measurement preparing one of the
basis states. ((Indeed, it would suffice to instal a L\"uders
measurement with the rank one projection onto $\psi$ and the
projection operator onto the orthogonal complement of $\psi$.))
\medskip

Let $\psi = \psi^{\textsc{AB}}$ be an element of the basis
$\{ \psi_j^{\textsc{AB}} \}$,
\begin{equation} \label{telebasis}
\psi \in \{ \psi_1^{\textsc{AB}}, \psi_2^{\textsc{AB}}, \dots,
\psi_n^{\textsc{AB}} \} , \quad n = d^{\kA} d^{\kB}   \; ,
\end{equation}
and assume that the measurement reports that just this state is
prepared. Then, similar to (\ref{EPR1}),
\begin{equation} \label{tele7}
(|\psi \rangle\langle \psi |
\otimes \1^{\kC}) \, \varphi^{\rm in}
= \psi \otimes \phi^{\rm out} , \quad
\varphi^{\rm in} = \phi^{\rm in} \otimes \varphi \; .
\end{equation}
\begin{thm}[composition law]
There is a unique linear map
\begin{equation} \label{tele8}
 \cH^{\kA}  \to \cH^{\kC}
\end{equation}
defined by (\ref{tele7}) and expressed as in (\ref{tele4}) by
\begin{equation} \label{tele9}
\phi^{\rm in} \, \to \, \phi^{\rm out} =
 s_{\varphi}^{cb} \, s_{\psi}^{ba} \phi^{\rm in}
\equiv t^{ca}_{\varphi, \psi} \phi^{\rm in}  \; ,
\end{equation}
\end{thm}
The theorem and the teleportation map is in \cite{Uh00a}.\\
Proof: From (\ref{tele7}) one infers: The vector
$\phi^{\rm out} \in \cH^{\kC}$ is uniquely determined by the
vectors $\phi^{\rm in}$, $\varphi$, and $\psi$, The vector and
$\phi^{\rm out}$ {\em depends linearly} on $\phi^{\rm in}$ and
on $\varphi$. Its dependence on $\psi$ is antilinear.\\
Choosing in $\cH^{\kB}$ a basis
$\{\psi_1^{\kB}, \psi_2^{\kB}, \dots\}$ one may write
\begin{displaymath}
\varphi =
\sum \psi_j^{\kB} \otimes s_{\varphi}^{cb}\psi_j^{\kB}
\end{displaymath}
to resolve the left side of (\ref{tele7}) into
\begin{equation} \label{tele10}
(|\psi \rangle\langle \psi | \otimes \1^{\kC}) \,
\phi^{\rm in} \otimes \sum_j
\psi_j^{\kB} \otimes s_{\varphi}^{cb}\psi_j^{\kB} =
\sum_j |\psi \rangle\langle \psi | (\phi^{\rm in} \otimes
\psi_j^{\kB} \otimes s_{\varphi}^{cb}\psi_j^{\kB} ) \,.
\end{equation}
Now consider the expressions
\begin{displaymath}
|\psi \rangle\langle \psi| \,
(\phi^{\rm in} \otimes \psi_j^{\kB}) =
\langle \psi ,  \phi^{\rm in} \otimes \psi_j^{\kB} \rangle
\, \psi   \; .
\end{displaymath}
Substitute into (\ref{tele10}) yields
\begin{displaymath}
(|\psi \rangle\langle \psi | \otimes \1^{\kC}) \,
(\phi^{\rm in} \otimes \varphi) = \sum_j
\langle \psi ,  \phi^{\rm in} \otimes \psi_j^{\kB} \rangle
\, \psi \otimes s_{\varphi}^{cb}\psi_j^{\kB} ) \,.
\end{displaymath}
On the right one has $\psi \otimes \phi^{\rm out}$. Hence
\begin{equation} \label{tele11}
\phi^{\rm out} = s_{\varphi}^{cb} \sum_j
\langle \phi^{\rm in} \otimes \psi_j^{\kB} , \psi \rangle
\, \psi_j^{\kB}
\end{equation}
By using (\ref{EPR6}) one gets
\begin{displaymath}
\langle \phi^{\rm in} \otimes \psi_j^{\kB} , \psi \rangle
= \langle \psi_j^{\kB} , s_{\varphi}^{ba} \phi^{\rm in}
\rangle \, \psi_j^{\kB} \; .
\end{displaymath}
Now, substituting it into (\ref{tele11}), one arrives at
\begin{displaymath}
\phi^{\rm out} = s_{\varphi}^{cb} \sum_j
\langle \psi_j^{\kB} , s_{\varphi}^{ba} \phi^{\rm in} \rangle
\, \psi_j^{\kB} = s_{\varphi}^{cb} s_{\varphi}^{ba}
\phi^{\rm in}
\end{displaymath}
and the assertion has been proved.

\noindent {\bf Remarks:} \\
1.) \, Comparing (\ref{tele7}) with (\ref{EPR1}) one gets the
identity
\begin{equation} \label{tele12}
t^{ca}_{\psi, \varphi}  \phi^{\rm in} =
s^{ ac, c}_{\chi} \psi , \quad
|\chi\rangle :=
\varphi^{\rm in} = \phi^{\rm in} \otimes \varphi \; .
\end{equation}
2.) \, Equation (\ref{tele9}) describes the possible result of a
measurement on the rather particular tripartite state
(\ref{tele7}), $\varphi^{\rm in} = \phi^{\rm in} \otimes \varphi$.
One may ask what happens in case of a general state vector.
To see it, one can choose any product decomposition
\begin{equation} \label{tele7a}
|\varphi^{\textsc A B C}\rangle := \sum_i
|\phi^{\textsc A}_i \rangle \otimes |\varphi^{\textsc B C}_i \rangle
\end{equation}
to get
\begin{displaymath}
( |\psi \rangle \langle
\psi| \otimes 1^{\textsc C} ) \, |\phi^{\textsc A B C} \rangle
= \sum_i |\psi\rangle \otimes t^{\textsc C A}_i
| \phi^{\textsc A}_i \rangle
\end{displaymath}
or, by the composition law, and
abbreviating the s--map belonging to
$|\varphi^{\textsc B C}_i\rangle$ by $s^{\textsc C B}_i$,
\begin{equation} \label{tele9a}
( |\psi \rangle \langle
\psi| \otimes 1^{\textsc C} ) \, |\phi^{\textsc A B C} \rangle
= |\psi\rangle \otimes \sum_i s^{\textsc C B}_i
\cdot s^{\textsc B A}_{\psi} \, | \phi^{\textsc A}_i \rangle
\end{equation}

\subsubsection{The trace norm} \label{C10.1.1}
If the Hilbert spaces are infinite dimensional the maps
of type $s^{xy}_{\psi}$ are Hilbert Schmidt ones.
Therefore, the teleportation maps must be of trace class.
Thus, also in finite dimensions, it seems quite
natural to use the trace norm to estimate them. It is,
\cite{CrUh09},
\begin{equation} \label{tele13}
\parallel t^{ca}_{\psi, \varphi} \parallel_1 := {\rm tr} \,
[ (t^{ca}_{\psi, \varphi})^{\dag} t^{ca}_{\psi, \varphi} ]^{1/2}
\end{equation}
Let us call $\rho_{\psi}$ the reduction of
$|\psi\rangle\langle\psi|$ to $\cH^{\textsc B}$, and
$\rho_{\varphi}$ the reduced density operator of
$|\varphi\rangle\langle\varphi|$ to the same Hilbert space.
The aim is to prove that (\ref{tele13}) depends on these data
only, i.~e. by knowing these two density operators on
$\cH^{\textsc B}$ the trace norm in question can be computed.
The result is the (square root) fidelity which is the square
root of the generalized transition probability:
\begin{equation} \label{teleTP}
\parallel t^{ca}_{\psi, \varphi} \parallel_1 =
\T \, [ \sqrt{\varrho_{\psi}} \, \varrho_{\varphi} \,
\sqrt{\varrho_{\psi}} ]^{1/2} \equiv
F(\varrho_{\psi}, \varrho_{\varphi})
\end{equation}
For the proof the case of equal dimensions of the three Hilbert
spaces and of maximal Schmidt rank of $\psi$ and $\varrho$ is
assumed. One gets by (\ref{tele9}) and by the help of
(\ref{EPRP3a})
\begin{displaymath}
(t^{ca}_{\psi, \varphi})^{\dag} t^{ca}_{\psi, \varphi} =
s_{\psi}^{ab} s_{\varphi}^{bc} s_{\varphi}^{cb} s_{\psi}^{ba}
= s_{\psi}^{ab} \rho_{\varphi} s_{\psi}^{ba}
\end{displaymath}
Again, using (\ref{EPRP3a}) appropriately, one obtains
\begin{displaymath}
(t^{ca}_{\psi, \varphi})^{\dag} t^{ca}_{\psi, \varphi} =
= j_{\psi}^{ab} ( \sqrt{\rho_{\psi}} \rho_{\varphi}
\sqrt{\rho_{\psi}} ) j_{\psi}^{ba} \; .
\end{displaymath}
The assumptions imply that $j_{\psi}^{ab}$ and $j_{\psi}^{ba}$
are isometries, and one of them is the inverse of the other.
Therefore
\begin{displaymath}
j_{\psi}^{ba} (t^{ac}_{\psi, \varphi} t^{ca}_{\psi, \varphi})
j_{\psi}^{ab}  = \sqrt{\rho_{\psi}} \rho_{\varphi}
\sqrt{\rho_{\psi}} \; .
\end{displaymath}
This equation proves an even stronger result than (\ref{tele13}):
\begin{lem}
The singular values of $t^{ac}_{\psi, \varphi}$ and those of
$t^{ca}_{\psi, \varphi}$ coincide with the eigenvalues of
$(\sqrt{\rho_{\psi}} \rho_{\varphi} \sqrt{\rho_{\psi}})^{1/2}$.
\end{lem}
Indeed, the 1- or trace norm of $t^{ac}_{\psi, \varphi}$ is the
is the sum of their singular values so that the lemma
establishes (\ref{teleTP}).

Another estimate, seen from the proof above, reads
\begin{equation} \label{telePa}
\parallel \phi^{\rm out}\parallel \leq
\parallel (\sqrt{\rho_{\psi}} \rho_{\varphi}
\sqrt{\rho_{\psi}})^{1/2} \parallel_1 \cdot
\parallel \phi^{\rm in} \parallel
\end{equation}
A comparison of different entanglement measures in quantum
teleportation is in \cite{SABP12}. The authors allow
$\varphi^{\textsc{BC}}$ to be mixed.

\subsection{Distributed measurements} \label{C10.2}
Looking at the quantum teleportation composition law
(\ref{tele9}) one may ask whether there is a similar
structure in multi-partite systems. An obvious ansatz is the
following: Assume $\cH$ is the direct product of $n+1$ Hilbert
spaces $\cH^j$. Then any $\varphi_{k+1,k} \in \cH^k \otimes
\cH^{k+1}$ corresponds uniquely to an antilinear map $s^{k+1, k}$
from $\cH^k$ into $\cH^{k+1}$. Hence there is a map $t^{n+1,1}$
defined by $t^{n+1,1} = s^{n+1,n} \cdots s^{2,1}$. This is a
composition of $n$ antilinear maps.

$t^{n+1,1}$ is linear and ``teleportation-like'' if $n$ is even.
It is antilinear and ``EPR-like'' for  $n$ odd. Together with
suitable measurements one gets something like a
{\em distributed teleportation} or a {\em distributes EPR scheme.}

The case $n=4$ has been treated in \cite{Uh00b} and
\cite{CrUh09} and will be outlined below. In \cite{CrUh09} and,
more recently in \cite{BNT12}, the case $n=3$, in which $\cH$
consists of four parts, has been considered.

\subsubsection{The case of five subsystems}
The quantum system in question is
\begin{equation} \label{dt1}
\cH = \cH^1 \otimes \cH^2 \otimes \cH^3 \otimes
\cH^4 \otimes \cH^5 \, .
\end{equation}
The input is an unknown vector $\phi^1 \in \cH^1$, the
ancillary vectors are selected from the $23$- and the
$45$-system,
\begin{equation} \label{dt2}
\varphi^{2,3} \in \cH^2 \otimes \cH^3 , \quad
\varphi^{4,5} \in \cH_4 \otimes \cH_5,
\end{equation}
and the vector of the total system we are starting with is
\begin{equation} \label{dt3}
 \varphi^{1,2,3,4,5} = \phi^1 \otimes
\varphi^{2,3} \otimes \varphi^{4,5} \, .
\end{equation}
The channel is triggered by measurements in the $1,2$- and
in the $3,4$-system.
Suppose these measurements  prepare successfully the vector
states
\begin{equation} \label{dt4}
\psi^{1,2} \in  \cH^1 \otimes \cH^2 , \quad
\psi^{3,4} \in \cH^3 \otimes \cH^4 \, .
\end{equation}
Then we get the relation
\begin{equation} \label{dt5}
( |\psi^{1,2} \rangle\langle \psi^{1,2}| \otimes
 |\psi^{3,4} \rangle\langle \psi^{3,4}| \otimes \1^5 )
\varphi^{1,2,3,4,5}
= \psi^{1,2} \otimes \psi^{3,4} \otimes \phi^5
\end{equation}
and the vector $\phi^1$ is mapped onto $\phi^5$,
$\phi^5 = t^{5,1} \phi^1$.
Introducing the maps $s^{k,k+1}$ corresponding to the vectors
\begin{displaymath}
\psi^{1,2}, \quad \varphi^{2,3}, \quad \psi^{3,4}, \quad
\varphi^{4,5},
\end{displaymath}
the {\em factorization rule} becomes
\begin{equation} \label{dt6}
\phi^5 = t^{5,1} \phi^1, \quad
{\bf t}^{5,1} = s^{54} \, s^{4,3} \,
s^{3,2} \, s^{2,1} \, .
\end{equation}

 \subsubsection{The EPR--like case of four subsystems}
Remaining within the previous setting, ignoring however the
Hilbert space $\cH^1$. The following is an extended EPR
protocol. Instead of giving an (unknown) input vector out of
$\cH^2$, there is a measurement on the system carried in $\cH^2$.
The state vector carrying the entanglement reads
\begin{displaymath}
 \varphi^{2,3,4,5} = \varphi^{2,3} \otimes \varphi^{4,5}
\in \cH^2 \otimes \cH^3 \otimes \cH^4 \otimes \cH^5 \; .
 \end{displaymath}
A test (a measurement) is performed to check whether $\psi^{3,4}$
is prepared or not.

Let the answer be YES. Then the subsystems 2,3 and 4,5 become
disentangled both. The state of the 3,4 system changes to
 $\psi^{3,4}$. The  previously unentangled systems 2,5  will
become entangled. Indeed, the newly prepared state is
 \begin{equation} \label{dt7}
 \chi^{2,3,4,5} :=
 (\1^2 \otimes |\psi^{3,4} \rangle\langle \psi^{3,4}|
 \otimes \1^5 ) \, \varphi^{2,3,4,5} \, .
 \end{equation}
If there is a decomposition
\begin{displaymath}
 \psi^{3,4} = \sum \lambda_j \phi^3_j \otimes \phi^4_j
\end{displaymath}
of $\psi^{3,4}$, one obtains
\begin{displaymath}
 \chi^{2,3,4,5} = \sum \lambda_j \lambda_k
 [(\1^2 \otimes |\phi^3_j \rangle\langle \phi^3_j |) \varphi^{2,3}]
 \otimes
 [( |\phi^4_j \rangle\langle \phi^4_j | \otimes \1^5) \varphi^{4,5}]
 \, .
\end{displaymath}
Let $s^{2,3}$ and $s^{4,5}$ denote the antilinear mappings
defined by $\varphi^{2.3}$ and $\varphi^{3,4}$  respectively.
They allow to rewrite $\chi^{2,3,4,5}$ as
\begin{displaymath}
 \chi^{2,3,4,5} =  \sum \lambda_j \lambda_k (s^{2,3} \phi^3_k
\otimes \phi^3_j)  \otimes  (\phi^4_j \otimes s^{5,4} \phi^4_k)
\end{displaymath}
 which is equal to
 \begin{equation} \label{dt8}
 \chi^{2,3,4,5} = \sum \lambda_k (s^{2,3} \phi^3_k) \otimes
 \psi^{3,4} \otimes ({\bf s}^{5,4} \phi^4_k) \, .
 \end{equation}
 The Hilbert space $\cH^3 \otimes \cH^4$ is decoupled
 from $\cH^2$ and $\cH^5$. The vector state of the latter
 can be characterized by a map from $\cH^3 \otimes \cH^4$
 into $\cH^2 \otimes \cH^5$.
 \begin{equation} \label{dt9}
 \varphi^{23} := ( s^{2,3} \otimes s^{5,4} ) \,
 \psi^{3,4} \: ,
 \end{equation}
 indicating how the entanglement within the 2,5-system is
produced by entanglement swapping, and how the three vectors
involved come together to achieve it.

\section{Appendix: Antilinear operator spaces} \label{CA}
The subspaces of $\cB(\cH_{\rc})$ and their relations to
completely copositive maps are topics calling for attention
and research. Though there are many similarities to spaces of
linear operators, (see \cite{Paulsen02} for an introduction
to operator spaces), there are remarkable differences also:
There does not exist a substitute for the identity $\1$ in linear
spaces of antilinear operators. There is, however, the canonical
Hermitian form, see \ref{C2.4}, which does {\em not} depend on
the scalar product of $\cH$. Further one may hope for interesting
factorizations of operator spaces as products of antilinear ones.
Here only a first impression can be gained. (To my
knowledge there is no systematic exploration presently.)

Following \cite{Paulsen02} a complex-linear subspace of
$\cB(\cH)_{\rc}$ will be called an {\em antilinear operator space}
or an {\em AO-space} for short. If the operator space contains
with any $\vartheta$ also $\vartheta^{\dag}$, it is an
{\em antilinear operator system} or an {\em AO-system.}
\medskip

Let $\cM$ be an AO-space. Its {\em canonical form}, denoted by
$(.,.)_M$, is the restriction onto $\cM$ of the canonical form
(\ref{Hform1}). There are decompositions
\begin{displaymath}
{\cM = \cM^+ + \cM^- + \cM^0}, \eqno(a1)
\end{displaymath}
such that the restriction of the canonical form onto $\cM^+$
is positive definite, onto $\cM^-$ is negative definite, and
is vanishing on $\cM^0$. Their dimensions are the inertia, see
\cite{Horn90}, of the Hermitian form $(.,.)_M$, Hence
\begin{displaymath}
\dim \cM = \dim \cM^+ + \dim \cM^- + \dim \cM^0 \; . \eqno(a2)
\end{displaymath}
A peculiarity is that the inertia of $(.,.)_M$ do not depend on
the scalar product of $\cH$. Indeed the canonical form
(\ref{Hform1}) posses just that property.

Fixing an arbitrary Hilbert scalar product on $\cM$ there are
$\dim \cM$ mutual orthogonal elements $\vartheta_k \in \cM$
fulfilling

a) The first $\dim \cM^+$ elements constitute a basis of $\cM^+$,

b) The next $\dim \cM^-$ elements generate $\cM^-$,

c) The remaining $\dim \cM^0$ elements span $\cM^0$,

More can be said about the decomposition if $\cM_T$ is an
antilinear operator system. In this case, $\cM_T$
contains with $\vartheta$ necessarily $\vartheta^{\dag}$ and,
hence, $\vartheta \pm \vartheta^{\dag}$. This simple observation
results in the AO--systems
\begin{displaymath}
\cM_T = \cM_T^{+} \oplus \cM_T^{-},         \eqno(a3)
\end{displaymath}
and in the nice property
\begin{displaymath}
\cM_T^{+} = \cM_T \cap \cB(\cH)^{+} , \quad
\cM_T^{-} = \cM_T \cap \cB(\cH)^{-} \; .         \eqno(a4)
\end{displaymath}
\medskip

If an AO-space carries a scalar product $\langle. , .\rangle_T$,
it will be called {\em antilinear operator Hilbert space} or
{\em AOH-space} for short.

\noindent {\bf Lemma A.1.}\\
Let $\cM$ be an AOH-space and $\vartheta_1,\dots,\vartheta_m$,
$m= \dim \cM$, be a basis of $\cM$ with respect to the scalar
product $\langle .,. \rangle_T$. Then the map
\begin{displaymath}
X \to T(X) = \sum_{j=1}^m \vartheta_j X^{\dag} \vartheta_j^{\dag}
\eqno{a5}
\end{displaymath}
does not depend on the choice of the basis.
\medskip

Proof: Let $\vartheta'_1, \dots$ be another basis. There is a
unitary matrix $u_{jk}$ such that $\vartheta_j = \sum_k u_{jk}
\vartheta'_k$. Due to this relation the sum in (a1) is
replaced by
\begin{displaymath}
\sum_{j} \sum_{kl}
u_{jk} \vartheta_k X^{\dag} u_{jl} \vartheta_l^{\dag}
\end{displaymath}
because the Hermitian adjoint acts linearly on antilinear
operators. By placing $u_{jl}$ on the left changes it to
$u_{jl}^{*}$. Now $\sum_j u_{jk} u_{jl}^{\dag} = \delta_{kl}$
proves the assertion.

$T$ in eq. (a5) is a completely copositive map with length
$\dim \cM$. On the other hand, given such a map $T$ as in
eq. (a5), there is a uniquely associated AOH-space $\cM_T$.
It is the AO-space $\cM$ generated by the antilinear operators
$\vartheta_j$, $j=1, \dots, m$ in a representation eq. (a5).
Requiring the $\vartheta_j$ to become an orthonormal basis fixes
a scalar product which makes $\cM$ an AOH-space $\cM_T$, Hence

\noindent {\bf Proposition A.2.}\\
There is a bijection between AOH-spaces and completely copositive
maps.
\begin{displaymath}
 T \,\Longleftrightarrow \,\{ \cM_T, \,\langle.,.\rangle_T\}\:.
\eqno(a6)
\end{displaymath}
If $\vartheta_j$, $j= 1, \dots \dim \cM_T$ is an Hilbert
basis of $\cM_T$, then $T$ as in eq. (a5) is uniquely
associated to the given AOH-space and vice vera.

The correspondence mimics similar constructs for completely
positive maps.

\noindent {\bf Examples} \\
(1) Let $\cM_T = \cB(\cH)_{\rc}^{+}$.  The relevant scalar
product is the restriction to $\cB(\cH)_{\rc}^{+}$ of the
canonical form (\ref{Hform1}), i.~e.
\begin{displaymath}
\langle \vartheta'', \vartheta' \rangle_T := \T \, \vartheta'
\vartheta'' \equiv (\vartheta'', \vartheta') \; .
\end{displaymath}
$T^{+} := T_M$ is defined by eq. (a5). According to lemma
A.1., $T^{+}$  can be computed with any basis of
$\cB(\cH)_{\rc}^{+}$. One obtains
\begin{displaymath}
 T^{+}(X) =  \frac{(\T \, X) \1 + X}{2} \; , \eqno(a7)
\end{displaymath}
$T^{+}$ is simultaneously completely copositive and completely
positive. Notice also
\begin{displaymath}
\T \, T^{+}(X) = \frac{d+1}{2} \T \, X , \quad
T^{+}(\1) =  \frac{d+1}{2} \1 \; . \eqno(a8)
\end{displaymath}
(2) Let $\cM = \cB(\cH)_{\rc}^{-}$ and
\begin{displaymath}
\langle \vartheta'', \vartheta' \rangle_T = - \T \, \vartheta'
\vartheta'' \; .
\end{displaymath}
The length of any basis is $d(d-1)/2$. Let $T_M = T^{-}$ then
\begin{displaymath}
 T^{-}(X) = \frac{(\T \, X) \1 - X}{2}  \; . \eqno(a9)
\end{displaymath}
This completely copositive map is not even 2-positive. One knows
by the work of M.-D. Choi that one can construct examples of
k-positive maps by convexly combining $T^{-}$ and $T^{+}$ for
all relevant $k$.\\
(3) Let $\phi_1, \dots, \phi_d$ be a basis of $\cH$. Let
$\cM$ be spanned by the operators
\begin{displaymath}
|\phi_{2m} \rangle \langle \phi_{2n+1}|_{\rc}, \quad
2m \leq d, \quad 2n+1 \leq d \;.  \eqno(a10)
\end{displaymath}
Then, as see from (\ref{mranka}), $\cM$ is of type $\cM^{0}$.
If $d$ is even, then $\dim \cM = d^2/4$. It is
$\dim \cM = (d^2-1)/4$ for $d$ odd.

$\cM$, as defined above, is an algebra. The same is with
$\cM^{\dag}$. Clearly $\cM \cap \cM^{\dag}$ consists of the null
operator only. The product spaces $\cM \, \cM^{\dag}$ and
$\cM^{\dag} \cM$ are "almost" operator systems: They do not
contain the identity operator.
\medskip

\underline{Acknowledgement:} I like to thank {\em Bernd Crell} and
{\em Meik Hellmund} for helpful remarks and support,
{\em Stephan R.~Garcia} and  {\em Mihai Putinar}
for calling my attention to the ``Finish School'',
and, last but not least, {\em ShaoMing Fei} for invitating me to publish in
SCIENCE CHINA Physics, Mechanics $\&$ Astronomy.

\end{document}